\documentclass{article}

\pdfoutput=1
\usepackage{graphicx,amsmath,amssymb,epsf} \usepackage{latexsym,bm,
  slashed} \usepackage{xcolor} \definecolor{dark}{rgb}{0.10,0.2,0.3}
\definecolor{magenta}{rgb}{0.7,0.1,0.3}
\definecolor{purpure}{rgb}{0.5,0.15,0.3}
\usepackage[font=small,format=plain,labelfont=bf,up,textfont=it,up]{caption}
\usepackage{hyperref, cite} \hypersetup{colorlinks, citecolor=blue,
  filecolor=blue, linkcolor=magenta,
  urlcolor=purpure,hyperfootnotes=true,pdftex} 

\usepackage{subcaption}

\title{Multiperipheral final states \\in crowded twin-jet events at the LHC} 
\author{N. Bethencourt de Le{\'o}n$^{1}$, G. Chachamis$^2$, A. Sabio Vera$^{1,3}$\\ 
\\
\small $^1$ Instituto de F{\'\i}sica Te{\'o}rica UAM/CSIC, Nicol{\'a}s Cabrera 15, E-28049 Madrid, Spain.\\
\small $^2$ Laborat{\' o}rio de Instrumenta\c{c}{\~ a}o e F{\' \i}sica Experimental de Part{\' \i}culas (LIP),\\
\small Av. Prof. Gama Pinto, 2, P-1649-003 Lisboa, Portugal.\\
\small $^3$ Theoretical Physics Department, Universidad Aut{\' o}noma de Madrid, E-28049 Madrid, Spain.\\} 

\begin{document} 

\maketitle 

\begin{abstract}
The 13 TeV run of the LHC has provided a unique opportunity to explore multi-jet final states with 
unprecedented accuracy. An interesting region for study is that of events where 
one jet is tagged in the forward direction and another one in the backward direction and a plethora of low energy mini-jets populate the possibly large rapidity span in between them. Since the number of these events is very high, it is possible to introduce stringent constraints on the transverse momentum of the two leading jets which can be kept in small windows not very different from each other, defining what we call ``twin jets". The associated ``crowd" of mini-jets can also have a restricted span in transverse momentum. The study of these events for a fixed multiplicity is an ideal playground to investigate different models of multi-particle production in hadron-hadron collisions. We set up an exploratory analysis by using an ancient model of Chew and Pignotti to describe the gross features one can expect for the structure of single and double differential-in-rapidity cross sections and for particle-particle rapidity correlations when the longitudinal phase space completely decouples from the transverse degrees of freedom. 
\end{abstract}

\section{Introduction}

A fascinating open problem in Quantum Chromodynamics (QCD) is the correct understanding of multi-particle events generated in the interaction at high center-of-mass energies of leptons with hadrons or hadrons with hadrons since its theoretical description is quite cumbersome when higher order quantum corrections are taken into account. Experimental data on multi-particle hadroproduction in the last 5-6 decades 
have continuously reinforced the concept that the final state particles tend to merge into correlated clusters~\cite{Dremin:1977wc}.
The investigation of exclusive quantities and correlations between final state particles has been
offering important information about the strong force in high-energy interactions studied in collider experiments, well before the advent of QCD. Evidently, high-order correlations among emitted particles have been found in all types of hadron collisions as manifestations of multiplicity fluctuations. Furthermore, particle correlations summarize important properties of jets without being too sensitive
to missing soft particles in the jet\cite{Tannenbaum:2005by} (see also the discussion in the introduction of Ref.~\cite{SanchisLozano:2008te}).

There are many different effective approaches to multi-particle hadroproduction which are mainly based on the resummation of the leading contributions to the scattering amplitudes associated to these type of events (e.g. DGLAP~\cite{Gribov:1972ri,Gribov:1972rt,Lipatov:1974qm,Altarelli:1977zs,Dokshitzer:1977sg}, BFKL~\cite{Kuraev:1977fs,Kuraev:1976ge,Fadin:1975cb,Lipatov:1976zz,Lipatov:1985uk}, CCFM~\cite{Ciafaloni:1987ur,Catani:1989yc}, Linked Dipole Chain~\cite{Gustafson:1986db,Gustafson:1987rq,Andersson:1995ju}, Lund Model~\cite{Andersson:1998tv}). The correct understanding of these 
issues is very important since it is related to the reduction of  the theoretical uncertainty in the 
parton distribution functions of hadrons at small values of Bjorken $x$ and also to Standard Model backgrounds for new physics searches with high multiplicity at the Large Hadron Collider (LHC). 

The latest 13 TeV run of the LHC has generated a large amount of data which is needed to understand in detail. In the present work we would like to highlight a type of final state configurations which are very important to fix the range of applicability of multi-particle production 
models. They correspond to events with several jets where the two outermost in rapidity jets are found one in the forward and one in the backward regions and can be cleanly tagged since their transverse momentum is relatively large. 
In order to pin down events with low momentum transfer we propose to focus on events where these two jets have similar transverse momentum in what we call ``twin-jet" configurations. Together with these twin-jets a set of low energy ``mini-jets", with their $p_T$ also constrained within a limited window and which we can refer to as the ``crowd", are also present. This corresponds to a subset of the so-called Mueller-Navelet jets~\cite{Mueller:1986ey}. We require a very similar $p_T$ for the twin-jets (although not within overlapping range) to enforce a {\it de facto} symmetry on their influence upon the crowd mini-jets.

One of the advantages of the 
13 TeV run of the LHC is that these constrained events are numerous and their statistical 
analysis can reach a good level of accuracy. To improve their study at future LHC runs with higher energies it will be important to have some periods of dedicated low luminosity runs available. 
  
Our task in the present work is to investigate some of the gross features which can be expected in these processes. Since the most striking characteristics of the final states in multiperipheral models and within the cluster concept are present in the rapidity space, with their origin stemming from the decoupling of transverse coordinates from longitudinal ones due to kinematical reasons, we will focus this first exploratory analysis on differential cross sections and particle-particle correlations in rapidity space. 

We introduce the method of analysis in Section 2, together with the description of single differential distributions. An expansion on a finite basis of Chebyshev polynomials is introduced 
whose corresponding coefficients could be compared with a similar analysis of the LHC data. 
In Section 3 double differential distributions and their corresponding finite expansions on products of Chebyshev polynomials are discussed in detail. These are compared with equivalent ones in the case of completely independent particle production to fix the region in phase space where particle-particle correlations can be expected. These correlations are explored in more detail in Section 4. In Section 5 we further motivate the study of the Chew-Pignotti model by comparing it with a collinear limit of the BFKL equation and discuss the kinematical cuts needed at the LHC to measure the  proposed observables. After this we draw some Conclusions.

\section{Single differential distributions}

In a simplified analysis we work with a Chew-Pignotti type of multiperipheral model~\cite{Chew:1968fe} following the discussion by DeTar~\cite{Detar:1971qw} which will allow us to generate some analytic results that can be easily compared to the experimental data\footnote{For a review on multiperipheral models and the cluster concept in multiple hadron production see Ref.~\cite{Dremin:1977wc} and references therein.}. In these types of models the dependence on transverse coordinates factorizes from the rapidity dependence and we can write the cross section for the production of  $N+2$ particles (the ``2" refers to the twin-jets which in each event are considered to have rapidities $\pm \frac{Y}{2}$) in the simple form
\begin{eqnarray}
\sigma_{N+2} &=& \alpha^{N+2} \int_{0}^{Y} \prod_{i=1}^{N+1} dz_i \delta \left(Y-
\sum_{s=1}^{N+1} z_s \right) \nonumber\\
&=&  \alpha^{N+2}
\int_{-\frac{Y}{2}}^{\frac{Y}{2}} dy_N \int_{-\frac{Y}{2}}^{y_N} dy_{N-1} \cdots \int_{-\frac{Y}{2}}^{y_3} dy_2 
\int_{-\frac{Y}{2}}^{y_2} dy_1\, = \,  \alpha^2 
\frac{\left(\alpha Y\right)^N}{N!} \, .
\label{sigma_N2}
\end{eqnarray}
This trivially leads to a total cross section $\sigma_{\rm total} = 
\sum_{N=0}^\infty \sigma_{N+2} = \alpha^2 e^{\alpha Y}$ whose growth with $Y$ can be tamed 
only by introducing a non trivial dynamics in transverse space, something beyond the simple scope of the work here presented. We assign a rapidity 
$y_l$, with $l=0, \dots, N+1$, to each of the final-state particles. At $y_0=-\frac{Y}{2}$ and $y_{N+1}=\frac{Y}{2}$ we position our ``twin-jets" with a  simplified jet vertex equal to $\alpha$, the strong coupling constant\footnote{We would like to make clear at this point that we choose to work with limits $y_0=-\frac{Y}{2}$ and $y_{N+1}=\frac{Y}{2}$ to treat the forward and backward rapidity direction in a symmetric way. One could in principle also work with limits $y_0=0$ and $y_{N+1}=Y$. It does not mean that an event in experimental data with e.g. $y_0=0.1$ and $y_{N+1}=6.5$ should be excluded. It rather implies that that event will enter an experimental analysis with shifted $y_0=-3.2$ and $y_{N+1}=3.2$ to match our conventions here.}. The set of mini-jets with $l=1, \dots N$ is what we denote as ``the crowd" 
for which $y_l = -\frac{Y}{2}+\sum_{j=1}^l z_j$. 

\begin{figure}
\begin{center}
\begin{flushleft}
\hspace{1.cm}\includegraphics[width=7cm]{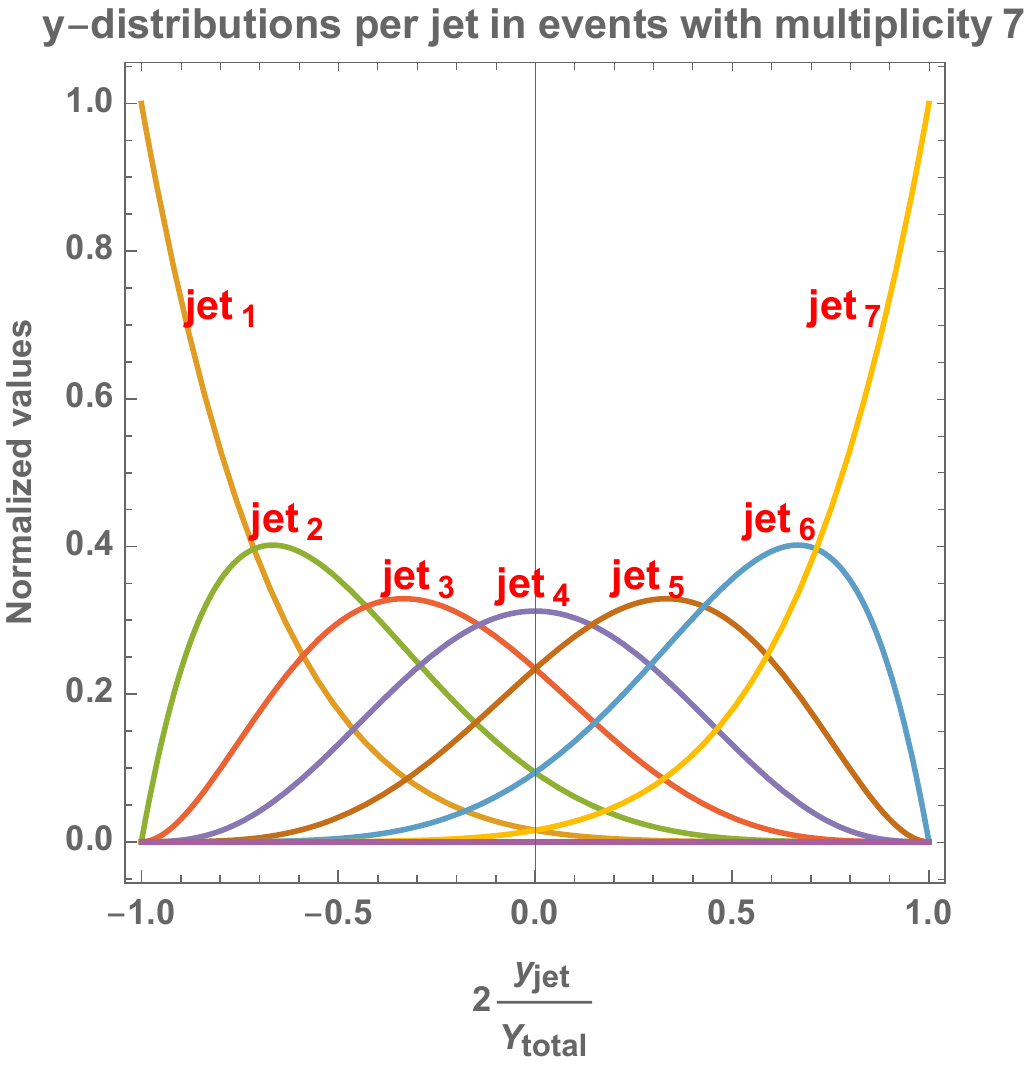}
\end{flushleft}
\vspace{-7.6cm}
\begin{center}
\hspace{8cm}\includegraphics[width=7.cm]{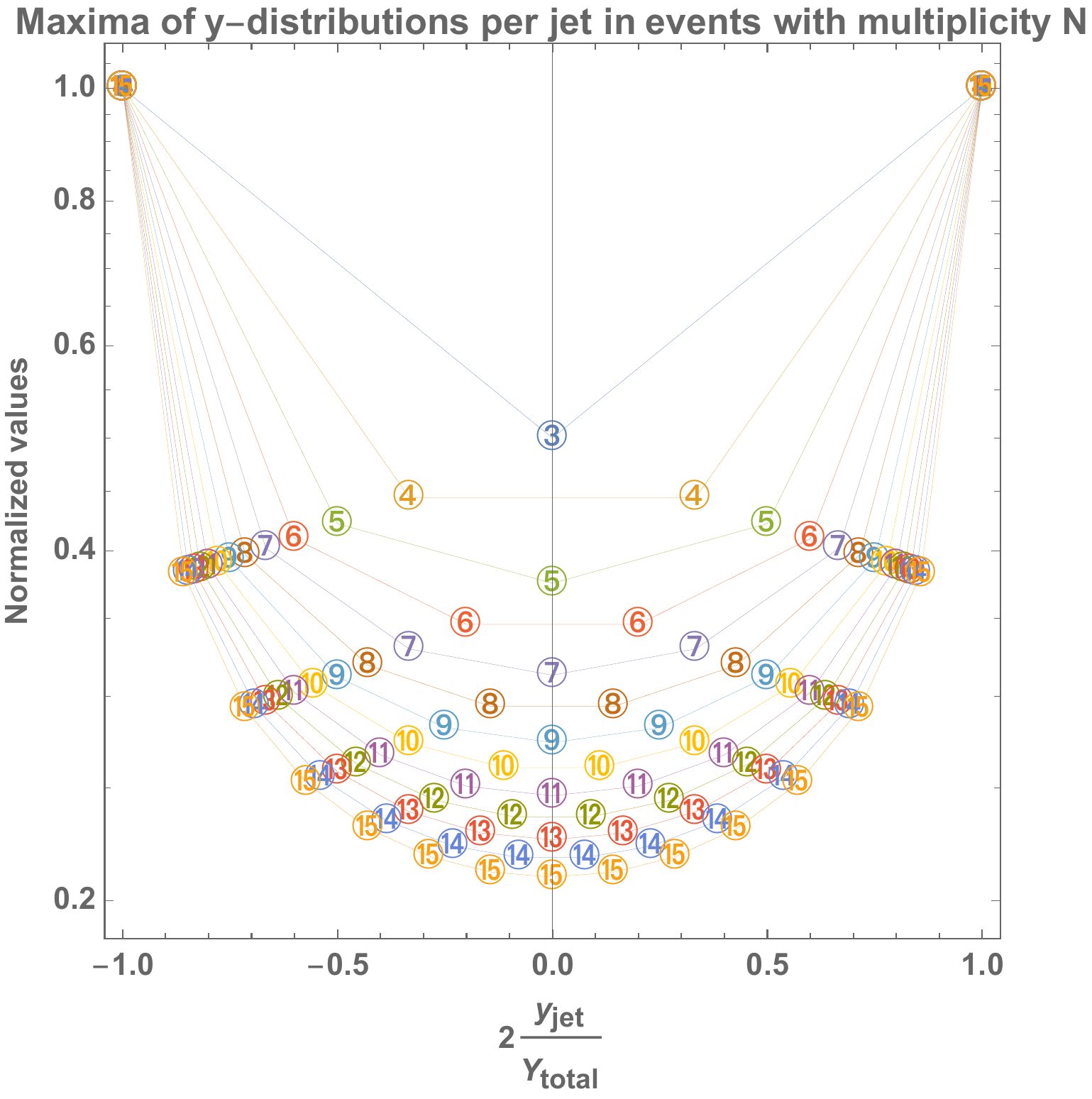}
\end{center}
\vspace{-.5cm}
\caption{Rapidity distributions for each of the jets in a final state with seven mini-jets (left).  Their maxima are indicated by the symbol \textcircled{\tiny 7} (right). The positions of the $y$-distribution maxima in configurations with multiplicity $N$ are marked by \textcircled{\tiny N} (right).}
\label{Multiplicity7yjets}
\end{center}
\end{figure}

More than in the total cross section, we are interested in a qualitative description of the differential distributions for each final state in events with fixed multiplicity. Multiplicity in our discussion is defined as the number of mini-jets (or even jets) in the crowd assuming that an IR safe jet clustering algorithm has been applied to the final state particles. One can argue that the multiplicity is not a unique number for a final state and that it depends on the lower transverse momentum cutoff and the resolution radius in the rapidity-azimuthal angle plane one considers for the jet algorithm. Nevertheless, we expect that our discussion in the following holds as long as there is a well defined mechanism that decides the multiplicity of a final state. This is actually a strong statement in itself from the simple model we are considering. It implies that if an event which was initially classified as having multiplicity $N_1+2$ complies with the $N_1+2$ differential distributions then it will continue to comply with the $N_2+2$ differential distributions assuming that in the latter case a different set of parameters is chosen for the jet clustering algorithm which results in a shifted number of crowd mini-jets from $N_1$ to $N_2$. Evidently, the crowd mini-jets contribute to the $N+2$ particle production cross section with a differential distribution in rapidity which can be derived from Eq.~(\ref{sigma_N2}) and is of the form
\begin{eqnarray}
\frac{d \sigma_{N+2}^{(l)}}{d y_l} &=&  \alpha^{N+2} \int_0^Y 
  \prod_{i=1}^{N+1} dz_i \delta \left(Y-\sum_{s=1}^{N+1} z_s \right) 
\delta \left(y_l +\frac{Y}{2}- \sum_{j=1}^l z_j\right) \nonumber\\
&=& \alpha^{N+2} \int_{y_l}^\frac{Y}{2} dy_N \int_{y_l}^{y_N} dy_{N-1} \cdots \int_{y_l}^{y_{l+2}} dy_{l+1}
\int_{-\frac{Y}{2}}^{y_l} dy_{l-1}  \cdots \int_{-\frac{Y}{2}}^{y_3} dy_2 
\int_{-\frac{Y}{2}}^{y_2} dy_1\nonumber\\
&=&  \alpha^{N+2} \frac{\left(\frac{Y}{2}-y_l \right)^{N-l}}{(N-l)!} \frac{\left(y_l+\frac{Y}{2}\right)^{l-1}}{(l-1)!}  \, .
\label{dsdy}
\end{eqnarray}
In the very large multiplicity limit this converges to an asymptotic Poisson distribution as can be seen, {\it e.g.}, in the region $y \simeq - \frac{Y}{2}$ with $y = \left(\frac{\lambda}{N}-\frac{1}{2}\right) Y$ where 
\begin{eqnarray}
\lim_{N \to \infty}{N-1 \choose l-1}   
\left(1-\frac{\lambda}{N}\right)^{N-l} 
\left(\frac{\lambda}{N}\right)^{l-1}
&=& e^{- \lambda} \frac{\lambda^{l-1}}{(l-1)!} \, .
\end{eqnarray}
Considering the limit $l \to 1$ and $y_l \to - \frac{Y}{2}$ in 
Eq.~(\ref{dsdy}) we can obtain a normalized universal distribution for each $N$ when plotted versus $2y/Y$. We show it for multiplicity seven in Fig.~\ref{Multiplicity7yjets} (left). The notation jet$_{i=1,2, \dots, N}$ is introduced for jets ordered with rapidities $y_1 < y_2 < \dots < y_N$. It can be seen that they reflect the characteristic cluster structure in the multiperipheral region of 
phase space.
Each of these normalized $y$-distributions spans an area of $\frac{2}{N}$ and has a maximum at $y=\frac{2l-N-1}{2(N-1)} Y$ where its value is
\begin{eqnarray}
{N-1 \choose l-1} \frac{(l-1)^{l-1}}{(N-1)^{N-1}} \left(N- l\right)^{N-l} \, .
\end{eqnarray}
These maxima are shown for increasing multiplicities in Fig.~\ref{Multiplicity7yjets} (right). 

It is possible to make more concrete predictions to be eventually compared with the experimental data if we use the following expansion of Eq.~(\ref{dsdy}) in terms of a finite number of Chebyshev polynomials $T_n$ (this is a standard procedure, see, {\it e.g.}~\cite{Bzdak:2012tp}):
\begin{eqnarray}
\frac{d \sigma_{N+2}^{(l)}}{d y_l} \, = \,    \alpha^{N+2} 
\left(\frac{Y}{2}\right)^{N-1} 
\frac{\left(1-x_l \right)^{N-l}}{(N-l)!} \frac{\left(1+x_l\right)^{l-1}}{(l-1)!} \, = \,  
\alpha^{N+2} \left(\frac{Y}{2}\right)^{N-1} 
 \sum_{s=0}^{N-1}  {\cal C}_{N+2,s}^{(l)} \, T_s (x_l) \, ,
\end{eqnarray}
where $x_l = 2 y_l / Y$. The coefficients
\begin{eqnarray}
 {\cal C}_{N+2,s}^{(l)} &=& \frac{2-\delta_s^0 }{\pi }
 \int_{-1}^1 \frac{dx_l}{\sqrt{1- x_l^2}} T_s (x_l) 
\frac{\left(1-x_l \right)^{N-l}}{(N-l)!} \frac{\left(1+x_l\right)^{l-1}}{(l-1)!} 
\label{coeffs}
\end{eqnarray}
can be obtained from the corresponding fit to the LHC data and can be compared to the 
values generated by Eq.~(\ref{coeffs}). For multiplicity $3+2$ their numerical values are
\begin{center}
\begin{tabular}{llll}
  $y_1 : $
    &${\cal C}_{5,0}^{(1)} =  \frac{3}{4}$ & ${\cal C}_{5,1}^{(1)} =  -1 $ & 
  ${\cal C}_{5,2}^{(1)} =  \frac{1}{4}$  \\
  $y_2 : $
    &${\cal C}_{5,0}^{(2)} =  \frac{1}{2}$ & ${\cal C}_{5,1}^{(2)} =  0$ & 
  ${\cal C}_{5,2}^{(2)} =  -\frac{1}{2}$  \\
    $y_3 : $
    &${\cal C}_{5,0}^{(3)} =  \frac{3}{4}$ & ${\cal C}_{5,1}^{(3)} =  1$ & 
  ${\cal C}_{5,2}^{(3)} =   \frac{1}{4}$  \\
\end{tabular}
\end{center}
For multiplicity $4+2$:
\begin{center}
\begin{tabular}{lllll}
  $y_1 : $
    &${\cal C}_{6,0}^{(1)} =  \frac{5}{12}$ & ${\cal C}_{6,1}^{(1)} =  
  -\frac{5}{8}$ & 
  ${\cal C}_{6,2}^{(1)} =  \frac{1}{4}$ & ${\cal C}_{6,3}^{(1)} =  
  -\frac{1}{24}$ \\
    $y_2 : $
    &${\cal C}_{6,0}^{(2)} =   \frac{1}{4}$ & ${\cal C}_{6,1}^{(2)} =  
    -\frac{1}{8}$ & 
  ${\cal C}_{6,2}^{(2)} =  -\frac{1}{4}$ & ${\cal C}_{6,3}^{(2)} =  
  \frac{1}{8}$ \\
    $y_3 : $
    &${\cal C}_{6,0}^{(3)} =   \frac{1}{4}$ & ${\cal C}_{6,1}^{(3)} =  
    \frac{1}{8}$ & 
  ${\cal C}_{6,2}^{(3)} =  -\frac{1}{4}$ & ${\cal C}_{6,3}^{(3)} =  
  -\frac{1}{8}$ \\    
  $y_4 : $
    &${\cal C}_{6,0}^{(4)} =   \frac{5}{12}$ & ${\cal C}_{6,1}^{(4)} =  
  \frac{5}{8}$ & 
  ${\cal C}_{6,2}^{(4)} =  \frac{1}{4}$ & ${\cal C}_{6,3}^{(4)} =  
  \frac{1}{24}$ \\\end{tabular}
\end{center}
As a final example, with $5+2$, we find
\begin{center}
\begin{tabular}{llllll}
  $y_1 : $
    &${\cal C}_{7,0}^{(1)} =  \frac{35}{192}$ 
  & ${\cal C}_{7,1}^{(1)} =  -\frac{7}{24}$ 
  & ${\cal C}_{7,2}^{(1)} =  \frac{7}{48}$ 
  & ${\cal C}_{7,3}^{(1)} =  -\frac{1}{24}$
  & ${\cal C}_{7,4}^{(1)} =  \frac{1}{192}$ \\
    $y_2 : $
    &${\cal C}_{7,0}^{(2)} =  \frac{5}{48}$ 
  & ${\cal C}_{7,1}^{(2)} =  -\frac{1}{12}$ 
  & ${\cal C}_{7,2}^{(2)} =  -\frac{1}{12}$ 
  & ${\cal C}_{7,3}^{(2)} =  \frac{1}{12}$
  & ${\cal C}_{7,4}^{(2)} =  -\frac{1}{48}$ \\
      $y_3 : $
    &${\cal C}_{7,0}^{(3)} =  \frac{3}{32}$ 
  & ${\cal C}_{7,1}^{(3)} = 0$ 
  & ${\cal C}_{7,2}^{(3)} =  - \frac{1}{8}$ 
  & ${\cal C}_{7,3}^{(3)} =  0$
  & ${\cal C}_{7,4}^{(3)} =  \frac{1}{32}$ \\
        $y_4 : $
    &${\cal C}_{7,0}^{(4)} =  \frac{5}{48}$ 
  & ${\cal C}_{7,1}^{(4)} =  \frac{1}{12}$ 
  & ${\cal C}_{7,2}^{(4)} =  -\frac{1}{12}$ 
  & ${\cal C}_{7,3}^{(4)} =  -\frac{1}{12}$
  & ${\cal C}_{7,4}^{(4)} =  -\frac{1}{48}$ \\
          $y_5 : $
    &${\cal C}_{7,0}^{(5)} =  \frac{35}{192}$ 
  & ${\cal C}_{7,1}^{(5)} =  \frac{7}{24}$ 
  & ${\cal C}_{7,2}^{(5)} =  \frac{7}{48}$ 
  & ${\cal C}_{7,3}^{(5)} =  \frac{1}{24}$
  & ${\cal C}_{7,4}^{(5)} =  \frac{1}{192}$ \\
\end{tabular}
\end{center}
These are predictions associated to the clustering structure of the final state 
stemming from this simple analysis. It will be interesting to compare them 
with similar ones obtained from different models of multi-particle production. 

\section{Double differential distributions}

Let us now derive similar expressions for double differential rapidity distributions for pairs 
of ordered in rapidity jets, {\it i.e.}
\begin{eqnarray}
\frac{d^2 \sigma_{N+2}^{(l,m)}}{d y_l d y_m} &=&  \alpha^{N+2} 
\int_{0}^{Y}
  \prod_{i=1}^{N+1} dz_i \delta \left(Y-\sum_{s=1}^{N+1} z_s \right) 
\delta \left(y_l +\frac{Y}{2}- \sum_{j=1}^l z_j\right)
\delta \left(y_m +\frac{Y}{2}- \sum_{k=1}^m z_k\right) \nonumber\\
&=&  \alpha^{N+2} \frac{\left(\frac{Y}{2}-y_l \right)^{N-l}}{(N-l)!}
\frac{(y_l-y_m)^{l-m-1}}{(l-m-1)!} 
\frac{\left(y_m+\frac{Y}{2}\right)^{m-1}}{(m-1)!}  \, .
\label{d2sdydy}
\end{eqnarray}
The expansion in the Chebyshev basis takes now place via pairs of polynomials:
\begin{eqnarray}
\frac{d^2 \sigma_{N+2}^{(l,m)}}{d y_l d y_m} &=&    \alpha^{N+2} 
\left(\frac{Y}{2}\right)^{N-2}
\frac{\left(1-x_l \right)^{N-l}}{(N-l)!}
\frac{(x_l-x_m)^{l-m-1}}{(l-m-1)!} 
\frac{\left(1+x_m\right)^{m-1}}{(m-1)!}  \nonumber\\ 
&=&   \alpha^{N+2} 
\left(\frac{Y}{2}\right)^{N-2} 
 \sum_{r,s=0}^{N-2}  {\cal D}_{N+2,r,s}^{(l,m)} \, T_r (x_l)\, T_s (x_m) 
\end{eqnarray}
with coefficients
\begin{eqnarray}
   {\cal D}_{N+2,r,s}^{(l,m)} &=&  \frac{(2-\delta_r^0)(2-\delta_s^0 ) }{\pi^2 (N-l)! (l-m-1)! (m-1)!}
\nonumber\\
&\times&   
 \int_{-1}^1 \frac{dx_l \, dx_m \, (1-x_l)^{N-l} T_r (x_l) \,  \left(1+x_m\right)^{m-1} T_s (x_m)}{\sqrt{1- x_l^2}\sqrt{1- x_m^2}}  (x_l-x_m)^{l-m-1} \, .
\end{eqnarray}
The two-dimensional gradient of these distributions is zero at 
the point $(x_l,x_m)_n^{\rm max}= \left(\frac{2 (l-1)-n}{n-2}, \frac{2 m-n}{n-2}\right)$ which is where they have a maximum value.

\begin{figure}
\begin{flushleft}
\hspace{1cm}\includegraphics[width=6cm]{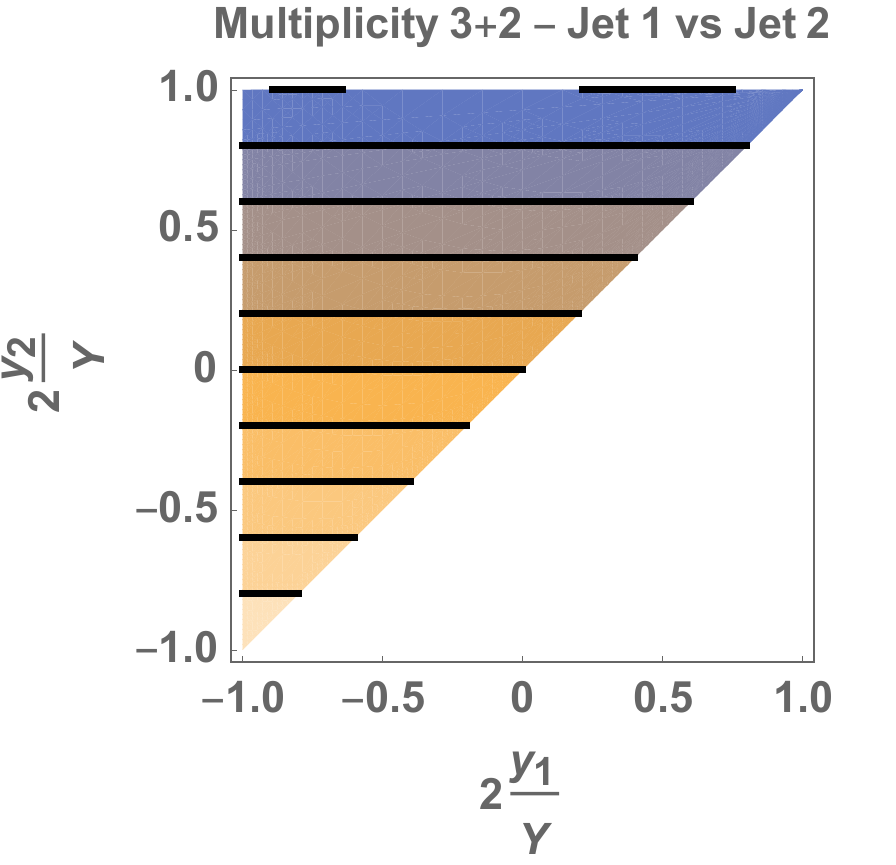}
\end{flushleft}
\vspace{-5.8cm}
\begin{flushleft}
\hspace{5.5cm}\includegraphics[width=.7cm]{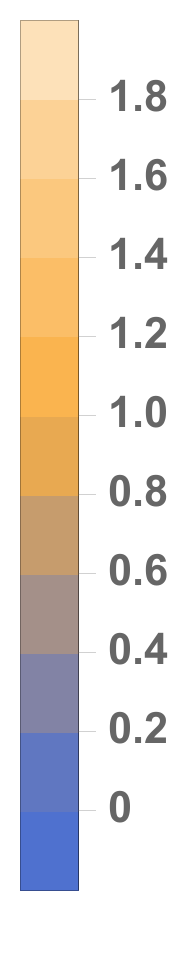}
\end{flushleft}
\vspace{-5.5cm}
\begin{center}
\hspace{7cm}\includegraphics[width=6cm]{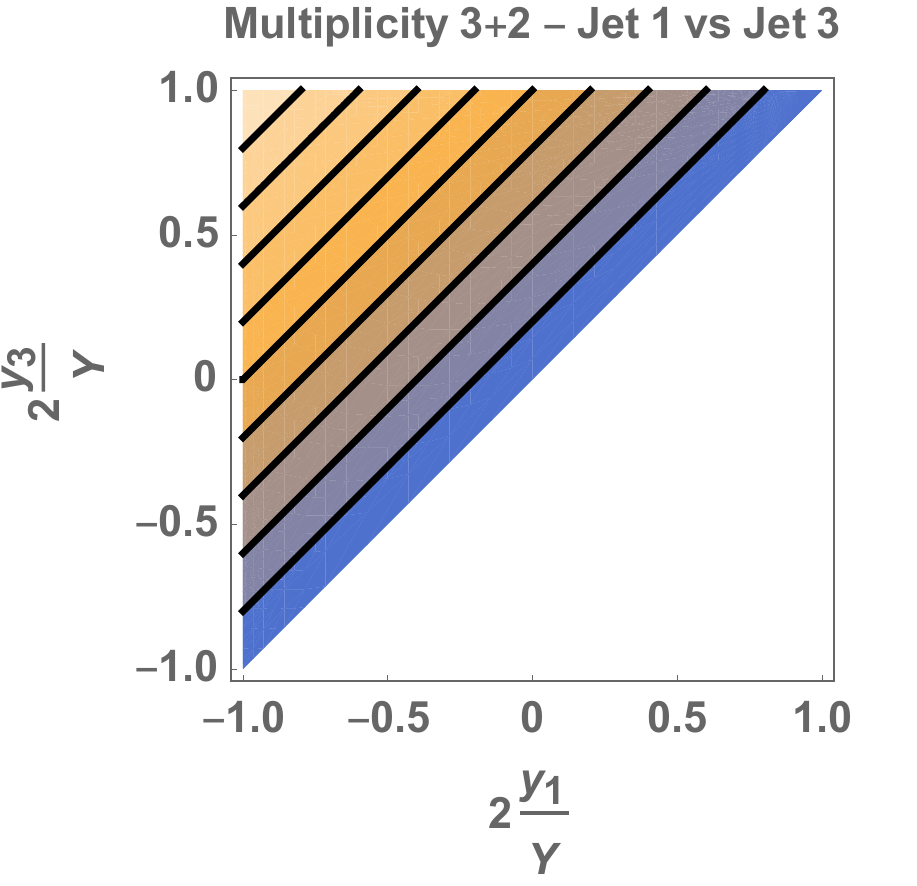}
\end{center}
\vspace{-5.8cm}
\begin{center}
\hspace{10.8cm}\includegraphics[width=.7cm]{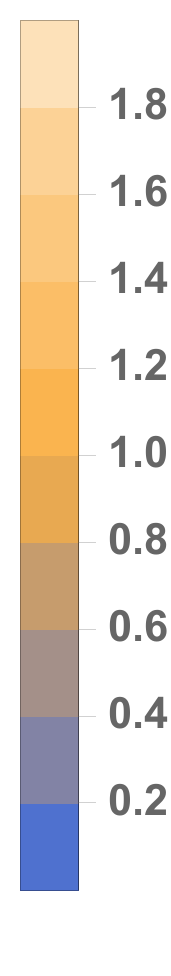}
\end{center}
\vspace{.4cm}
\caption{Left: Double differential cross section $
 \frac{2}{\alpha^5 Y}
\frac{d^2 \sigma_{3+2}^{(2,1)}}{d y_2 d y_1} = 1 - x_2  = 
T_0 (x_2) T_0 (x_1) -  T_1 (x_2) T_0 (x_1)$ (the related distribution by $x_2 \to - x_2$ is $\frac{2}{\alpha^5 Y}
\frac{d^2 \sigma_{3+2}^{(3,2)}}{d y_3 d y_2} =  1+x_2 = 
 T_0 (x_3) T_0 (x_2) +  T_1 (x_3) T_0 (x_2)$).  
 Right: Double differential cross section $
 \frac{2}{\alpha^5 Y}
\frac{d^2 \sigma_{3+2}^{(3,1)}}{d y_3 d y_1} =  x_3 - x_1 = 
 T_1 (x_3) T_0 (x_1) -  T_0 (x_3) T_1 (x_1)$. In both $x_L = 2 y_L/Y$.}
\label{DSigma3-1223}
\end{figure}

\begin{figure}
\begin{flushleft}
\hspace{1cm}\includegraphics[width=6cm]{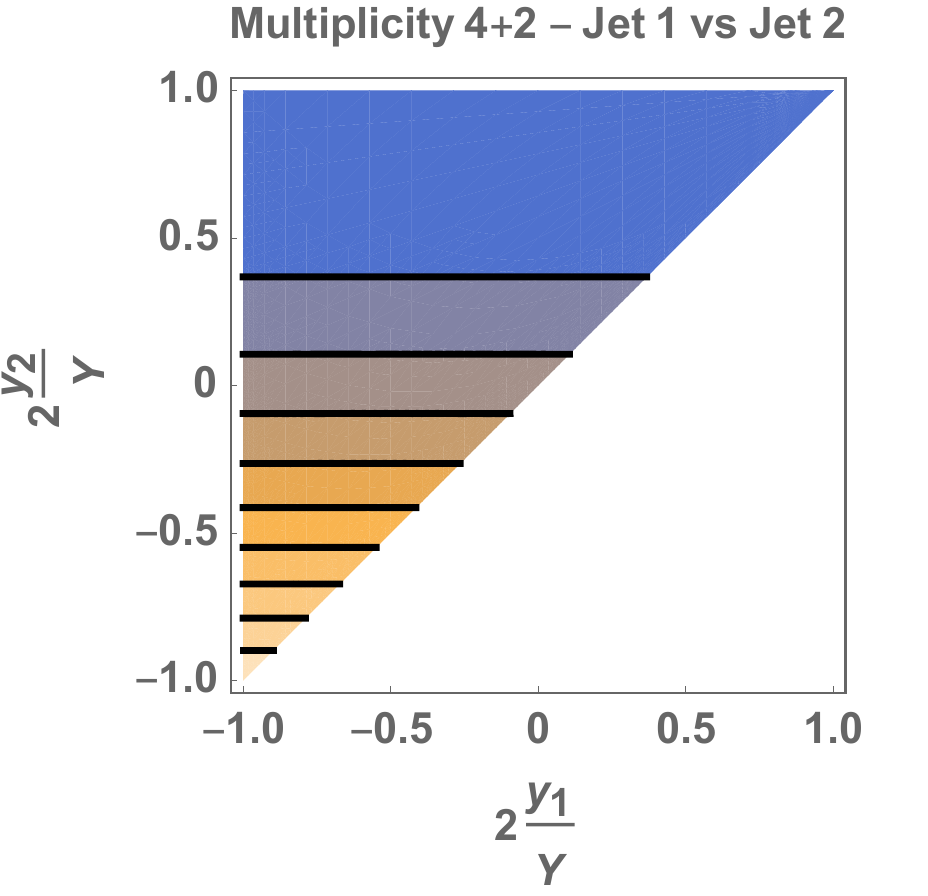}
\end{flushleft}
\vspace{-5.8cm}
\begin{flushleft}
\hspace{5.5cm}\includegraphics[width=.7cm]{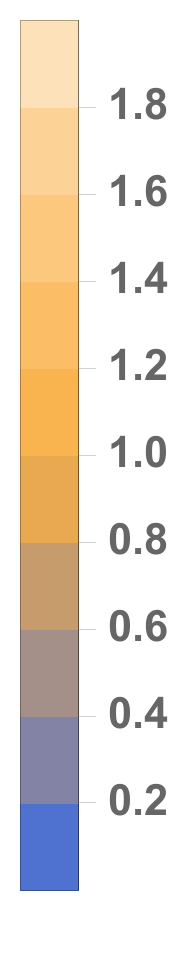}
\end{flushleft}
\vspace{-5.5cm}
\begin{center}
\hspace{7cm}\includegraphics[width=6cm]{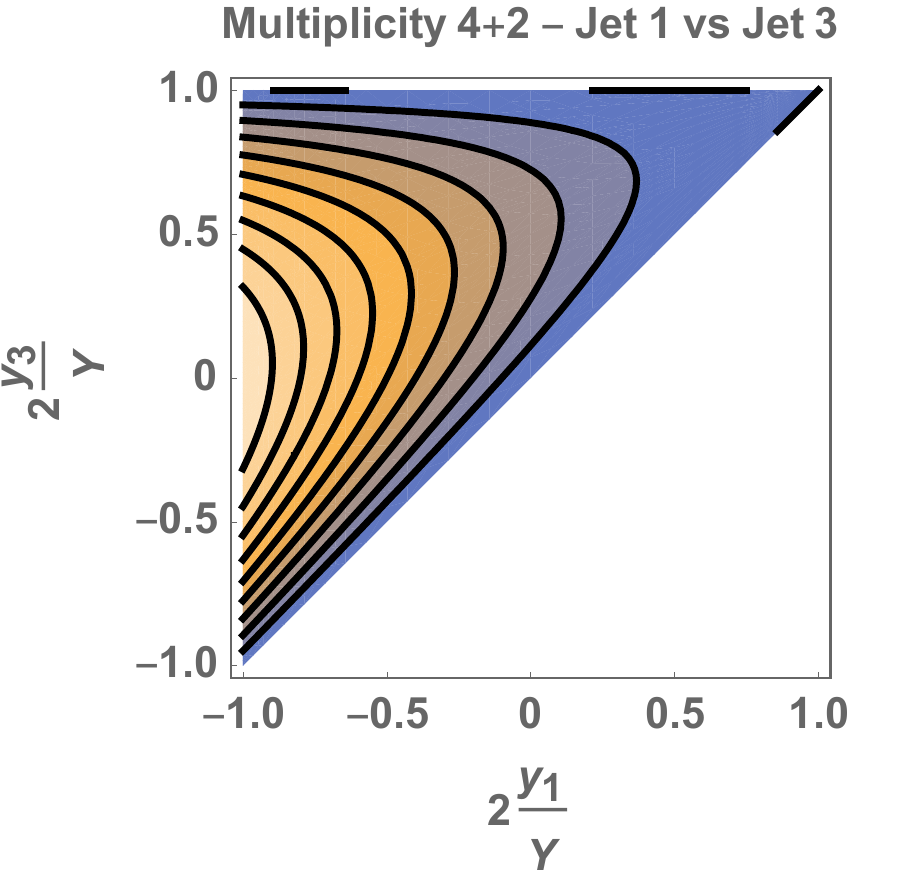}
\end{center}
\vspace{-5.8cm}
\begin{center}
\hspace{10.8cm}\includegraphics[width=.7cm]{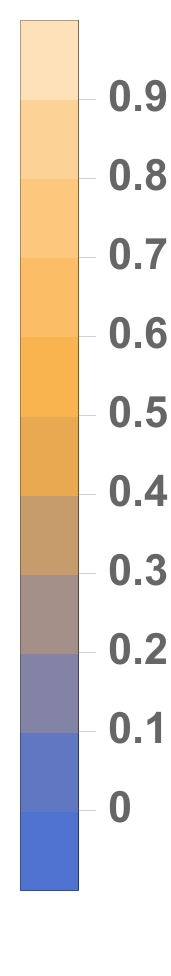}
\end{center}
\vspace{.4cm}
\caption{Left: Double differential cross section $
 \frac{4}{\alpha^6 Y^2}
\frac{d^2 \sigma_{4+2}^{(2,1)}}{d y_2 d y_1} = 
   \frac{1}{2} \left( 1 - x_2\right)^2  = 
  \frac{3}{4} T_0 (x_2) T_0 (x_1)
 -  T_1 (x_2) T_0 (x_1)
 + \frac{1}{4} T_2 (x_2)T_0 (x_1)$ (the related distribution generated by 
 $x_2 \to -x_3$ is $
 \frac{4}{\alpha^6 Y^2}
\frac{d^2 \sigma_{4+2}^{(4,3)}}{d y_4 d y_3} = 
 \frac{1}{2}   \left( 1+x_3\right)^2 = 
  \frac{3}{4} T_0 (x_4) T_0 (x_3)
 +  T_0 (x_4) T_1 (x_3)
 + \frac{1}{4} T_0(x_4)T_2(x_3)$). Right: 
 Double differential cross section $
 \frac{4}{\alpha^6 Y^2}
\frac{d^2 \sigma_{4+2}^{(3,1)}}{d y_3 d y_1} = 
   \left( 1 - x_3\right) \left( x_3 - x_1\right) = 
  -\frac{1}{2} T_0 (x_3) T_0 (x_1)
 +  T_1 (x_3) T_0 (x_1)
 -\frac{1}{2} T_2 (x_3)T_0 (x_1)
 - T_0 (x_3)T_1(x_1)+T_1 (x_3)T_1(x_1) + \frac{1}{4} T_2 (x_2)T_0 (x_1)$ 
 (the related distribution generated by 
 $x_3 \to -x_2$, $x_1 \to -x_4$ is  $\frac{4}{\alpha^6 Y^2}
\frac{d^2 \sigma_{4+2}^{(4,2)}}{d y_4 d y_2} = 
 \left(x_4-x_2 \right)    \left( 1+x_2\right) = 
 - \frac{1}{2} T_0 (x_4) T_0 (x_2)
 +  T_1 (x_4) T_0 (x_2)
 - \frac{1}{2} T_0(x_4)T_2(x_2)
 -  T_0(x_4)T_1(x_2) + T_1(x_4)T_1(x_2)$). In both $x_L = 2 y_L/Y$.}
 \label{DSigma6-1234}
\end{figure}

\begin{figure}
\begin{flushleft}
\hspace{1cm}\includegraphics[width=6cm]{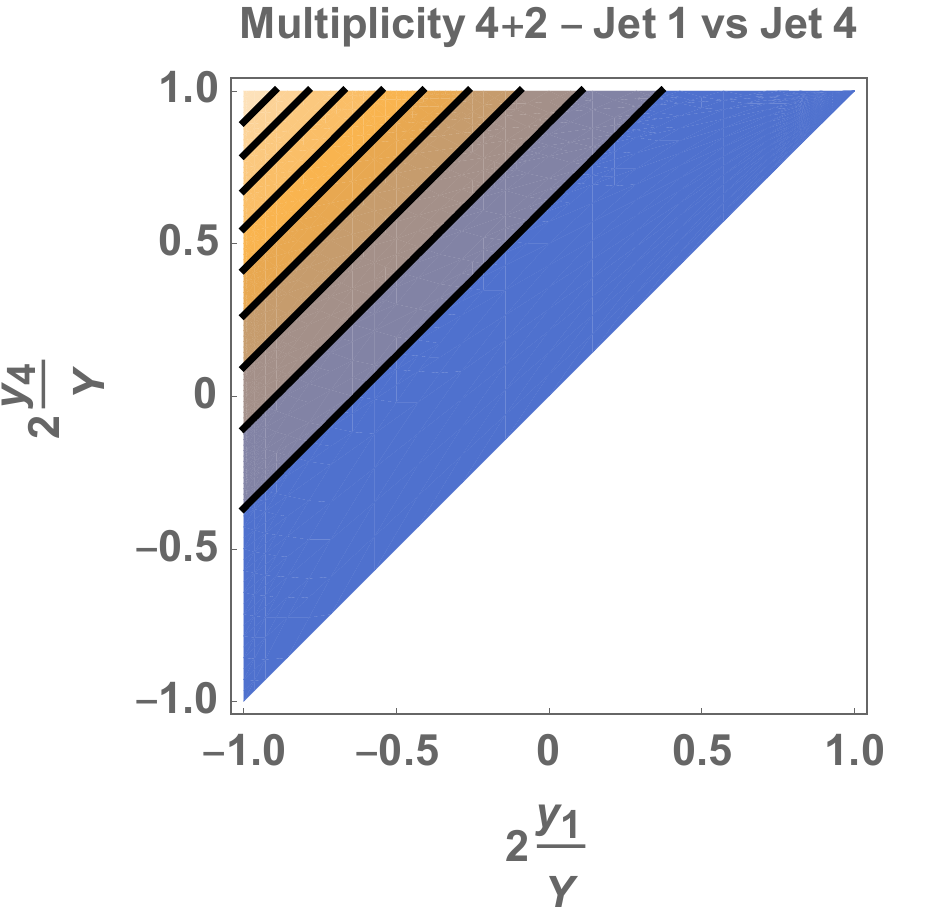}
\end{flushleft}
\vspace{-5.8cm}
\begin{flushleft}
\hspace{5.5cm}\includegraphics[width=.7cm]{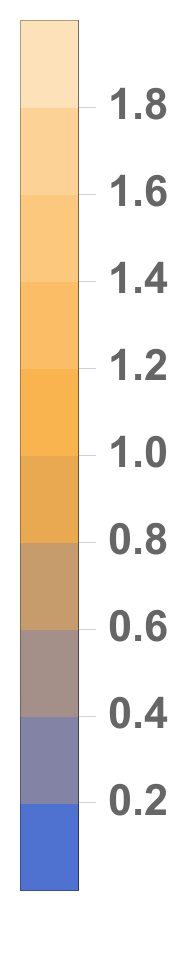}
\end{flushleft}
\vspace{-5.5cm}
\begin{center}
\hspace{7cm}\includegraphics[width=6cm]{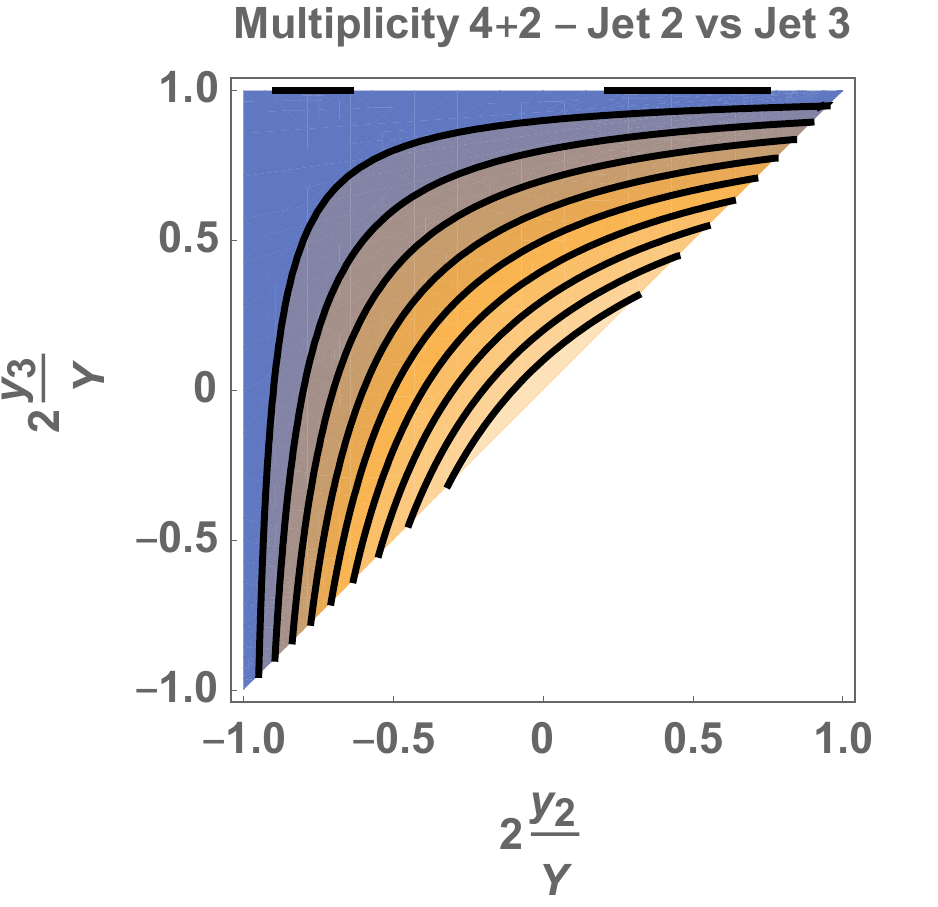}
\end{center}
\vspace{-5.8cm}
\begin{center}
\hspace{10.8cm}\includegraphics[width=.7cm]{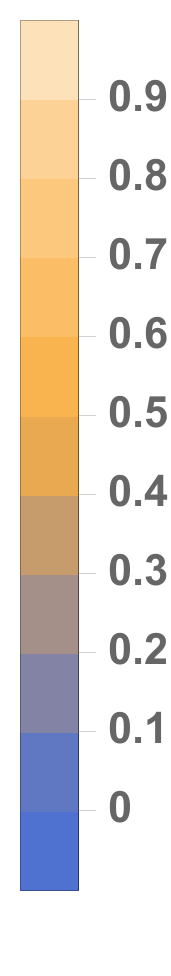}
\end{center}
\vspace{.4cm}
\caption{Left: Double differential cross section $
 \frac{4}{\alpha^6 Y^2}
\frac{d^2 \sigma_{4+2}^{(4,1)}}{d y_4 d y_1} = 
 \frac{1}{2}   \left( x_1 - x_4\right)^2 = 
  \frac{1}{2} T_0 (x_4) T_0 (x_1)
 +  \frac{1}{4} T_2 (x_4) T_0 (x_1)
 - T_1(x_4)T_1(x_1)
 +\frac{1}{4} T_0(x_4)T_2(x_1)$. 
 Right: Double differential cross section $
 \frac{4}{\alpha^6 Y^2}
\frac{d^2 \sigma_{4+2}^{(3,2)}}{d y_3 d y_2} = 
   \left( 1 + x_2\right) \left( 1 - x_3\right) = 
  T_0 (x_3) T_0 (x_2)
 -  T_1 (x_3) T_0 (x_2)
 + T_0(x_3)T_1(x_2)
 - T_1 (x_3)T_1(x_2)$. In both $x_L = 2 y_L/Y$. }
\label{DSigma6-1423}
\end{figure}

\begin{figure}
\begin{flushleft}
\hspace{1cm}\includegraphics[width=6cm]{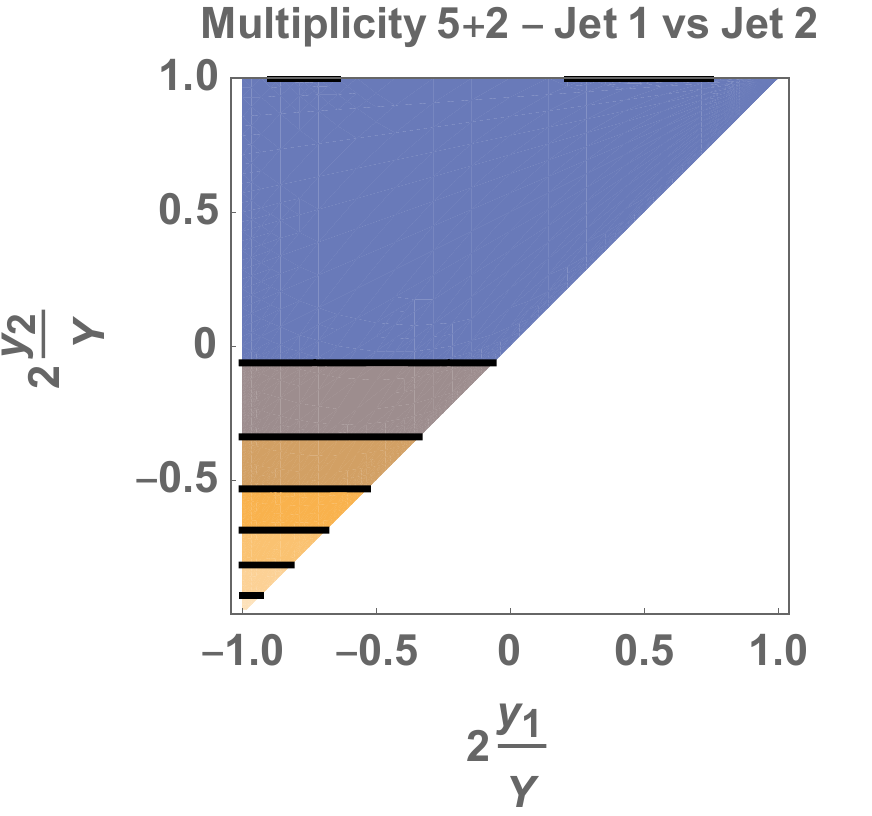}
\end{flushleft}
\vspace{-5.6cm}
\begin{flushleft}
\hspace{5.5cm}\includegraphics[width=.7cm]{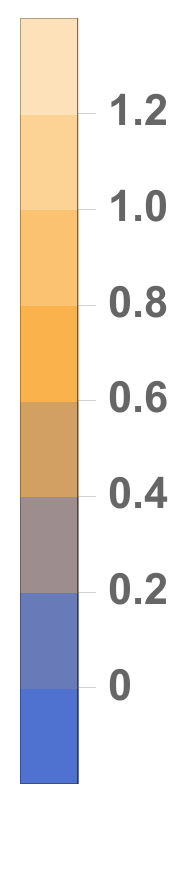}
\end{flushleft}
\vspace{-5.2cm}
\begin{center}
\hspace{7cm}\includegraphics[width=6cm]{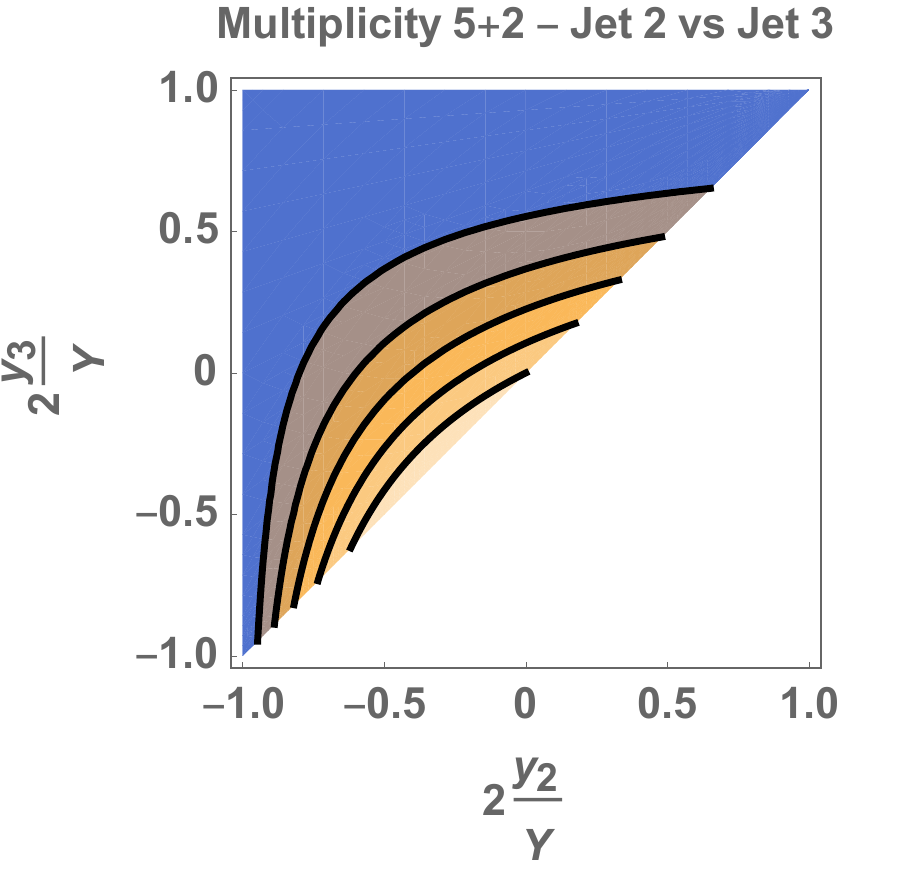}
\end{center}
\vspace{-5.4cm}
\begin{center}
\hspace{10.8cm}\includegraphics[width=.7cm]{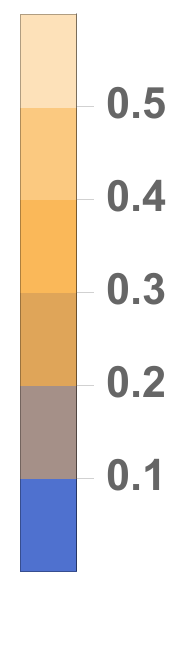}
\end{center}
\vspace{1.4cm}
\caption{Left: Double differential cross section $
 \frac{8}{\alpha^7 Y^3}
\frac{d^2 \sigma_{5+2}^{(2,1)}}{d y_2 d y_1} = 
 \frac{1}{6}   \left(1 - x_2\right)^3 = 
  \frac{5}{12} T_0 (x_2) T_0 (x_1)
 -  \frac{5}{8} T_1 (x_2) T_0 (x_1)
 +\frac{1}{4} T_2(x_2)T_0(x_1)
 -\frac{1}{24} T_3(x_2)T_0(x_1)$ (the related distribution with 
 $x_2 \to - x_4$ is $
  \frac{8}{\alpha^7 Y^3}
\frac{d^2 \sigma_{5+2}^{(5,4)}}{d y_5 d y_4} = 
 \frac{1}{6}   \left(1 + x_4\right)^3 = 
  \frac{5}{12} T_0 (x_5) T_0 (x_4)
 +  \frac{5}{8} T_0 (x_5) T_1 (x_4)
 +\frac{1}{4} T_0(x_5)T_2(x_4)
 +\frac{1}{24} T_0(x_5)T_3(x_4)$). 
 Right: Double differential cross section $
 \frac{8}{\alpha^7 Y^3}
\frac{d^2 \sigma_{5+2}^{(3,2)}}{d y_3 d y_2} = 
 \frac{1}{2}   \left(1 + x_2\right) \left(1 - x_3\right)^2 = 
  \frac{3}{4} T_0 (x_3) T_0 (x_2)
 -   T_1 (x_3) T_0 (x_2)
 +\frac{1}{4} T_2(x_3)T_0(x_2)
 +\frac{3}{4} T_0(x_3)T_1(x_2)
 - T_1(x_3)T_1(x_2)
 +\frac{1}{4} T_2(x_3)T_1(x_2)$
 (the related distibution with $x_3 \to - x_3$ and $x_2 \to -x_4$ is $
  \frac{8}{\alpha^7 Y^3}
\frac{d^2 \sigma_{5+2}^{(4,3)}}{d y_4 d y_3} = 
 \frac{1}{2}   \left(1 + x_3\right)^2 \left(1 - x_4\right) = 
  \frac{3}{4} T_0 (x_4) T_0 (x_3)
 -  \frac{3}{4} T_1 (x_4) T_0 (x_3)
 + T_0(x_4)T_1(x_3)
  - T_1(x_4)T_1(x_3)
 +\frac{1}{4} T_0(x_4)T_2(x_3)
 - \frac{1}{4} T_1(x_4)T_2(x_3)$). In both $x_L = 2 y_L/Y$.. }
\label{DSigma7-1245}
\end{figure}

\begin{figure}
\begin{flushleft}
\hspace{1cm}\includegraphics[width=6cm]{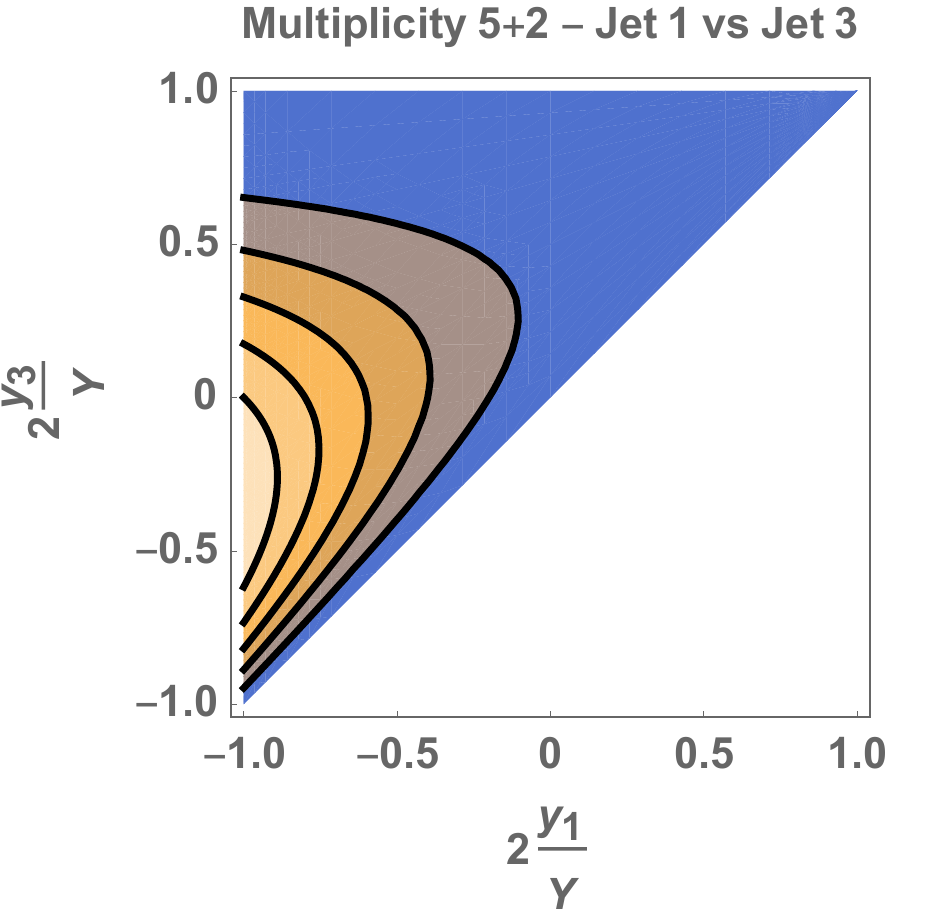}
\end{flushleft}
\vspace{-4.8cm}
\begin{flushleft}
\hspace{5.5cm}\includegraphics[width=.7cm]{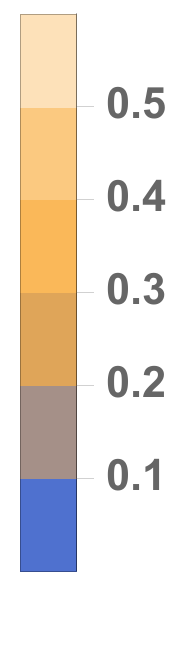}
\end{flushleft}
\vspace{-5.2cm}
\begin{center}
\hspace{7cm}\includegraphics[width=6.cm]{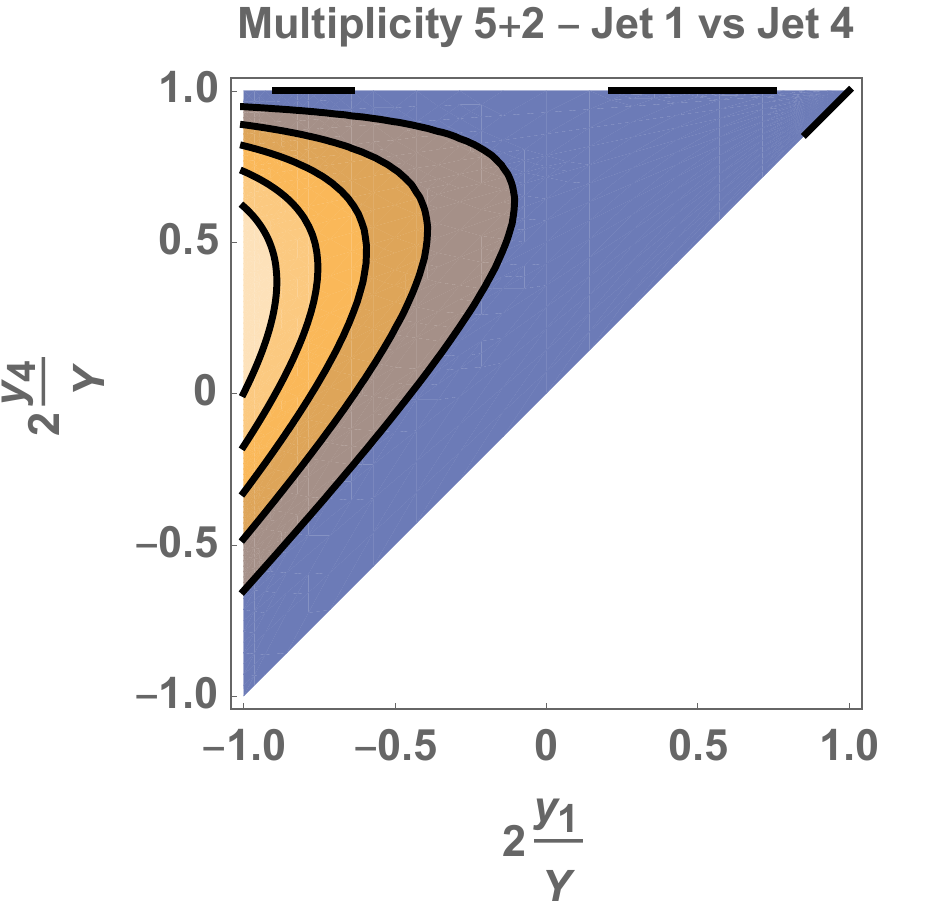}
\end{center}
\vspace{-5.cm}
\begin{center}
\hspace{10.8cm}\includegraphics[width=.7cm]{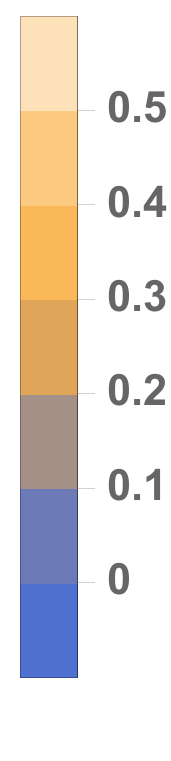}
\end{center}
\vspace{.6cm}
\caption{Left: Double differential cross section $
 \frac{8}{\alpha^7 Y^3}
\frac{d^2 \sigma_{5+2}^{(3,1)}}{d y_3 d y_1} = 
 \frac{1}{2}   \left(x_3 - x_1\right) \left(1 - x_3\right)^2 = 
 - \frac{1}{2} T_0 (x_3) T_0 (x_1)
 +  \frac{7}{8} T_1 (x_3) T_0 (x_1)
 -\frac{1}{2} T_2(x_3)T_0(x_1)
 +\frac{1}{8} T_3(x_3)T_0(x_1)
 - \frac{3}{4} T_0(x_3)T_1(x_1)
 + T_1(x_3)T_1(x_1)
 - \frac{1}{4} T_2(x_3)T_1(x_1)$ (the related by $x_3 \to -x_3$ and 
 $x_1 \to -x_5$  distribution is $
  \frac{8}{\alpha^7 Y^3}
\frac{d^2 \sigma_{5+2}^{(5,3)}}{d y_5 d y_3} = 
 \frac{1}{2}   \left(1 + x_3\right)^2 \left(x_5 - x_3\right) = 
 - \frac{1}{2} T_0 (x_5) T_0 (x_3)
 +  \frac{3}{4} T_1 (x_5) T_0 (x_3)
 +  \frac{1}{4} T_1 (x_5) T_2 (x_3)
 - \frac{1}{8} T_0(x_5)T_3(x_3)
  -\frac{7}{8} T_0(x_5)T_1(x_3)
 + T_1(x_5)T_1(x_3)
 - \frac{1}{2} T_0(x_5)T_2(x_3)$). Right: Double differential cross section $
 \frac{8}{\alpha^7 Y^3}
\frac{d^2 \sigma_{5+2}^{(4,1)}}{d y_4 d y_1} = 
 \frac{1}{2}   \left(1 - x_4\right) \left(x_1 - x_4\right)^2 = 
  \frac{1}{2} T_0 (x_4) T_0 (x_1)
 -  \frac{5}{8} T_1 (x_4) T_0 (x_1)
 +  \frac{1}{4} T_2 (x_4) T_0 (x_1)
 - \frac{1}{8} T_3(x_4)T_0(x_1)
  + \frac{1}{2}T_0(x_4)T_1(x_1)
 - T_1(x_4)T_1(x_1)
 + \frac{1}{2} T_2(x_4)T_1(x_1)
 +  \frac{1}{4} T_0(x_4)T_2(x_1)
 -  \frac{1}{4} T_1(x_4)T_2(x_1)$
(the related by $x_4 \to -x_2$ and $x_1 \to -x_5$ is $
  \frac{8}{\alpha^7 Y^3}
\frac{d^2 \sigma_{5+2}^{(5,2)}}{d y_5 d y_2} = 
 \frac{1}{2}   \left(1 + x_2\right) \left(x_5 - x_2\right)^2 = 
 \frac{1}{2} T_0 (x_5) T_0 (x_2)
 -  \frac{1}{2} T_1 (x_5) T_0 (x_2)
 +  \frac{1}{4} T_2 (x_5) T_0 (x_2)
 + \frac{5}{8} T_0(x_5)T_1(x_2)
  - T_1(x_5)T_1(x_2)
 + \frac{1}{4} T_2(x_5)T_1(x_2)
 + \frac{1}{4} T_0(x_5)T_2(x_2)
 -  \frac{1}{2} T_1(x_5)T_2(x_2)
 +  \frac{1}{8} T_0(x_5)T_3(x_2)$). In both $x_L = 2 y_L/Y$. }
\label{DSigma7-1335}
\end{figure}

\begin{figure}
\begin{flushleft}
\hspace{1cm}\includegraphics[width=6cm]{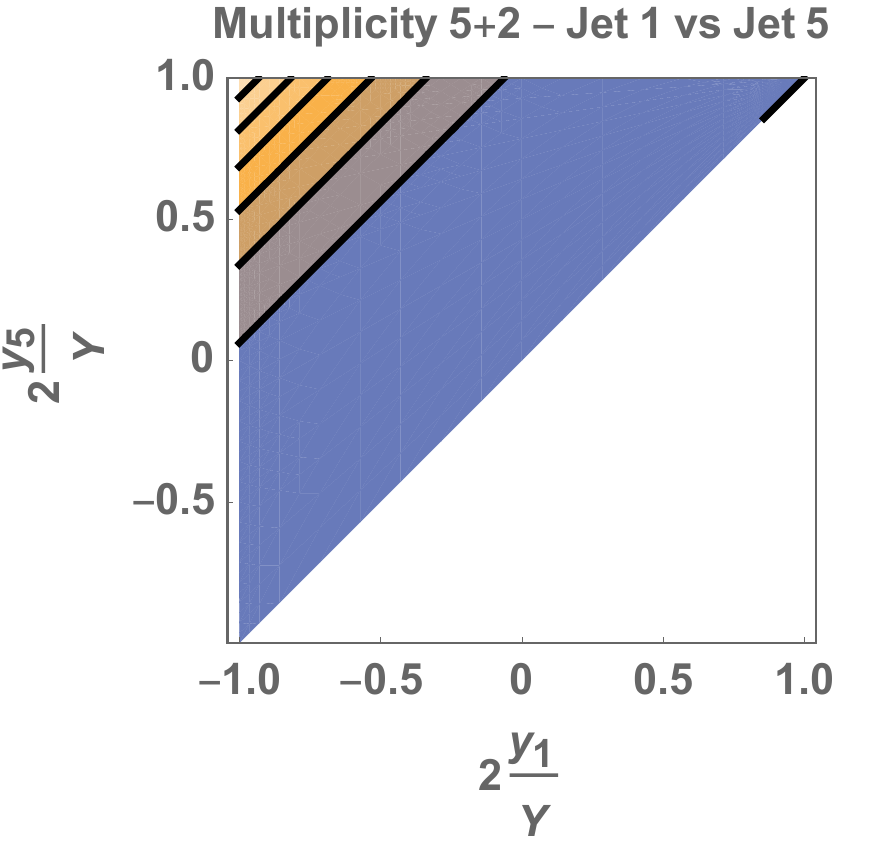}
\end{flushleft}
\vspace{-5.1cm}
\begin{flushleft}
\hspace{5.5cm}\includegraphics[width=.7cm]{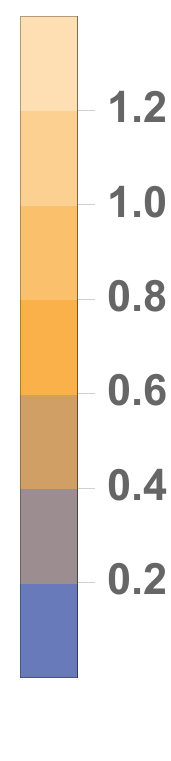}
\end{flushleft}
\vspace{-5.3cm}
\begin{center}
\hspace{7cm}\includegraphics[width=6.cm]{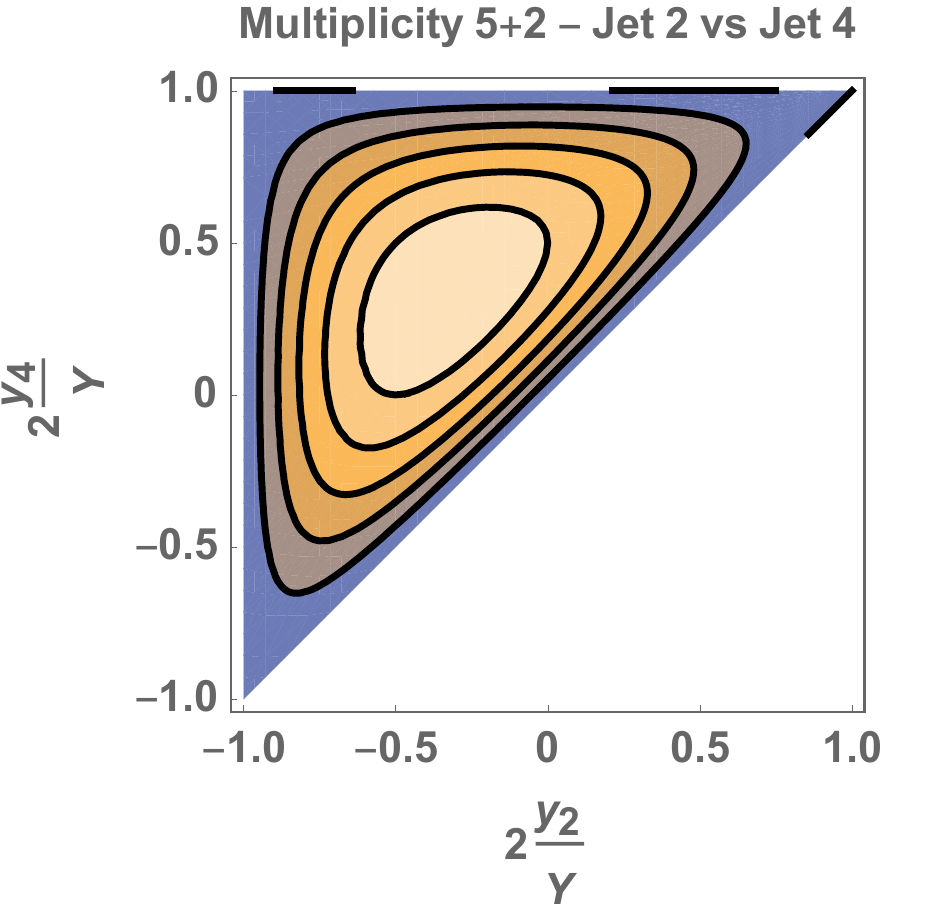}
\end{center}
\vspace{-5.cm}
\begin{center}
\hspace{10.8cm}\includegraphics[width=.7cm]{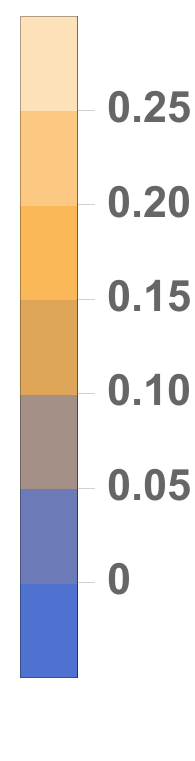}
\end{center}
\vspace{1.cm}
\caption{Left: Double differential cross section $
 \frac{8}{\alpha^7 Y^3}
\frac{d^2 \sigma_{5+2}^{(5,1)}}{d y_5 d y_1} = 
 \frac{1}{6}   \left(x_5 - x_1\right)^3 = 
   \frac{3}{8} T_1 (x_5) T_0 (x_1)
 +  \frac{1}{24} T_3 (x_5) T_0 (x_1)
 - \frac{3}{8} T_0(x_5)T_1(x_1)
  - \frac{1}{4}T_2(x_5)T_1(x_1)
 + \frac{1}{4} T_1(x_5)T_2(x_1)
 -  \frac{1}{24} T_0(x_5)T_3(x_1)$. 
 Right: Double differential cross section $
  \frac{8}{\alpha^7 Y^3}
\frac{d^2 \sigma_{5+2}^{(4,2)}}{d y_4 d y_2} = 
   \left(1 + x_2\right) \left(1 - x_4\right) \left(x_4 - x_2\right) = 
   -T_0 (x_4) T_0 (x_2)
  +  \frac{3}{2} T_1 (x_4) T_0 (x_2)
 -  \frac{1}{2} T_2 (x_4) T_0 (x_2)
 - \frac{3}{2} T_0(x_4)T_1(x_2)
  + 2 T_1(x_4)T_1(x_2)
 - \frac{1}{2} T_2(x_4)T_1(x_2)
 -  \frac{1}{2} T_0(x_4)T_2(x_1)
 +\frac{1}{2} T_1(x_4)T_2(x_1)$. In both $x_L = 2 y_L/Y$. }
\label{DSigma7-1524}
\end{figure}

In the case with multiplicity 2+2 we get as the only non-zero component 
${\cal D}_{4,0,0}^{(2,1)} =  1$ and we can formally write $
 \frac{1}{\alpha^4}
\frac{d^2 \sigma_{2+2}^{(2,1)}}{d y_2 d y_1} = 1 = 
  T_0 (x_2) T_0 (x_1)$. For multiplicity 3+2 there is more structure:
\begin{center}
\begin{tabular}{llllll}
 Fig.$~\ref{DSigma3-1223}; (x_2,x_1)_{n=3}^{\rm max}=(-1,-1)$:
  &${\cal D}_{5,0,0}^{(2,1)} =  1$ 
  & ${\cal D}_{5,1,0}^{(2,1)} =  -1$  
  &
 & \\
  Fig.$~\ref{DSigma3-1223}; (x_3,x_2)_{n=3}^{\rm max}=(1,1)$:
  &${\cal D}_{5,0,0}^{(3,2)} =  1$ 
  & ${\cal D}_{5,0,1}^{(3,2)} =  1$  
  &
 & \\
  Fig.$~\ref{DSigma3-1223}; (x_3,x_1)_{n=3}^{\rm max}=(1,-1)$:
  &${\cal D}_{5,1,0}^{(3,1)} =  1$ 
  & ${\cal D}_{5,0,1}^{(3,1)} =  -1$  
  &
 & \\
  \end{tabular}
\end{center}
We have plotted these three distributions in Fig.~\ref{DSigma3-1223}. 
For multiplicity 4+2 the non-zero cases are 
\begin{center}
\begin{tabular}{llllll}
 Fig.$~\ref{DSigma6-1234}; (x_2,x_1)_{n=4}^{\rm max}=(-1,-1)$:
  &${\cal D}_{6,0,0}^{(2,1)} =  \frac{3}{4}$ 
  & ${\cal D}_{6,1,0}^{(2,1)} =  -1$ 
  & ${\cal D}_{6,2,0}^{(2,1)} =  \frac{1}{4}$ 
  &
 & \\
           Fig.$~\ref{DSigma6-1234}; (x_4,x_3)_{n=4}^{\rm max}=(1,1)$:
    &${\cal D}_{6,0,0}^{(4,3)} =  \frac{3}{4}$
  & ${\cal D}_{6, 0,1}^{(4,3)} =  1$
  &  ${\cal D}_{6,0,2}^{(4,3)} =  \frac{1}{4}$
  & 
   &\\
    Fig.$~\ref{DSigma6-1234}; (x_3,x_1)_{n=4}^{\rm max}=(0,-1)$:
    &${\cal D}_{6,0,0}^{(3,1)} =  -\frac{1}{2}$
  & ${\cal D}_{6, 1,0}^{(3,1)} =  1$
  & ${\cal D}_{6, 2,0}^{(3,1)} =  -\frac{1}{2}$ 
  & ${\cal D}_{6,0,1}^{(3,1)} =  -1$\\
   & ${\cal D}_{6,1,1}^{(3,1)} =  1$\\
         Fig.$~\ref{DSigma6-1234}; (x_4,x_2)_{n=4}^{\rm max}=(1,0)$:
    &${\cal D}_{6,0,0}^{(4,2)} =  -\frac{1}{2}$
  & ${\cal D}_{6, 1,0}^{(4,2)} =  1$
  &  ${\cal D}_{6,0,2}^{(4,2)} =  -\frac{1}{2}$
  & ${\cal D}_{6,0,1}^{(4,2)} =  -1$\\
   &${\cal D}_{6,1,1}^{(4,2)} =  1$ \\
       Fig.$~\ref{DSigma6-1423}; (x_4,x_1)_{n=4}^{\rm max}=(1,-1)$:
    &${\cal D}_{6,0,0}^{(4,1)} =  \frac{1}{2}$
  & ${\cal D}_{6, 2,0}^{(4,1)} =  \frac{1}{4}$
  & ${\cal D}_{6, 1,1}^{(4,1)} =  -1$ 
  & ${\cal D}_{6,0,2}^{(4,1)} =  \frac{1}{4}$
   & \\
       Fig.$~\ref{DSigma6-1423}; (x_3,x_2)_{n=4}^{\rm max}=(0,0)$:
    &${\cal D}_{6,0,0}^{(3,2)} =  1$
  & ${\cal D}_{6, 1,0}^{(3,2)} =  -1$
  &   ${\cal D}_{6, 0,1}^{(3,2)} =  1$
  &  ${\cal D}_{6, 1,1}^{(3,2)} =  -1$
   & 
  \end{tabular}
\end{center}
The associated figures are shown in Figs.~\ref{DSigma6-1234} and~\ref{DSigma6-1423}.
As a final example, for multiplicity 5+2, we have
\begin{center}
\begin{tabular}{lllllllll}
   Fig.$~\ref{DSigma7-1245}; (x_2,x_1)_{n=5}^{\rm max}=(-1,-1)$:
    &${\cal D}_{7,0,0}^{(2,1)} =  \frac{5}{12}$ 
  & ${\cal D}_{7,1,0}^{(2,1)} =  -\frac{5}{8}$ 
  & ${\cal D}_{7,2,0}^{(2,1)} =  \frac{1}{4}$ 
  & ${\cal D}_{7,3,0}^{(2,1)} =  -\frac{1}{24}$
  &
          \\
   Fig.$~\ref{DSigma7-1245}; (x_5,x_4)_{n=5}^{\rm max}=(1,1)$:
    &${\cal D}_{7,0,0}^{(5,4)} =  \frac{5}{12}$
   &${\cal D}_{7,0,1}^{(5,4)} =  \frac{5}{8}$
   &${\cal D}_{7,0,2}^{(5,4)} =  \frac{1}{4}$
   &${\cal D}_{7,0,3}^{(5,4)} =  \frac{1}{24}$\\
 Fig.$~\ref{DSigma7-1245}; (x_3,x_2)_{n=5}^{\rm max}=\left(-\frac{1}{3},-\frac{1}{3}\right)$:
    &${\cal D}_{7,0,0}^{(3,2)} =  \frac{3}{4}$
 &${\cal D}_{7,1,0}^{(3,2)} =  -1$
  &${\cal D}_{7,2,0}^{(3,2)} =  \frac{1}{4}$
   &${\cal D}_{7,0,1}^{(3,2)} =  \frac{3}{4}$ \\
   &${\cal D}_{7,1,1}^{(3,2)} =  -1$
   &${\cal D}_{7,2,1}^{(3,2)} =  \frac{1}{4}$\\
   Fig.$~\ref{DSigma7-1245}; (x_4,x_3)_{n=5}^{\rm max}=\left(\frac{1}{3},\frac{1}{3}\right)$:
    &${\cal D}_{7,0,0}^{(4,3)} =  \frac{3}{4}$
   &${\cal D}_{7,1,0}^{(4,3)} =  -\frac{3}{4}$
     &${\cal D}_{7,0,1}^{(4,3)} =  1$
       &${\cal D}_{7,1,1}^{(4,3)} =  -1$\\
       &${\cal D}_{7,0,2}^{(4,3)} =  \frac{1}{4}$
       &${\cal D}_{7,1,2}^{(4,3)} =  -\frac{1}{4}$\\
         Fig.$~\ref{DSigma7-1335}; (x_3,x_1)_{n=5}^{\rm max}=\left(-\frac{1}{3},-1\right)$:
    &${\cal D}_{7,0,0}^{(3,1)} =  -\frac{1}{2}$
 & ${\cal D}_{7,1,0}^{(3,1)} =  \frac{7}{8}$
 & ${\cal D}_{7,2,0}^{(3,1)} = - \frac{1}{2}$
 & ${\cal D}_{7,3,0}^{(3,1)} =  \frac{1}{8}$\\
 & ${\cal D}_{7,0,1}^{(3,1)} =  -\frac{3}{4}$
 & ${\cal D}_{7,1,1}^{(3,1)} =  1$
 & ${\cal D}_{7,2,1}^{(3,1)} =  -\frac{1}{4}$\\
   Fig.$~\ref{DSigma7-1335}; (x_5,x_3)_{n=5}^{\rm max}=\left(1,\frac{1}{3}\right)$:
    &${\cal D}_{7,0,0}^{(5,3)} =  -\frac{1}{2}$
   &${\cal D}_{7,1,0}^{(5,3)} =  \frac{3}{4}$
    &${\cal D}_{7,0,2}^{(5,3)} =  -\frac{1}{2}$
     &${\cal D}_{7,0,3}^{(5,3)} =  -\frac{1}{8}$\\
      &${\cal D}_{7,0,1}^{(5,3)} =  -\frac{7}{8}$
       &${\cal D}_{7,1,1}^{(5,3)} =  1$
        &${\cal D}_{7,1,2}^{(5,3)} =  \frac{1}{4}$\\
 Fig.$~\ref{DSigma7-1335}; (x_4,x_1)_{n=5}^{\rm max}=
 \left(\frac{1}{3},-1\right)$:
    &
 ${\cal D}_{7,0,0}^{(4,1)} =  \frac{1}{2}$
 &${\cal D}_{7,1,0}^{(4,1)} =  -\frac{5}{8}$
 &${\cal D}_{7,2,0}^{(4,1)} =  \frac{1}{4}$
 &${\cal D}_{7,3,0}^{(4,1)} =  -\frac{1}{8}$\\
 &${\cal D}_{7,0,1}^{(4,1)} =  \frac{1}{2}$
 &${\cal D}_{7,1,1}^{(4,1)} =  -1$
 &${\cal D}_{7,2,1}^{(4,1)} =  \frac{1}{2}$
 &${\cal D}_{7,0,2}^{(4,1)} =  \frac{1}{4}$\\
 &${\cal D}_{7,1,2}^{(4,1)} =  -\frac{1}{4}$\\
           Fig.$~\ref{DSigma7-1335}; (x_5,x_2)_{n=5}^{\rm max}=\left(1,-\frac{1}{3}\right)$:
    &${\cal D}_{7,0,0}^{(5,2)} =  \frac{1}{2}$
    &${\cal D}_{7,1,0}^{(5,2)} =  -\frac{1}{2}$
      &${\cal D}_{7,2,0}^{(5,2)} =  \frac{1}{4}$
        &${\cal D}_{7,0,1}^{(5,2)} =  \frac{5}{8}$\\
          &${\cal D}_{7,1,1}^{(5,2)} =  -1$
            &${\cal D}_{7,2,1}^{(5,2)} =  \frac{1}{4}$
              &${\cal D}_{7,0,2}^{(5,2)} =  \frac{1}{4}$
               &${\cal D}_{7,1,2}^{(5,2)} =  -\frac{1}{2}$\\
   & ${\cal D}_{7,0,3}^{(5,2)} =  \frac{1}{8}$\\
   Fig.$~\ref{DSigma7-1524}; (x_5,x_1)_{n=5}^{\rm max}=(1,-1)$:
    &${\cal D}_{7,1,0}^{(5,1)} =  \frac{3}{8}$
  &${\cal D}_{7,3,0}^{(5,1)} =  \frac{1}{24}$
 &${\cal D}_{7,0,1}^{(5,1)} =  -\frac{3}{8}$
 &${\cal D}_{7,2,1}^{(5,1)} =  -\frac{1}{4}$\\
 &${\cal D}_{7,1,2}^{(5,1)} =  \frac{1}{4}$
 &${\cal D}_{7,0,3}^{(5,1)} =  -\frac{1}{24}$\\
    Fig.$~\ref{DSigma7-1524}; (x_4,x_2)_{n=5}^{\rm max}=\left(\frac{1}{3},-\frac{1}{3}\right)$:
    &${\cal D}_{7,0,0}^{(4,2)} =  -1$
     &${\cal D}_{7,1,0}^{(4,2)} =  \frac{3}{2}$
      &${\cal D}_{7,2,0}^{(4,2)} =  -\frac{1}{2}$
       &${\cal D}_{7,0,1}^{(4,2)} =  -\frac{3}{2}$\\
       &${\cal D}_{7,1,1}^{(4,2)} =  2$
        &${\cal D}_{7,2,1}^{(4,2)} =  -\frac{1}{2}$
         &${\cal D}_{7,0,2}^{(4,2)} =  -\frac{1}{2}$
          &${\cal D}_{7,1,2}^{(4,2)} =  \frac{1}{2}$
   \end{tabular}
\end{center}
with the corresponding figures being Figs.~\ref{DSigma7-1245},~\ref{DSigma7-1335} and~\ref{DSigma7-1524}.

\section{Two-particle rapidity correlations}

In order to measure the degree of correlation among different mini-jets in the final state it is customary to evaluate the quantity
\begin{eqnarray}
{\cal R}_{N+2} \left(x_l,x_m\right) = \sigma_{N+2} 
\frac{ \frac{ d^2 \sigma_{N+2}^{(l,m)}}{d y_l d y_m} }{\frac{d \sigma_{N+2}^{(l)}}{d y_l} \frac{d \sigma_{N+2}^{(m)}}{d y_m}}-1 
 =  
\frac{2^N}{N!}\frac{(N-m)!(l-1)!}{(l-m-1)!}   
\frac{(x_l-x_m)^{l-m-1}}{\left(1+x_l\right)^{l-1}\left(1-x_m \right)^{N-m}}
 -1
\, ,
\label{R}
\end{eqnarray}
where $Y > y_l > y_m > 0$, $l>m$ and $x_J = 2 y_J / Y$. This compares the double differential cross sections to the totally uncorrelated case where these are obtained by simply multiplying two single differential cross sections corresponding to each the jets in the chosen pair. Before discussing Eq.~(\ref{R}) we have plotted the latter products in the form
\begin{eqnarray}
\alpha^{-2(N+2)} \left(\frac{Y}{2}\right)^{2(1-N)} 
\frac{d \sigma_{N+2}^{(l)}}{d y_l} 
\frac{d \sigma_{N+2}^{(m)}}{d y_m} \, = \,    
\frac{\left(1-x_l \right)^{N-l}}{(N-l)!} \frac{\left(1+x_l\right)^{l-1}}{(l-1)!} 
\frac{\left(1-x_m \right)^{N-m}}{(N-m)!} \frac{\left(1+x_m\right)^{m-1}}{(m-1)!} 
\end{eqnarray}
in several figures. The main message to take from them is that the maxima  are now desplaced to the point $(x_l,x_m)_n^{\rm max}= \left(\frac{2 l-1-n}{n-1}, \frac{2 m-n-1}{n-1}\right)$. Therefore, the lower order examples read
\begin{center}
\begin{tabular}{lllll||}
 Fig.$~\ref{DoubleSigma3-1223}$:
 & $(x_2,x_1)_{n=3}^{\rm max}=(0,-1);$
  &$(x_3,x_2)_{n=3}^{\rm max}=(1,0);$ 
 & $(x_3,x_1)_{n=3}^{\rm max}=(1,-1);$  
  \end{tabular}
\end{center}
\begin{center}
\begin{tabular}{llllll}
 Fig.$~\ref{DoubleSigma4-1234}$:
 & $(x_2,x_1)_{n=4}^{\rm max}=\left(-\frac{1}{3},-1\right)$;
 & $(x_4,x_3)_{n=4}^{\rm max}=\left(1,\frac{1}{3}\right)$;\\
    Fig.$~\ref{DoubleSigma4-1234}$:& $(x_3,x_1)_{n=4}^{\rm max}=\left(\frac{1}{3},-1\right)$;&
          $(x_4,x_2)_{n=4}^{\rm max}=\left(1,-\frac{1}{3}\right)$;\\
       Fig.$~\ref{DoubleSigma4-1423}$:& $(x_4,x_1)_{n=4}^{\rm max}=(1,-1)$; & $(x_3,x_2)_{n=4}^{\rm max}=\left(\frac{1}{3},-\frac{1}{3}\right)$;\\
  \end{tabular}
\end{center}
\begin{center}
\begin{tabular}{lllllllll}
   Fig.$~\ref{DoubleSigma7-1245}$:
   & $(x_2,x_1)_{n=5}^{\rm max}=\left(-\frac{1}{2},-1\right)$;
   & $(x_5,x_4)_{n=5}^{\rm max}=\left(1,\frac{1}{2}\right)$;\\
 Fig.$~\ref{DoubleSigma7-1245}$:&$(x_3,x_2)_{n=5}^{\rm max}=\left(0,-\frac{1}{2}\right);$&
   $ (x_4,x_3)_{n=5}^{\rm max}=\left(\frac{1}{2},0\right)$;\\
    Fig.$~\ref{DoubleSigma7-1425}$:
    &$(x_4,x_1)_{n=5}^{\rm max}=
 \left(\frac{1}{2},-1\right)$;& $(x_5,x_2)_{n=5}^{\rm max}=\left(1,-\frac{1}{2}\right)$;\\
         Fig.$~\ref{DoubleSigma7-1425}$:&$ (x_3,x_1)_{n=5}^{\rm max}=(0,-1)$;& $(x_5,x_3)_{n=5}^{\rm max}=
         (1,0)$;\\
   Fig.$~\ref{DoubleSigma7-1524}$:& $(x_5,x_1)_{n=5}^{\rm max}=(1,-1)$;& $(x_4,x_2)_{n=5}^{\rm max}=\left(\frac{1}{2},-\frac{1}{2}\right).$
   \end{tabular}
\end{center}
 \begin{figure}
\begin{flushleft}
\hspace{1cm}\includegraphics[width=6cm]{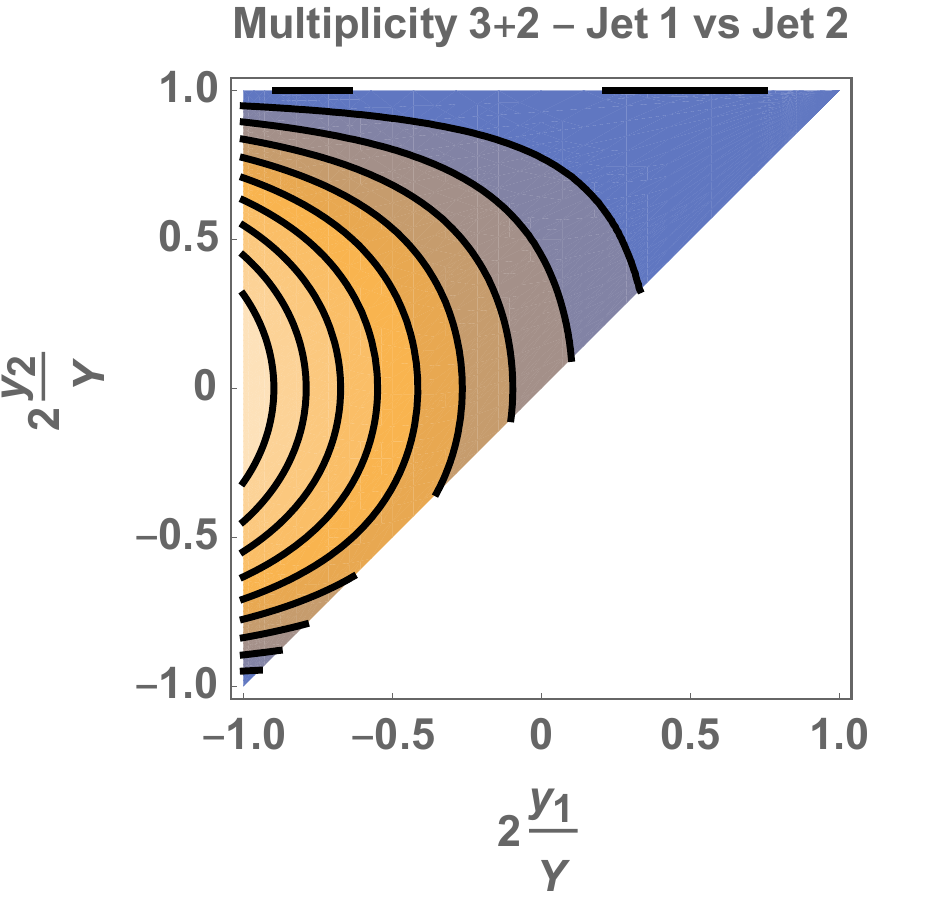}
\end{flushleft}
\vspace{-5.8cm}
\begin{flushleft}
\hspace{5.5cm}\includegraphics[width=.7cm]{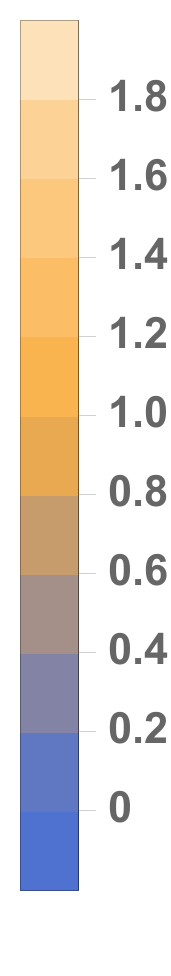}
\end{flushleft}
\vspace{-5.5cm}
\begin{center}
\hspace{7cm}\includegraphics[width=6.cm]{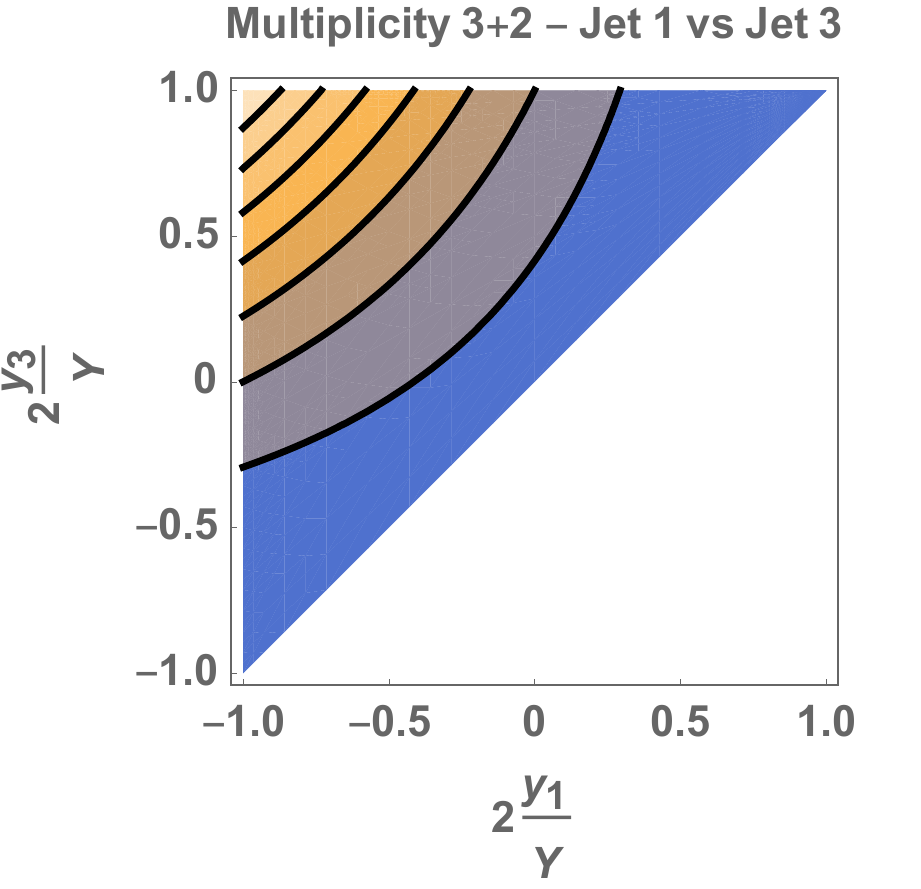}
\end{center}
\vspace{-5.8cm}
\begin{center}
\hspace{10.8cm}\includegraphics[width=.7cm]{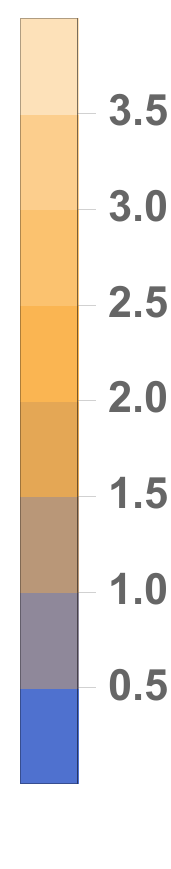}
\end{center}
\vspace{.6cm}
\caption{Left: Product of cross sections 
$\alpha^{-10} \left(\frac{Y}{2}\right)^{-4} 
\frac{d \sigma_{3+2}^{(2)}}{d y_2} 
\frac{d \sigma_{3+2}^{(1)}}{d y_1} \, = \,    
\frac{\left(1-x_1 \right)^{2}\left(1-x_2^2 \right)}{2}$ 
(the associated distribution with $x_1 \to -x_3$ 
is $\alpha^{-10} \left(\frac{Y}{2}\right)^{-4} 
\frac{d \sigma_{3+2}^{(3)}}{d y_3} 
\frac{d \sigma_{3+2}^{(2)}}{d y_2} \, = \,    
 \frac{\left(1-x_2^2 \right) \left(1+x_3\right)^{2}}{2}$). Right: Product of cross sections $
 \alpha^{-10} \left(\frac{Y}{2}\right)^{-4} 
\frac{d \sigma_{3+2}^{(3)}}{d y_3} 
\frac{d \sigma_{3+2}^{(1)}}{d y_1} \, = \,    
 \frac{\left(1-x_1 \right)^{2}\left(1+x_3\right)^{2}}{4}$. In both $x_L = 2 y_L/Y$. }
\label{DoubleSigma3-1223}
\end{figure}
 \begin{figure}
\begin{flushleft}
\hspace{1cm}\includegraphics[width=6cm]{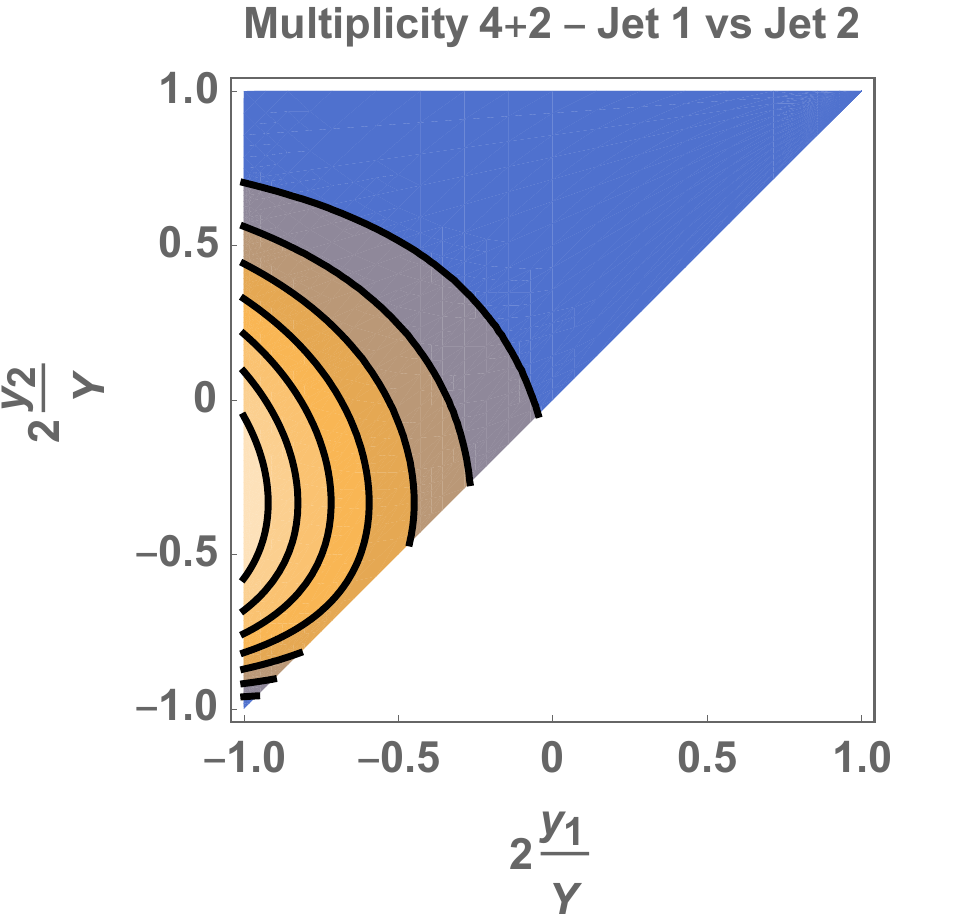}
\end{flushleft}
\vspace{-5.6cm}
\begin{flushleft}
\hspace{5.5cm}\includegraphics[width=.7cm]{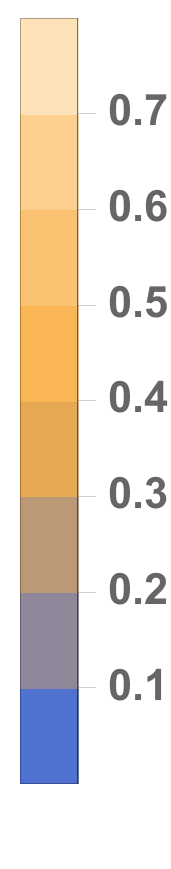}
\end{flushleft}
\vspace{-5.3cm}
\begin{center}
\hspace{7cm}\includegraphics[width=6.cm]{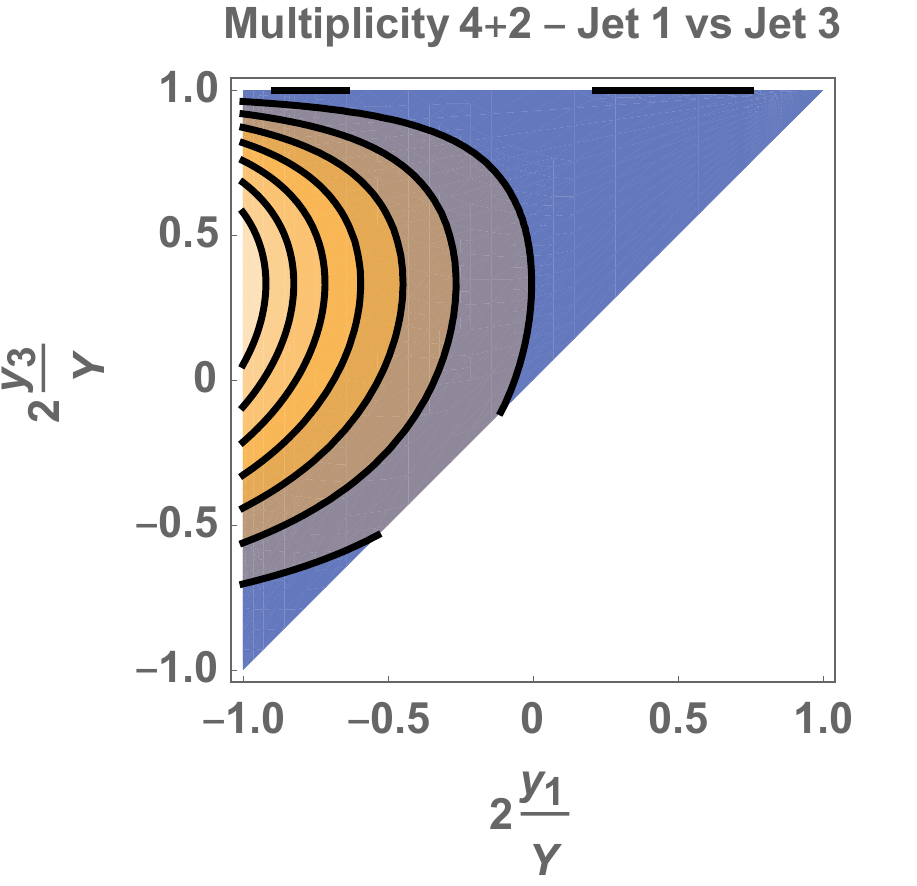}
\end{center}
\vspace{-5.4cm}
\begin{center}
\hspace{10.8cm}\includegraphics[width=.7cm]{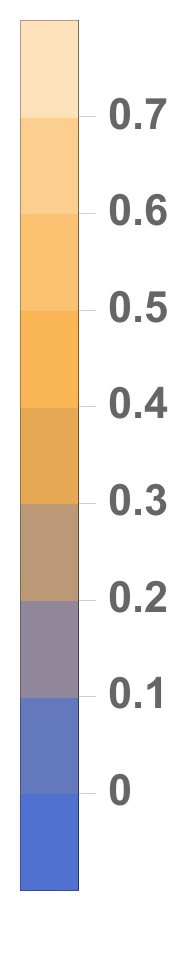}
\end{center}
\vspace{.2cm}
\caption{Left: Product of cross sections $
\alpha^{-12} \left(\frac{Y}{2}\right)^{-6} 
\frac{d \sigma_{4+2}^{(2)}}{d y_2} 
\frac{d \sigma_{4+2}^{(1)}}{d y_1} \, = \,    
\frac{\left(1-x_1 \right)^{3}\left(1-x_2 \right)^{2}\left(1+x_2\right)}{12}$
(the related by $x_2 \to -x_3$ and $x_1 \to -x_4$ distribution is 
$ \alpha^{-12} \left(\frac{Y}{2}\right)^{-6} 
\frac{d \sigma_{4+2}^{(4)}}{d y_4} 
\frac{d \sigma_{4+2}^{(3)}}{d y_3} \, = \,    
 \frac{\left(1-x_3 \right)\left(1+x_3\right)^{2}\left(1+x_4\right)^{3}}{12} $). 
 Right: Product of cross sections $
 \alpha^{-12} \left(\frac{Y}{2}\right)^{-6} 
\frac{d \sigma_{4+2}^{(3)}}{d y_3} 
\frac{d \sigma_{4+2}^{(1)}}{d y_1} \, = \,    
\frac{\left(1-x_3 \right) \left(1+x_3\right)^{2} \left(1-x_1 \right)^{3}}{12}  $
(the related by $x_3 \to -x_2$ and $x_1 \to -x_4$ distribution is 
$\alpha^{-12} \left(\frac{Y}{2}\right)^{-6} 
\frac{d \sigma_{4+2}^{(4)}}{d y_4} 
\frac{d \sigma_{4+2}^{(2)}}{d y_2} \, = \,    
\frac{\left(1+x_2\right)\left(1-x_2 \right)^{2} \left(1+x_4\right)^{3} }{12}$). In both $x_L = 2 y_L/Y$.}
\label{DoubleSigma4-1234}
\end{figure}
 \begin{figure}
\begin{flushleft}
\hspace{1cm}\includegraphics[width=6cm]{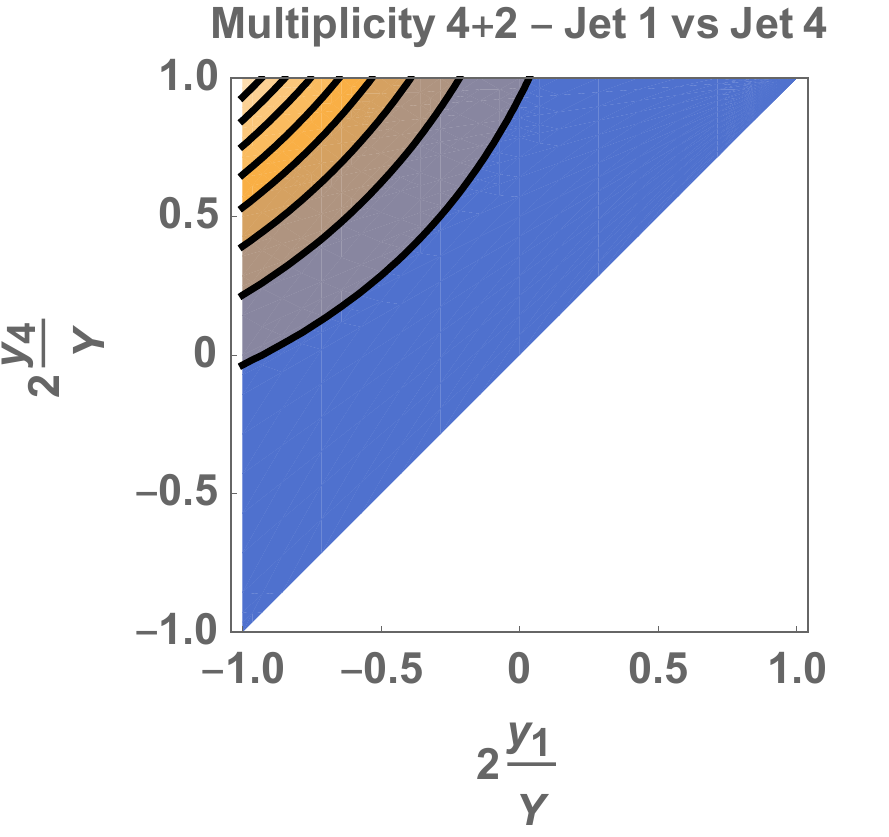}
\end{flushleft}
\vspace{-5.9cm}
\begin{flushleft}
\hspace{5.5cm}\includegraphics[width=.7cm]{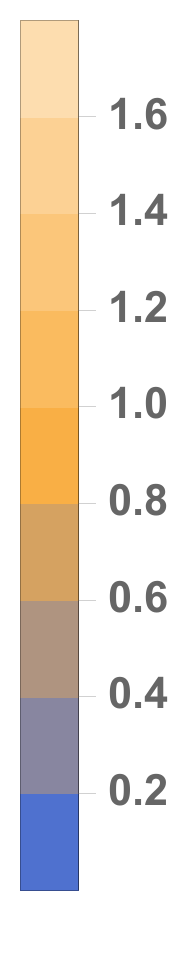}
\end{flushleft}
\vspace{-5.3cm}
\begin{center}
\hspace{7cm}\includegraphics[width=6.cm]{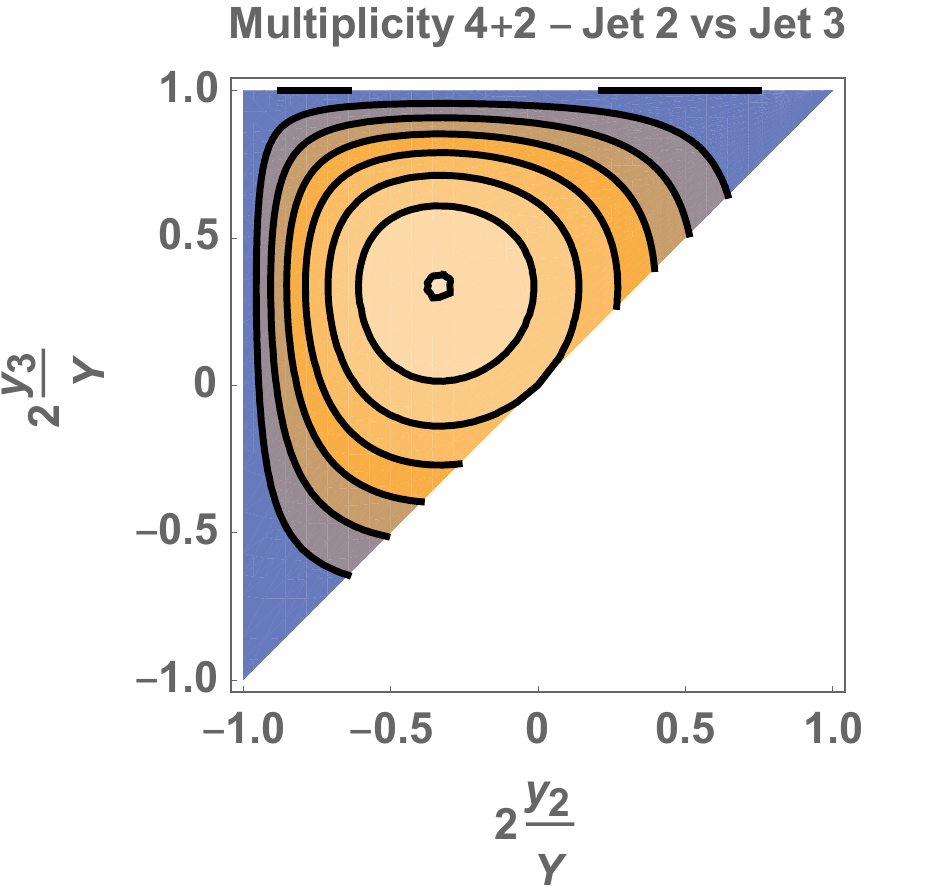}
\end{center}
\vspace{-5.4cm}
\begin{center}
\hspace{10.8cm}\includegraphics[width=.7cm]{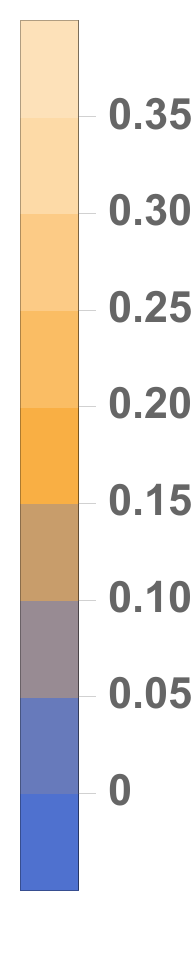}
\end{center}
\vspace{.6cm}
\caption{Left: Product of cross sections $
 \alpha^{-12} \left(\frac{Y}{2}\right)^{-6} 
\frac{d \sigma_{4+2}^{(4)}}{d y_4} 
\frac{d \sigma_{4+2}^{(1)}}{d y_1} \, = \,    
 \frac{\left(1-x_1 \right)^{3}\left(1+x_4\right)^{3}}{36} $. 
 Right: Product of cross sections $
 \alpha^{-12} \left(\frac{Y}{2}\right)^{-6} 
\frac{d \sigma_{4+2}^{(3)}}{d y_3} 
\frac{d \sigma_{4+2}^{(m)}}{d y_m} \, = \,    
 \frac{\left(1+x_2\right)\left(1-x_2 \right)^{2}\left(1-x_3 \right)\left(1+x_3\right)^{2}}{4}$. In both $x_L = 2 y_L/Y$. }
\label{DoubleSigma4-1423}
\end{figure}
 \begin{figure}
\begin{flushleft}
\hspace{1cm}\includegraphics[width=6.1cm]{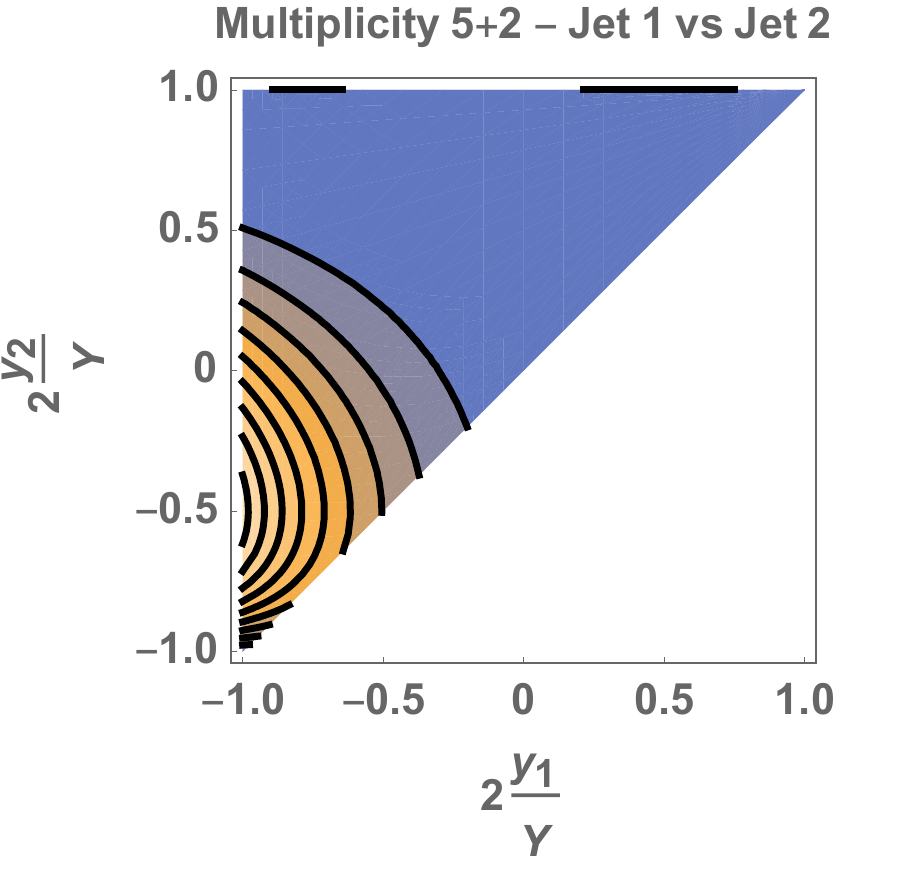}
\end{flushleft}
\vspace{-5.6cm}
\begin{flushleft}
\hspace{5.5cm}\includegraphics[width=.7cm]{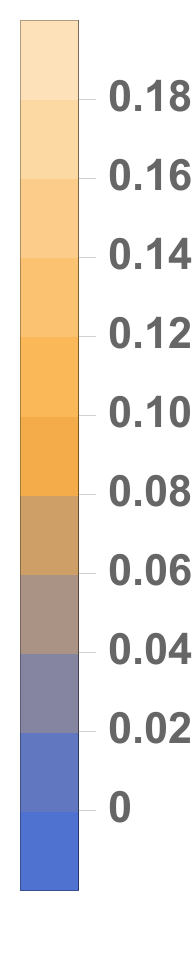}
\end{flushleft}
\vspace{-5.2cm}
\begin{center}
\hspace{7cm}\includegraphics[width=6.cm]{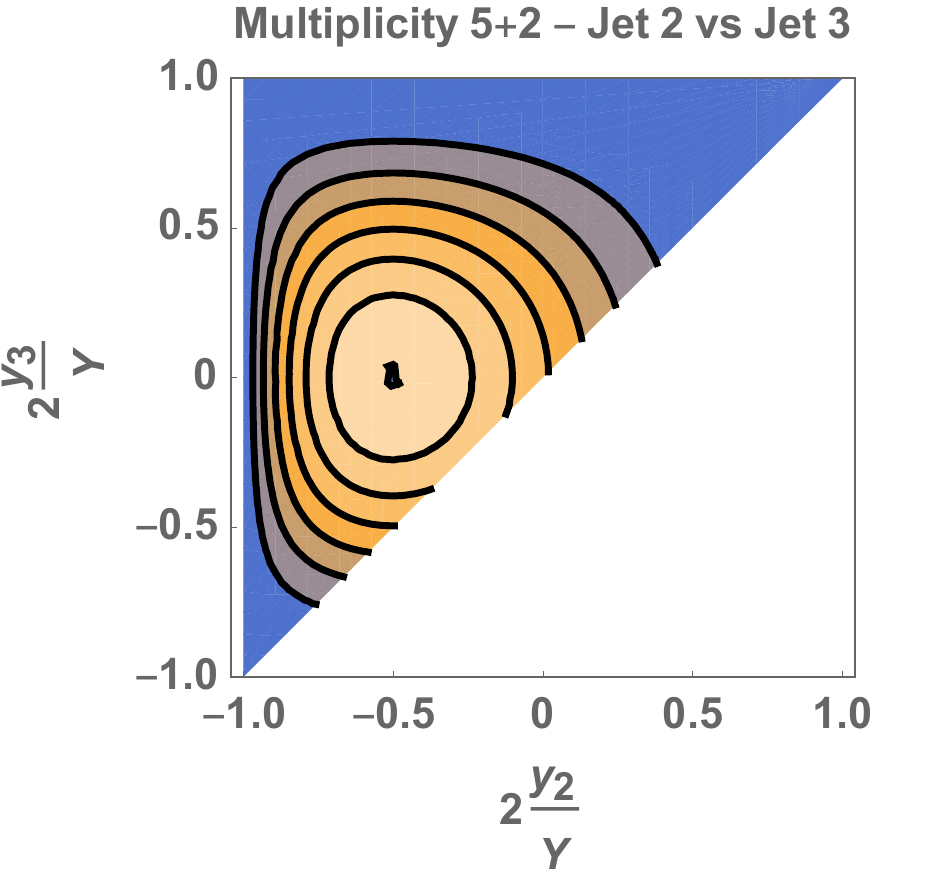}
\end{center}
\vspace{-5.4cm}
\begin{center}
\hspace{10.8cm}\includegraphics[width=.7cm]{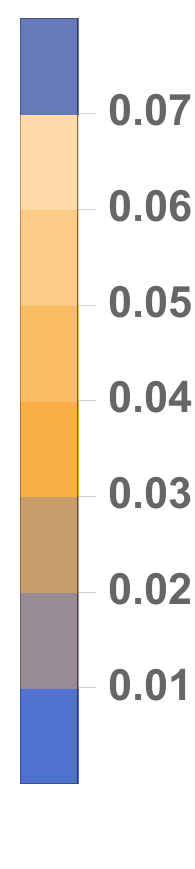}
\end{center}
\vspace{1.2cm}
\caption{Left: Product of cross sections $
\alpha^{-14} \left(\frac{Y}{2}\right)^{-8} 
\frac{d \sigma_{5+2}^{(2)}}{d y_2} 
\frac{d \sigma_{5+2}^{(1)}}{d y_1} \, = \,    
\frac{\left(1-x_1 \right)^{4}\left(1+x_2\right) \left(1-x_2 \right)^{3}}{144} $ 
(the related by $x_2 \to -x_4$ and $x_1 \to - x_5$ distribution is 
$\alpha^{-14} \left(\frac{Y}{2}\right)^{-8} 
\frac{d \sigma_{5+2}^{(5)}}{d y_5} 
\frac{d \sigma_{5+2}^{(4)}}{d y_4} \, = \,    
 \frac{\left(1-x_4 \right) \left(1+x_4\right)^{3}\left(1+x_5\right)^{4}}{144} 
$). Right: Product of cross sections $
\alpha^{-14} \left(\frac{Y}{2}\right)^{-8} 
\frac{d \sigma_{5+2}^{(3)}}{d y_3} 
\frac{d \sigma_{5+2}^{(2)}}{d y_2} \, = \,    
\frac{ \left(1+x_2\right) \left(1-x_2 \right)^{3}\left(1-x_3 \right)^{2}\left(1+x_3\right)^{2} }{24} $ 
(the related by $x_2 \to -x_4$ distribution is 
$\alpha^{-14} \left(\frac{Y}{2}\right)^{-8} 
\frac{d \sigma_{5+2}^{(4)}}{d y_4} 
\frac{d \sigma_{5+2}^{(3)}}{d y_3} \, = \,    
 \frac{\left(1-x_3 \right)^{2} \left(1+x_3\right)^{2} \left(1-x_4 \right)\left(1+x_4\right)^{3}}{24} $). In both $x_L = 2 y_L/Y$. }
\label{DoubleSigma7-1245}
\end{figure}
 \begin{figure}
\begin{flushleft}
\hspace{1cm}\includegraphics[width=6cm]{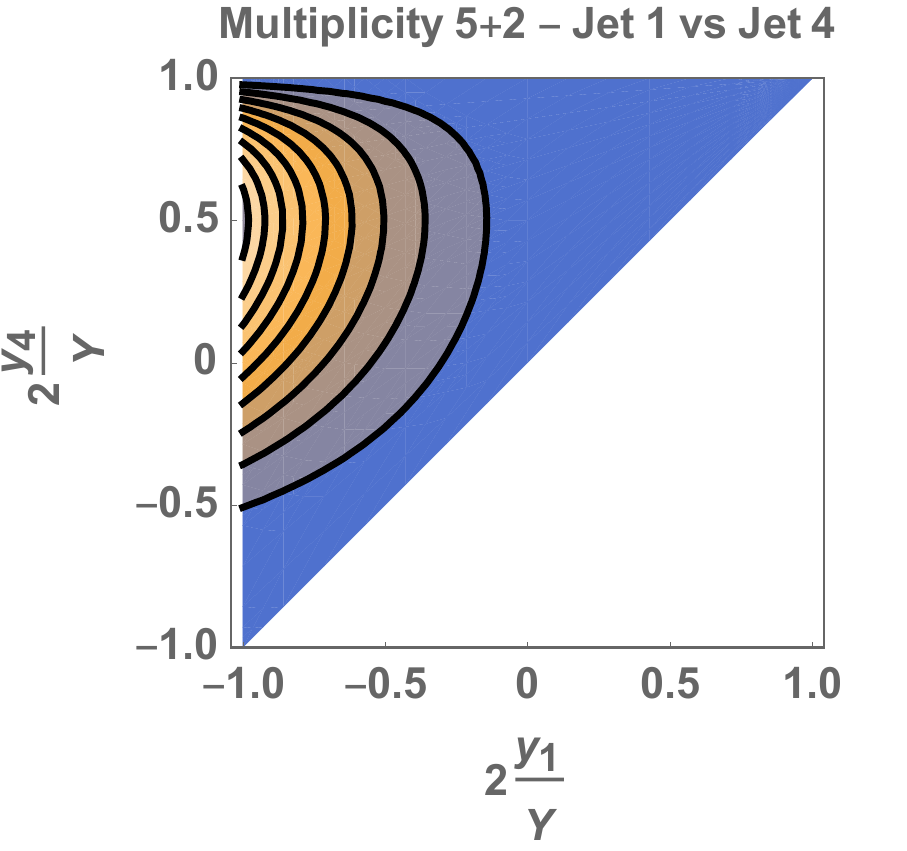}
\end{flushleft}
\vspace{-5.5cm}
\begin{flushleft}
\hspace{5.5cm}\includegraphics[width=.7cm]{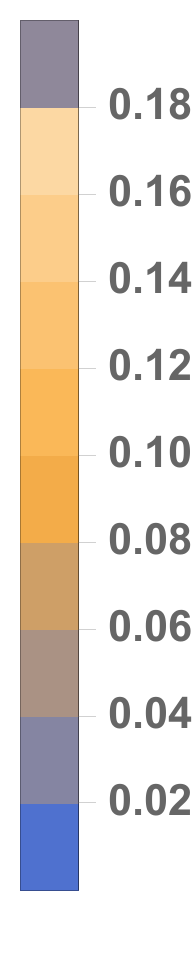}
\end{flushleft}
\vspace{-5.1cm}
\begin{center}
\hspace{7cm}\includegraphics[width=6.cm]{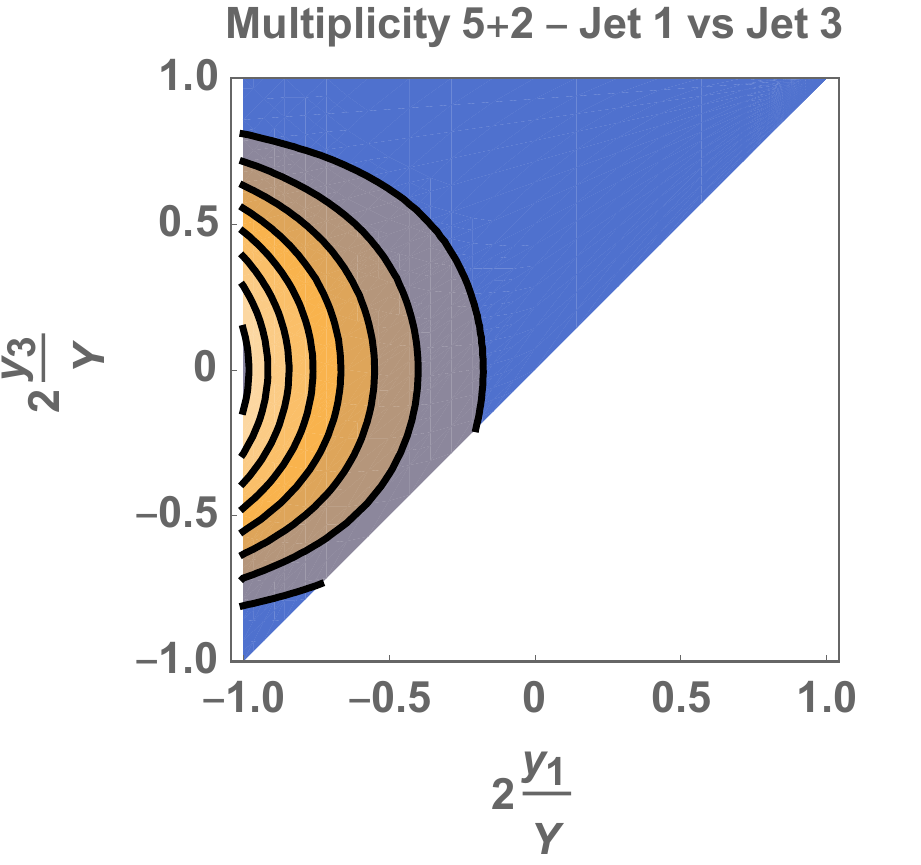}
\end{center}
\vspace{-5.5cm}
\begin{center}
\hspace{10.8cm}\includegraphics[width=.7cm]{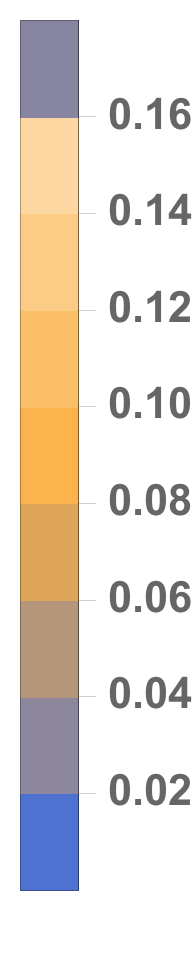}
\end{center}
\vspace{.6cm}
\caption{Left: Product of cross sections $
\alpha^{-14} \left(\frac{Y}{2}\right)^{-8} 
\frac{d \sigma_{5+2}^{(4)}}{d y_4} 
\frac{d \sigma_{5+2}^{(1)}}{d y_1} \, = \,    
 \frac{ \left(1-x_1 \right)^{4} \left(1-x_4 \right) \left(1+x_4\right)^{3}}{144}  $
 (the related by $x_1 \to - x_5$ and $x_4 \to -x_2$ distribution is 
 $ \alpha^{-14} \left(\frac{Y}{2}\right)^{-8} 
\frac{d \sigma_{5+2}^{(5)}}{d y_5} 
\frac{d \sigma_{5+2}^{(2)}}{d y_2} \, = \,    
 \frac{\left(1+x_2\right)\left(1-x_2 \right)^{3} \left(1+x_5\right)^{4}}{144}$. 
 Right: Product of cross sections $
\alpha^{-14} \left(\frac{Y}{2}\right)^{-8} 
\frac{d \sigma_{5+2}^{(3)}}{d y_3} 
\frac{d \sigma_{5+2}^{(1)}}{d y_1} \, = \,    
\frac{\left(1-x_1 \right)^{4} \left(1-x_3 \right)^{2} \left(1+x_3\right)^{2}}{96} $
(the related by $x_1 \to - x_5$ distribution is 
$\alpha^{-14} \left(\frac{Y}{2}\right)^{-8} 
\frac{d \sigma_{5+2}^{(5)}}{d y_5} 
\frac{d \sigma_{5+2}^{(3)}}{d y_3} \, = \,    
 \frac{\left(1-x_3 \right)^{2} \left(1+x_3\right)^{2}\left(1+x_5\right)^{4}}{96}$).
 In both $x_L = 2 y_L/Y$.  }
\label{DoubleSigma7-1425}
\end{figure}
 \begin{figure}
\begin{flushleft}
\hspace{1cm}\includegraphics[width=6cm]{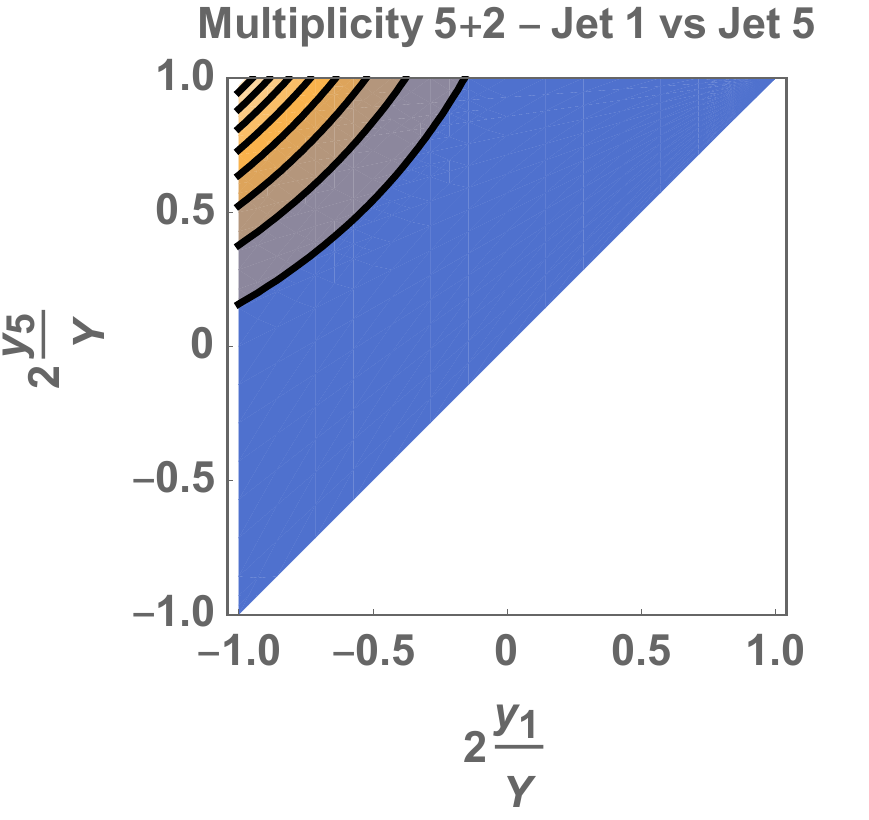}
\end{flushleft}
\vspace{-5.6cm}
\begin{flushleft}
\hspace{5.5cm}\includegraphics[width=.7cm]{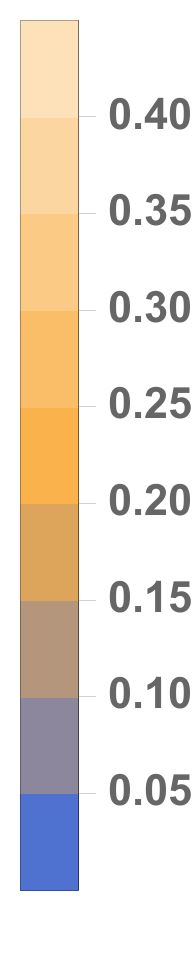}
\end{flushleft}
\vspace{-5.1cm}
\begin{center}
\hspace{7cm}\includegraphics[width=5.8cm]{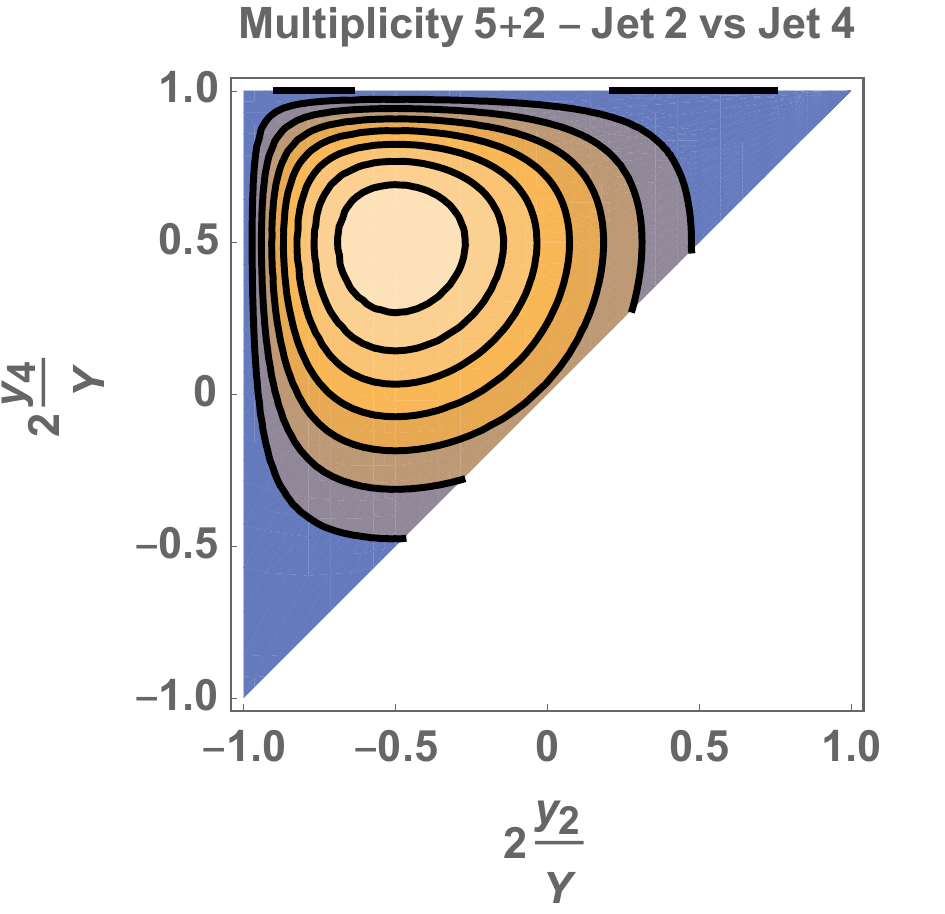}
\end{center}
\vspace{-5.1cm}
\begin{center}
\hspace{10.8cm}\includegraphics[width=.7cm]{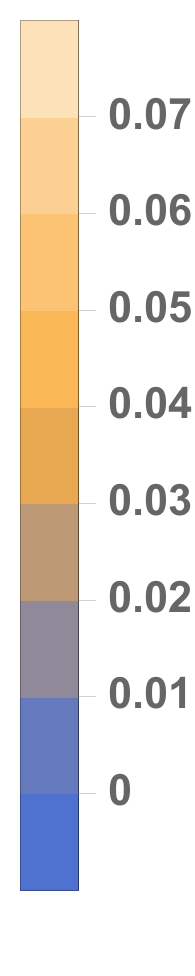}
\end{center}
\vspace{.6cm}
\caption{Left: Product of cross sections $
\alpha^{-14} \left(\frac{Y}{2}\right)^{-8} 
\frac{d \sigma_{5+2}^{(5)}}{d y_5} 
\frac{d \sigma_{5+2}^{(1)}}{d y_1} \, = \,    
 \frac{\left(1-x_1 \right)^{4} \left(1+x_5\right)^{4}}{576} $. 
 Right: Product of cross sections $
\alpha^{-14} \left(\frac{Y}{2}\right)^{-8} 
\frac{d \sigma_{5+2}^{(4)}}{d y_4} 
\frac{d \sigma_{5+2}^{(2)}}{d y_2} \, = \,    
 \frac{\left(1+x_2\right) \left(1-x_2 \right)^{3}
 \left(1-x_4 \right)
 \left(1+x_4\right)^{3}}{36} $. In both $x_L = 2 y_L/Y$. }
\label{DoubleSigma7-1524}
\end{figure}
It will be interesting to observe where the experimental data at the LHC are placed: are they closer to regions with some correlation in rapidity space or in totally decorrelated maxima? Note that the maxima in the previous section and these coincide only for those minijets at the furthest distance from each other in rapidity with  $l=n$ and $m=1$, {\it i.e.} $(x_n,x_1) = (1,-1)$ in this exploratory analysis. 

\begin{figure}
\begin{flushleft}
\hspace{1cm}\includegraphics[width=6cm]{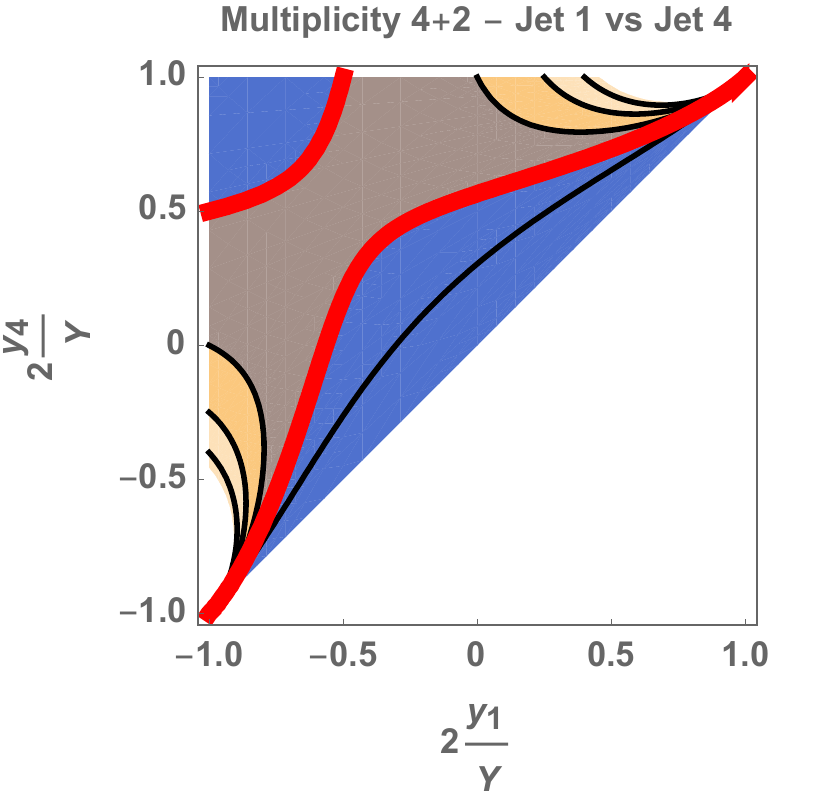}
\end{flushleft}
\vspace{-4.7cm}
\begin{flushleft}
\hspace{5.5cm}\includegraphics[width=.8cm]{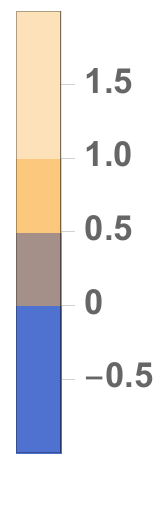}
\end{flushleft}
\vspace{-5.3cm}
\begin{center}
\hspace{7cm}\includegraphics[width=6.cm]{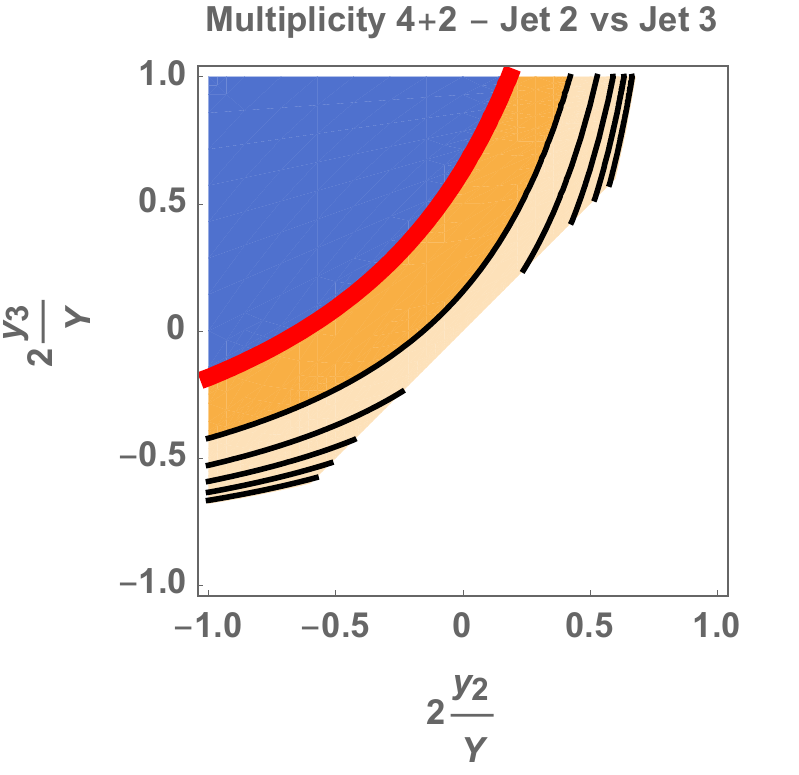}
\end{center}
\vspace{-5.4cm}
\begin{center}
\hspace{10.8cm}\includegraphics[width=.6cm]{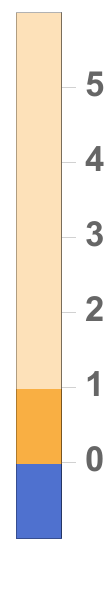}
\end{center}
\vspace{.6cm}
\caption{Left:  ${\cal R}_{4+2} \left(x_4,x_1\right) = \sigma_{4+2} 
\frac{ \frac{ d^2 \sigma_{4+2}^{(4,1)}}{d y_4 d y_1} }{\frac{d \sigma_{4+2}^{(4)}}{d y_4} \frac{d \sigma_{4+2}^{(1)}}{d y_1}}-1$. 
 Right: ${\cal R}_{4+2} \left(x_3,x_2\right) = \sigma_{4+2} 
\frac{ \frac{ d^2 \sigma_{4+2}^{(3,2)}}{d y_3 d y_2} }{\frac{d \sigma_{4+2}^{(3)}}{d y_3} \frac{d \sigma_{4+2}^{(2)}}{d y_2}}-1$. }
\label{R4-1423}
\end{figure}

\begin{figure}
\begin{flushleft}
\hspace{1cm}\includegraphics[width=6cm]{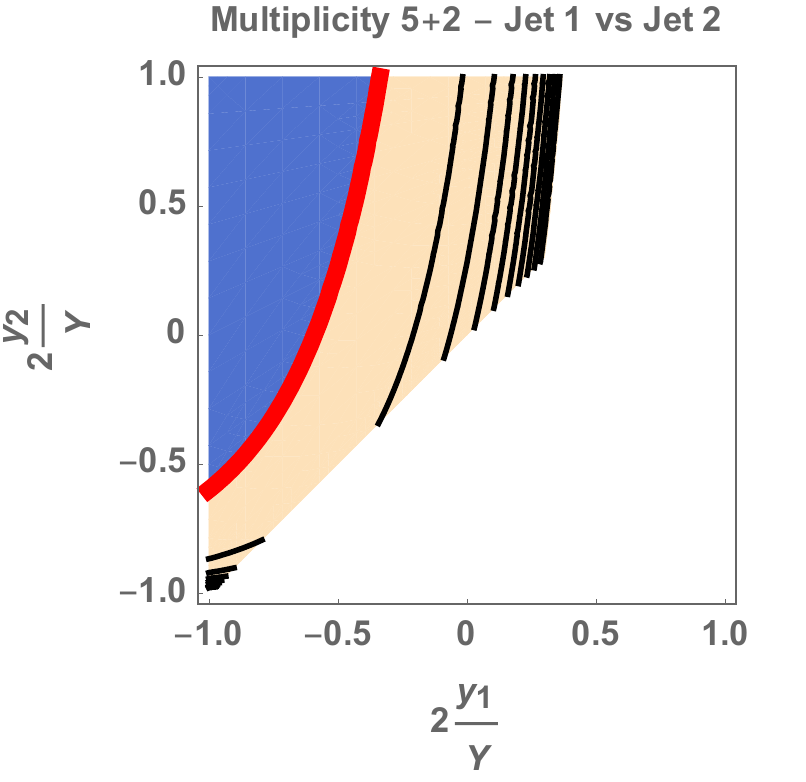}
\end{flushleft}
\vspace{-5.6cm}
\begin{flushleft}
\hspace{5.5cm}\includegraphics[width=.6cm]{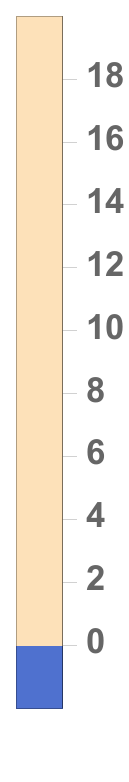}
\end{flushleft}
\vspace{-5.4cm}
\begin{center}
\hspace{7cm}\includegraphics[width=6.cm]{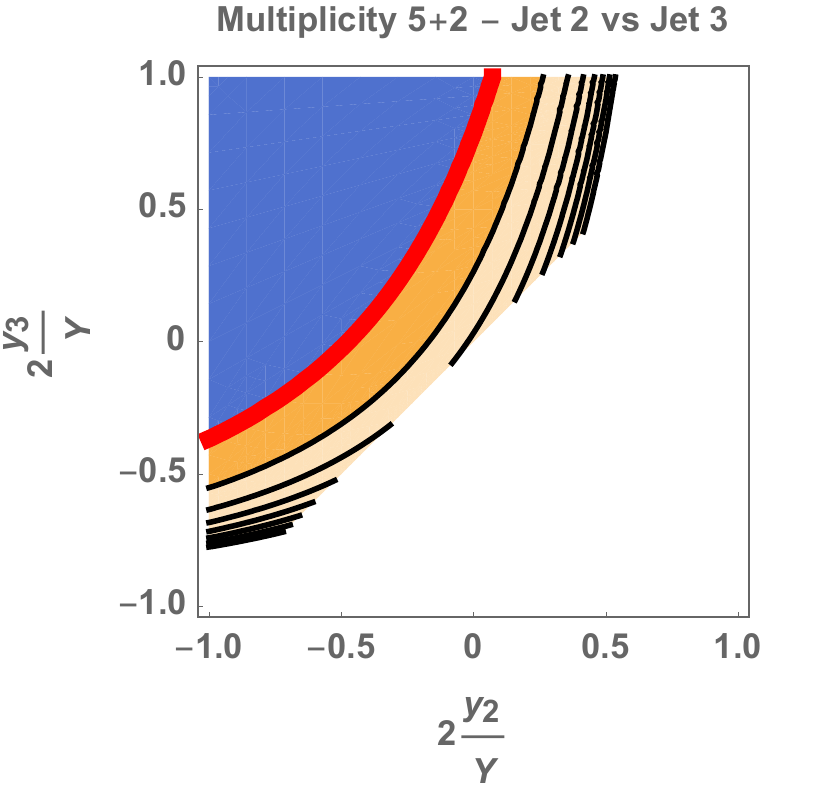}
\end{center}
\vspace{-6.cm}
\begin{center}
\hspace{10.8cm}\includegraphics[width=.6cm]{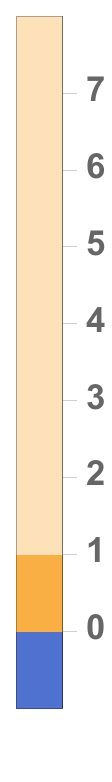}
\end{center}
\vspace{.4cm}
\caption{Left:  ${\cal R}_{5+2} \left(x_2,x_1\right) = \sigma_{5+2} 
\frac{ \frac{ d^2 \sigma_{5+2}^{(2,1)}}{d y_2 d y_1} }{\frac{d \sigma_{5+2}^{(2)}}{d y_2} \frac{d \sigma_{5+2}^{(1)}}{d y_1}}-1$ 
(the associated distribution ${\cal R}_{5+2} \left(x_5,x_4\right) = \sigma_{5+2} 
\frac{ \frac{ d^2 \sigma_{5+2}^{(5,4)}}{d y_5 d y_4} }{\frac{d \sigma_{5+2}^{(5)}}{d y_5} \frac{d \sigma_{5+2}^{(4)}}{d y_4}}-1$ is its mirror image wrt. 
the line $y_5=-y_4$). Right:  ${\cal R}_{5+2} \left(x_3,x_2\right) = \sigma_{5+2} 
\frac{ \frac{ d^2 \sigma_{5+2}^{(3,2)}}{d y_3 d y_2} }{\frac{d \sigma_{5+2}^{(3)}}{d y_3} \frac{d \sigma_{5+2}^{(2)}}{d y_2}}-1$ (the associated distribution ${\cal R}_{5+2} \left(x_4,x_3\right) = \sigma_{5+2} 
\frac{ \frac{ d^2 \sigma_{5+2}^{(4,3)}}{d y_4 d y_3} }{\frac{d \sigma_{5+2}^{(4)}}{d y_4} \frac{d \sigma_{5+2}^{(3)}}{d y_3}}-1$  is its mirror image wrt. 
the line $y_4=-y_3$).}
\label{R7-1245}
\end{figure}

\begin{figure}
\begin{flushleft}
\hspace{1cm}\includegraphics[width=6cm]{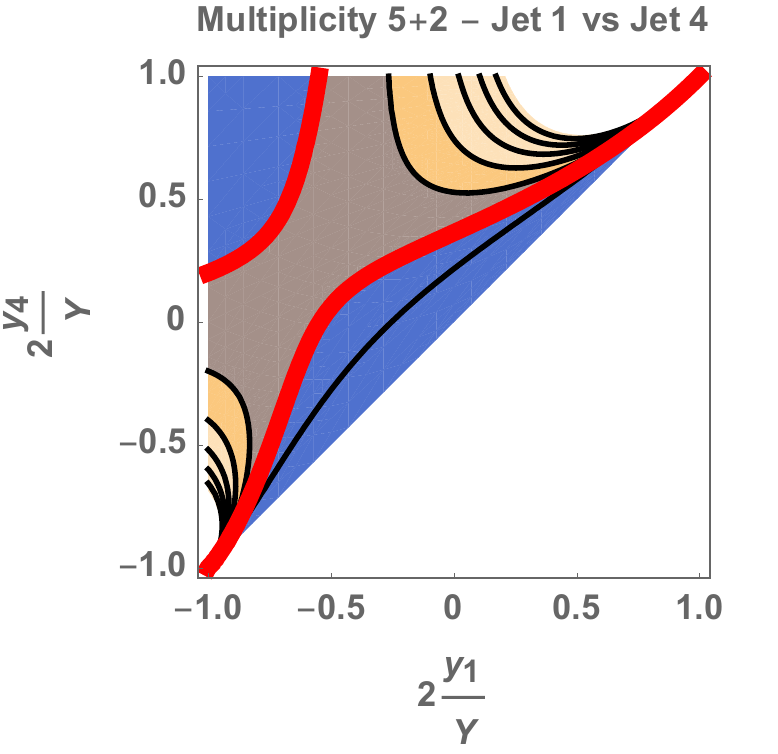}
\end{flushleft}
\vspace{-5.2cm}
\begin{flushleft}
\hspace{5.5cm}\includegraphics[width=.7cm]{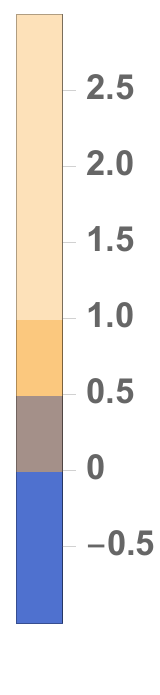}
\end{flushleft}
\vspace{-5.1cm}
\begin{center}
\hspace{7cm}\includegraphics[width=6.cm]{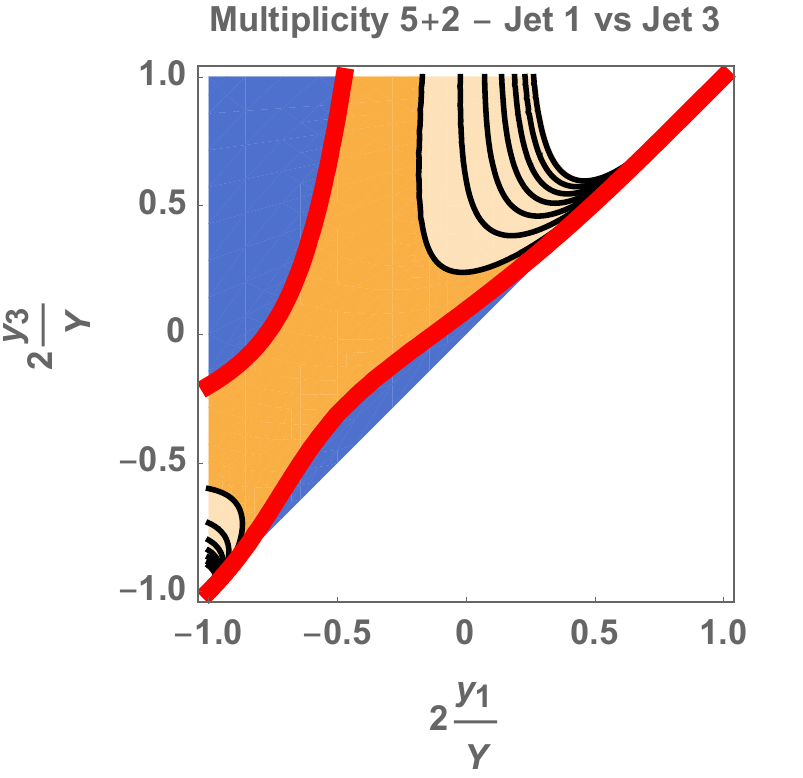}
\end{center}
\vspace{-5.5cm}
\begin{center}
\hspace{10.8cm}\includegraphics[width=.5cm]{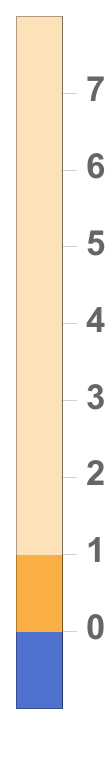}
\end{center}
\vspace{.6cm}
\caption{Left:  ${\cal R}_{5+2} \left(x_4,x_1\right) = \sigma_{5+2} 
\frac{ \frac{ d^2 \sigma_{5+2}^{(4,1)}}{d y_4 d y_1} }{\frac{d \sigma_{5+2}^{(4)}}{d y_4} \frac{d \sigma_{5+2}^{(1)}}{d y_1}}-1$ (the associated distribution ${\cal R}_{5+2} \left(x_5,x_2\right) = \sigma_{5+2} 
\frac{ \frac{ d^2 \sigma_{5+2}^{(5,2)}}{d y_5 d y_2} }{\frac{d \sigma_{5+2}^{(5)}}{d y_5} \frac{d \sigma_{5+2}^{(2)}}{d y_2}}-1$ is its mirror image wrt. the line $y_5=-y_2$). Right:  ${\cal R}_{5+2} \left(x_3,x_1\right) = \sigma_{5+2} 
\frac{ \frac{ d^2 \sigma_{5+2}^{(3,1)}}{d y_3 d y_1} }{\frac{d \sigma_{5+2}^{(3)}}{d y_3} \frac{d \sigma_{5+2}^{(1)}}{d y_1}}-1$  (the associated distribution ${\cal R}_{5+2} \left(x_5,x_3\right) = \sigma_{5+2} 
\frac{ \frac{ d^2 \sigma_{5+2}^{(5,3)}}{d y_5 d y_3} }{\frac{d \sigma_{5+2}^{(5)}}{d y_5} \frac{d \sigma_{5+2}^{(3)}}{d y_3}}-1$ is its mirror image wrt. the line $y_5=-y_3$).  }
\label{R7-1425}
\end{figure}

\begin{figure}
\begin{flushleft}
\hspace{1cm}\includegraphics[width=6cm]{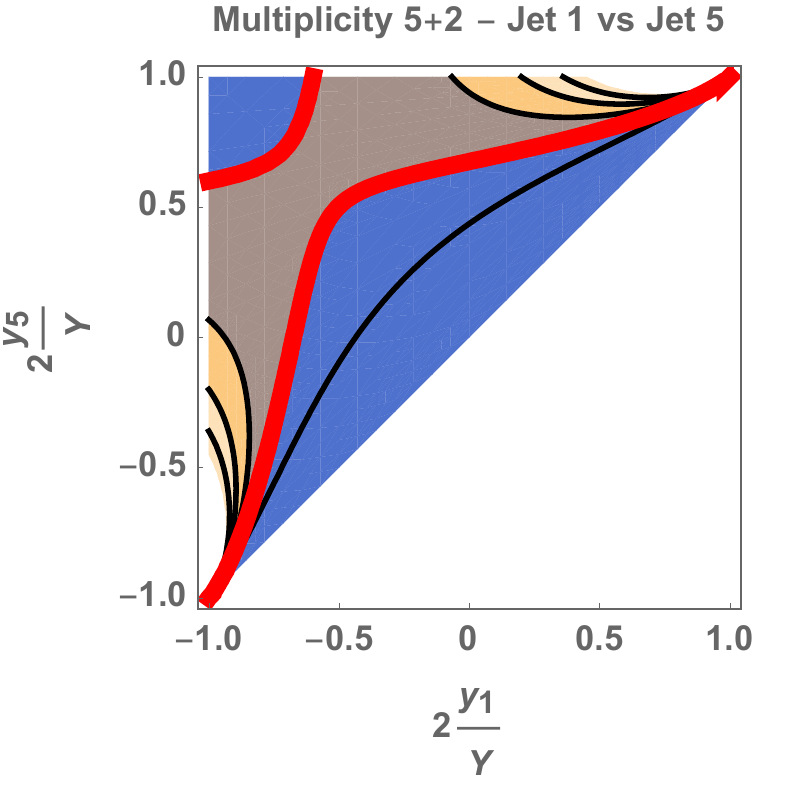}
\end{flushleft}
\vspace{-4.8cm}
\begin{flushleft}
\hspace{5.5cm}\includegraphics[width=.7cm]{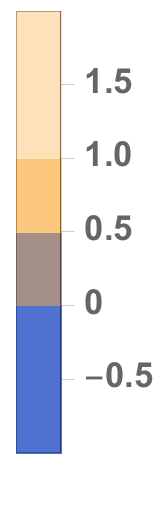}
\end{flushleft}
\vspace{-4.8cm}
\begin{center}
\hspace{7cm}\includegraphics[width=6.cm]{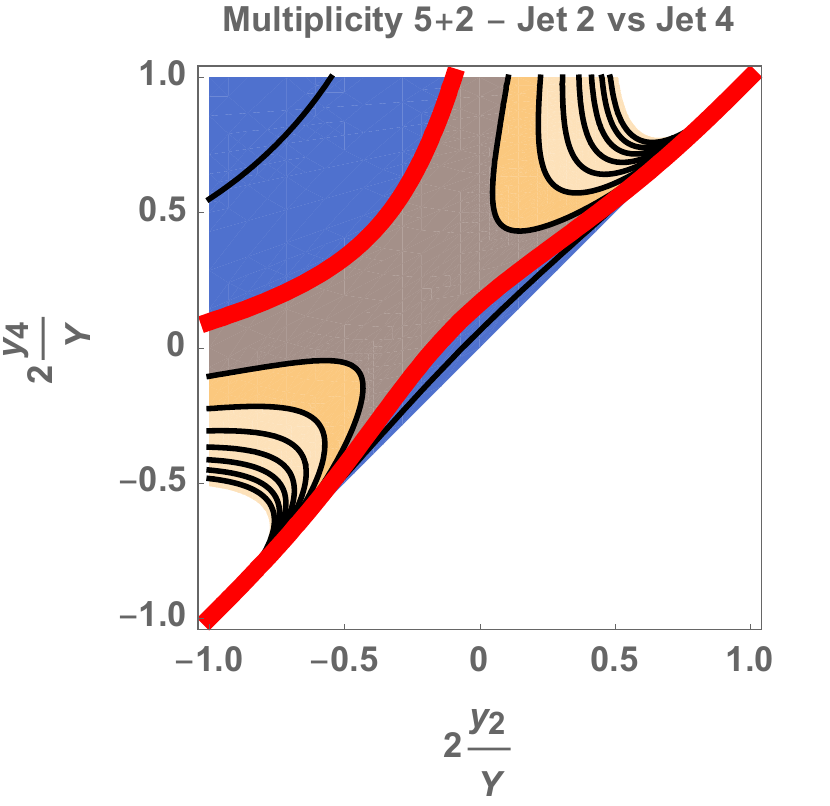}
\end{center}
\vspace{-5.3cm}
\begin{center}
\hspace{10.8cm}\includegraphics[width=.7cm]{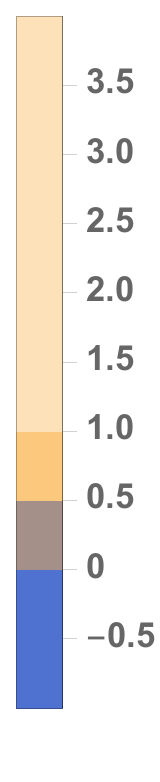}
\end{center}
\vspace{.6cm}
\caption{Left:  ${\cal R}_{5+2} \left(x_5,x_1\right) = \sigma_{5+2} 
\frac{ \frac{ d^2 \sigma_{5+2}^{(5,1)}}{d y_5 d y_1} }{\frac{d \sigma_{5+2}^{(5)}}{d y_5} \frac{d \sigma_{5+2}^{(1)}}{d y_1}}-1$. 
 Right: ${\cal R}_{5+2} \left(x_4,x_2\right) = \sigma_{5+2} 
\frac{ \frac{ d^2 \sigma_{5+2}^{(4,2)}}{d y_4 d y_2} }{\frac{d \sigma_{5+2}^{(4)}}{d y_4} \frac{d \sigma_{5+2}^{(2)}}{d y_2}}-1$. }
\label{R7-1524}
\end{figure}

We can finish our discussion investigating Eq.~(\ref{R}). We 
summarize some results in the following description of different figures where the red lines in each of them correspond to the 
values of the pair of rapidities for which ${\cal R}=0$ (the white regions  correspond to sectors of very rapid growth of ${\cal R}$). The closer the experimental results lie to these regions, the more decorrelated the mini-jet production will be. We provide the analytic formula for these lines. We again only show the lower multiplicities since these will be statistically richer in the recent LHC data where the number of events decreases rapidly with the multiplicity. In some cases there exist two branches for the lines of zeroes: 
\begin{center}
\begin{tabular}{llllll}
 Fig.$~\ref{R3-1223}$:
 & $x_2=\frac{8}{3 (1-x_1)^2}-1$;\\
  &$x_3 = \frac{2 \sqrt{2}}{\sqrt{3 (1-x_2)}}-1;$ \\
 Fig.$~\ref{R3-1223}$: 
 & $x_3 = \frac{\pm 4 \sqrt{1+3 {x_1} \left(1+ {x_1}- {x_1}^2\right)}-3
   ({x_1}-2) {x_1}+5}{3 (1-{x_1})^2};$ 
  &
 & \\
 Fig.$~\ref{R4-1234}$:
 & $x_2 = \frac{4}{(1-{x_1})^3}-1$;\\
 & $x_4 = \frac{2^{2/3}}{(1-{x_3})^{1/3}}-1$;\\
    Fig.$~\ref{R4-1234}$:
    & $x_3 = \frac{2 \left(\pm \sqrt{2} \sqrt{{x_1}^4-2 {x_1}^3+2
   {x_1}+1}+2\right)}{(1-{x_1})^3}-1$;&\\
       &   $x_4 = \frac{2}{1-x_2} \left(\frac{2}{3 h(x_2)}+h(x_2)\right)-1$; where $h(x) = \left(-\frac{1}{2}\right)^{1/3}
       \left(1-x_2^2 \pm 
       \sqrt{x_2^2 (x_2^2-2)-\frac{5}{27}}\right)^{1/3}$.\\
   \end{tabular}
\end{center}
For figures~\ref{R4-1423} to~\ref{R7-1524} the analytic expressions  
are more complicated. 
 \begin{figure}
\begin{flushleft}
\hspace{1cm}\includegraphics[width=6cm]{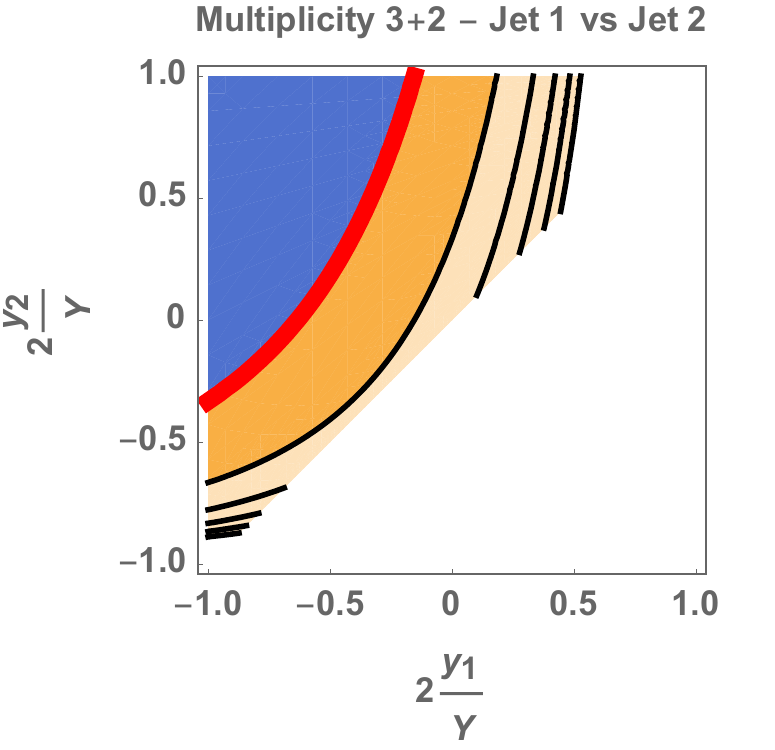}
\end{flushleft}
\vspace{-5.8cm}
\begin{flushleft}
\hspace{5.5cm}\includegraphics[width=.7cm]{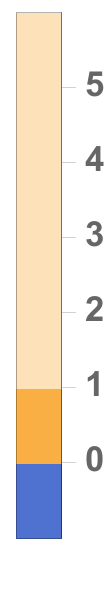}
\end{flushleft}
\vspace{-5.3cm}
\begin{center}
\hspace{7cm}\includegraphics[width=6.cm]{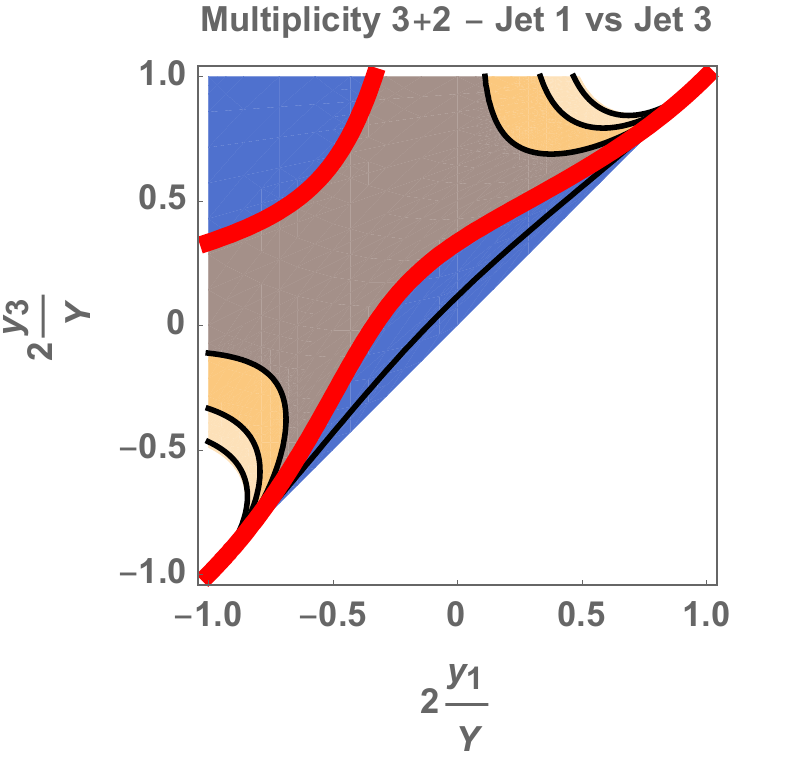}
\end{center}
\vspace{-5.cm}
\begin{center}
\hspace{10.8cm}\includegraphics[width=.7cm]{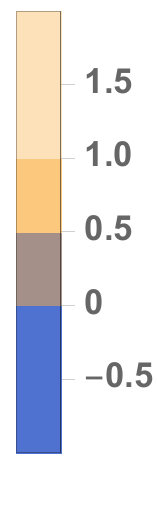}
\end{center}
\vspace{1.2cm}
\caption{Left:  ${\cal R}_{3+2} \left(x_2,x_1\right) = \sigma_{3+2} 
\frac{ \frac{ d^2 \sigma_{3+2}^{(2,1)}}{d y_2 d y_1} }{\frac{d \sigma_{3+2}^{(2)}}{d y_2} \frac{d \sigma_{3+2}^{(1)}}{d y_1}}-1$ 
(the associated distribution ${\cal R}_{3+2} \left(x_3,x_2\right) = \sigma_{3+2} 
\frac{ \frac{ d^2 \sigma_{3+2}^{(3,2)}}{d y_3 d y_2} }{\frac{d \sigma_{3+2}^{(3)}}{d y_3} \frac{d \sigma_{3+2}^{(2)}}{d y_2}}-1$ is its mirror image wrt. the $y_3 = - y_2$ line). Right: ${\cal R}_{3+2} \left(x_3,x_1\right) = \sigma_{3+2} 
\frac{ \frac{ d^2 \sigma_{3+2}^{(3,1)}}{d y_3 d y_1} }{\frac{d \sigma_{3+2}^{(3)}}{d y_3} \frac{d \sigma_{3+2}^{(1)}}{d y_1}}-1$ }
\label{R3-1223}
\end{figure}
 \begin{figure}
\begin{flushleft}
\hspace{1cm}\includegraphics[width=6cm]{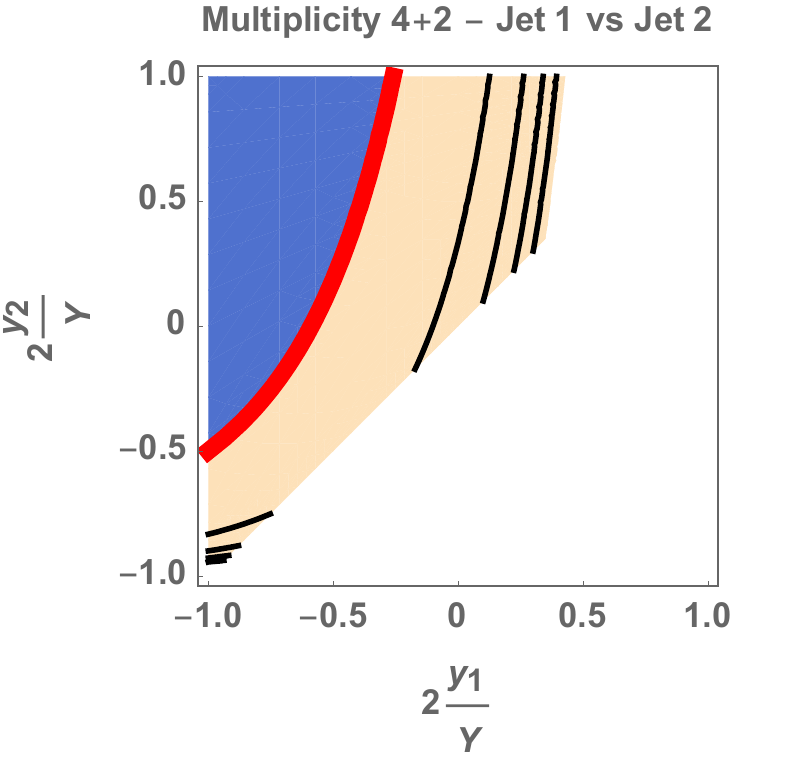}
\end{flushleft}
\vspace{-5.6cm}
\begin{flushleft}
\hspace{5.5cm}\includegraphics[width=.7cm]{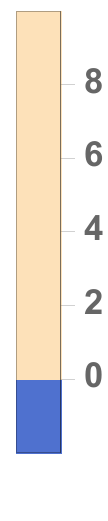}
\end{flushleft}
\vspace{-5.1cm}
\begin{center}
\hspace{7cm}\includegraphics[width=6.cm]{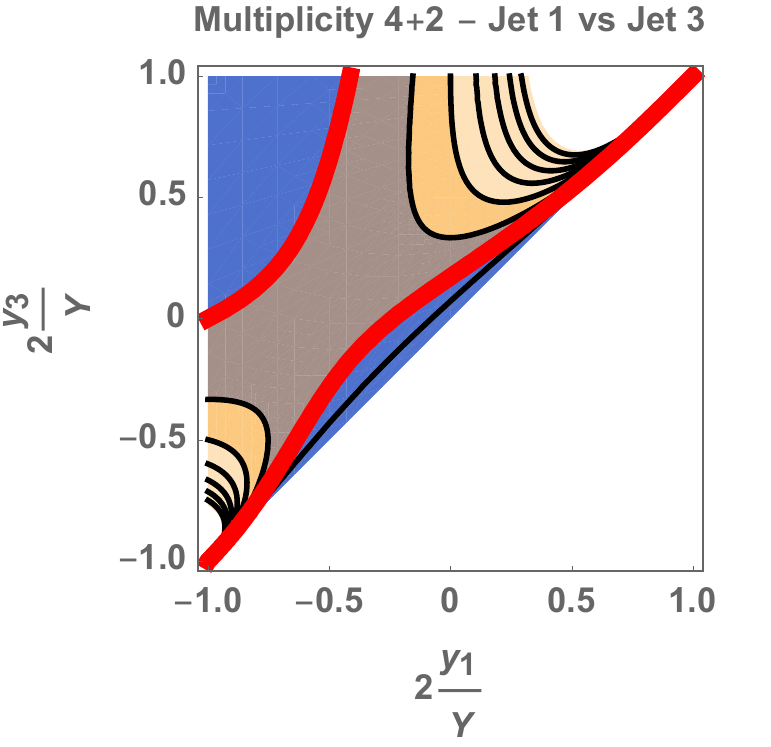}
\end{center}
\vspace{-5.4cm}
\begin{center}
\hspace{10.8cm}\includegraphics[width=.7cm]{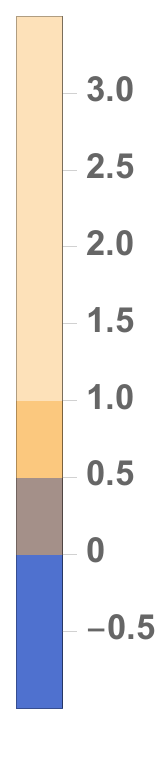}
\end{center}
\vspace{.6cm}
\caption{Left:  ${\cal R}_{4+2} \left(x_2,x_1\right) = \sigma_{4+2} 
\frac{ \frac{ d^2 \sigma_{4+2}^{(2,1)}}{d y_2 d y_1} }{\frac{d \sigma_{4+2}^{(2)}}{d y_2} \frac{d \sigma_{4+2}^{(1)}}{d y_1}}-1$ (the associated distribution  ${\cal R}_{4+2} \left(x_4,x_3\right) = \sigma_{4+2} 
\frac{ \frac{ d^2 \sigma_{4+2}^{(4,3)}}{d y_4 d y_3} }{\frac{d \sigma_{4+2}^{(4)}}{d y_4} \frac{d \sigma_{4+2}^{(3)}}{d y_3}}-1$ is its mirror image wrt.  to the line $y_4 = -y_3$). Right:  ${\cal R}_{4+2} \left(x_3,x_1\right) = \sigma_{4+2} 
\frac{ \frac{ d^2 \sigma_{4+2}^{(3,1)}}{d y_3 d y_1} }{\frac{d \sigma_{4+2}^{(3)}}{d y_3} \frac{d \sigma_{4+2}^{(1)}}{d y_1}}-1$ (the associated distribution ${\cal R}_{4+2} \left(x_4,x_2\right) = \sigma_{4+2} 
\frac{ \frac{ d^2 \sigma_{4+2}^{(4,2)}}{d y_4 d y_2} }{\frac{d \sigma_{4+2}^{(4)}}{d y_4} \frac{d \sigma_{4+2}^{(2)}}{d y_2}}-1$ is its mirror image wrt.  the line $y_4 = -y_2$).}
\label{R4-1234}
\end{figure}

\begin{figure}
\begin{subfigure}{.5\textwidth}
  \centering
  \includegraphics[width=.7\linewidth]{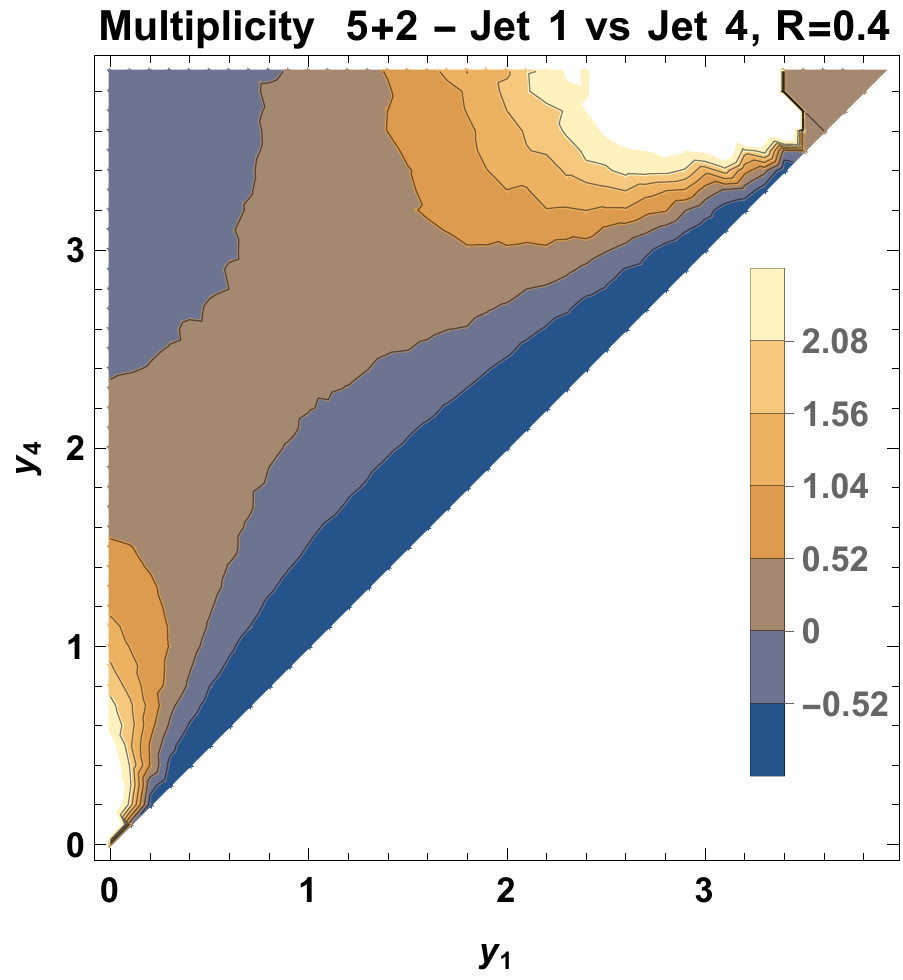}
  \caption{}
  \label{fig:sfig1}
\end{subfigure}%
\begin{subfigure}{.5\textwidth}
  \centering
  \includegraphics[width=.7\linewidth]{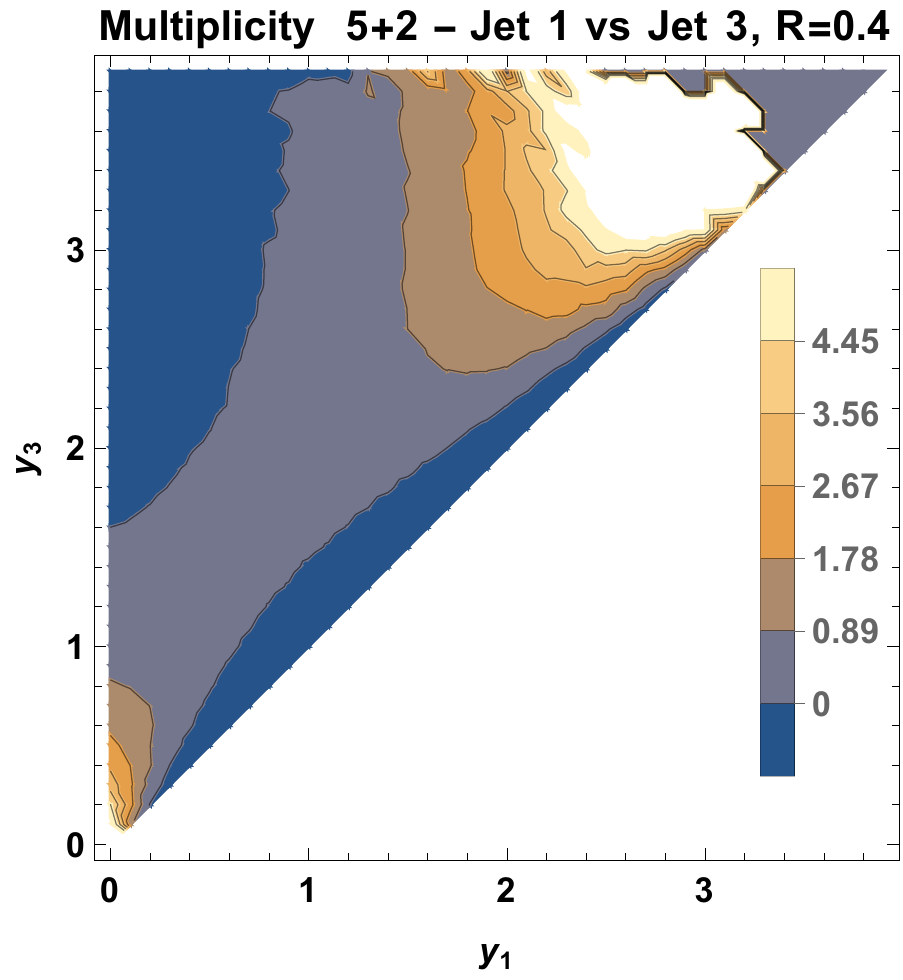}
  \caption{}
  \label{fig:sfig2}
\end{subfigure}
\\
\begin{subfigure}{.5\textwidth}
  \centering
  \includegraphics[width=.7\linewidth]{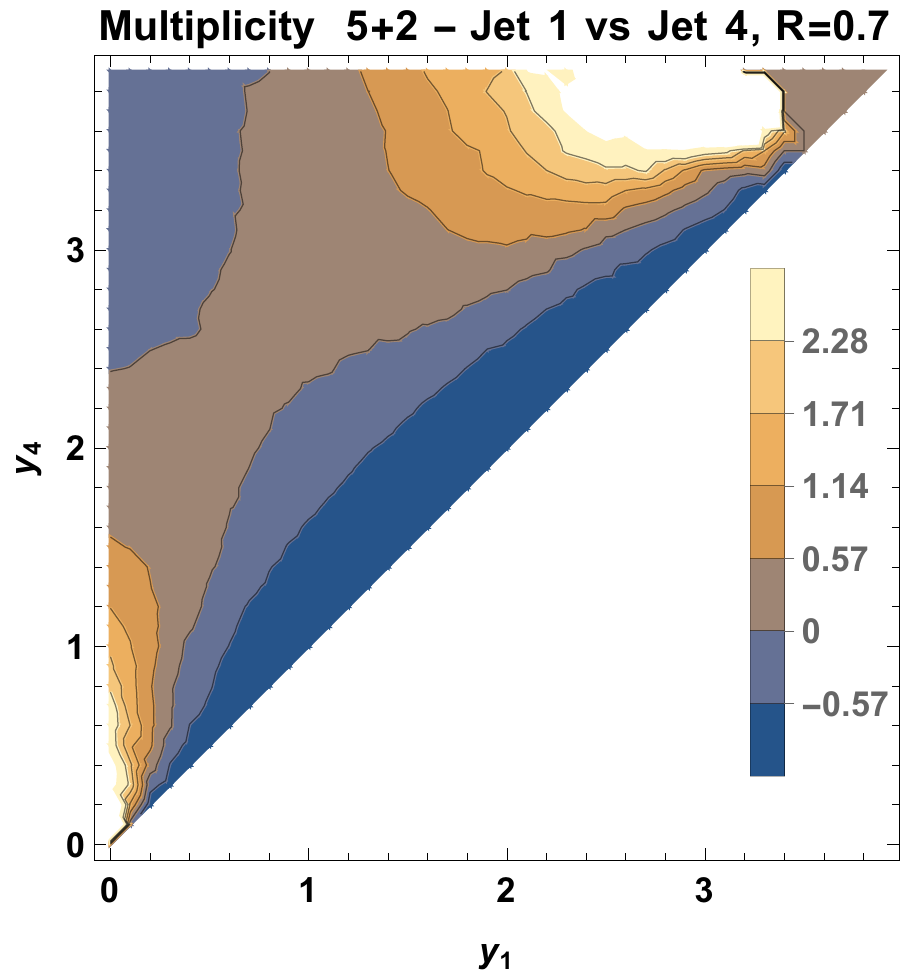}
  \caption{}
  \label{fig:sfig1}
\end{subfigure}%
\begin{subfigure}{.5\textwidth}
  \centering
  \includegraphics[width=.7\linewidth]{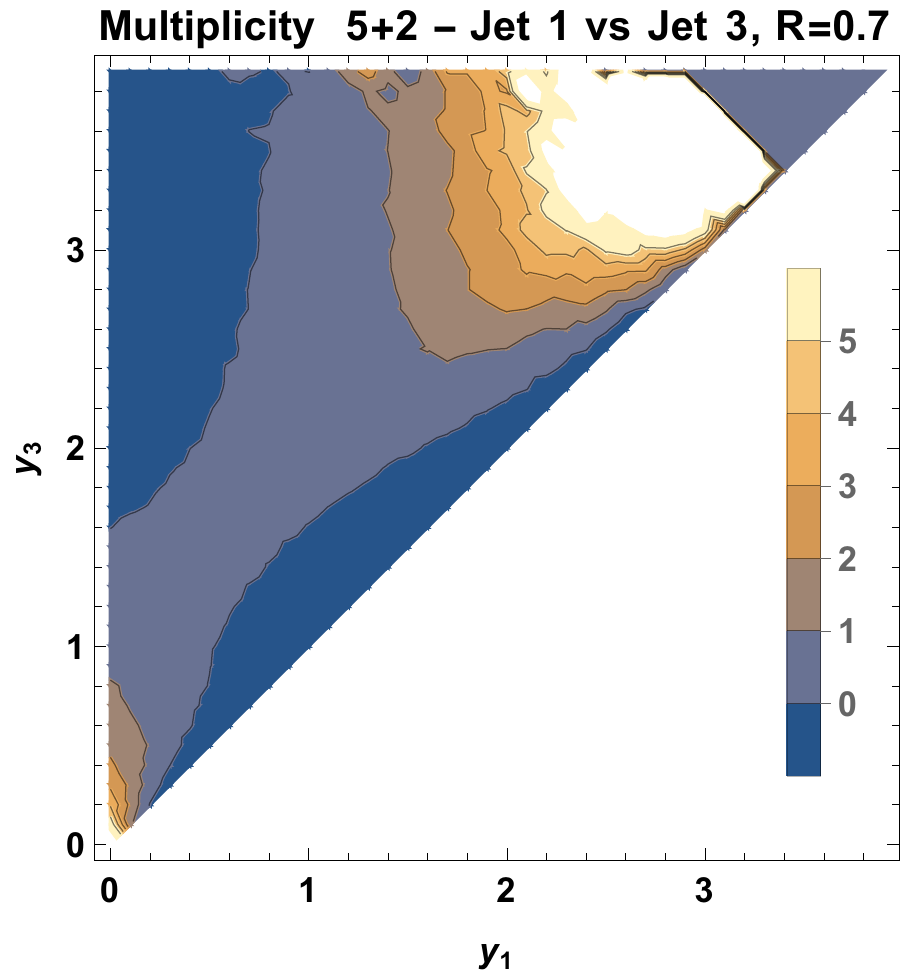}
  \caption{}
  \label{fig:sfig2}
\end{subfigure}
\caption{Top: The correlation functions of Fig.~\ref{R7-1425} with the collinear BFKL model and for jet radius $R = 0.4.$
Bottom: The same but for jet radius $R = 0.7$.}
\label{fig:coll1}
\end{figure}

\begin{figure}
\begin{subfigure}{.5\textwidth}
  \centering
  \includegraphics[width=.7\linewidth]{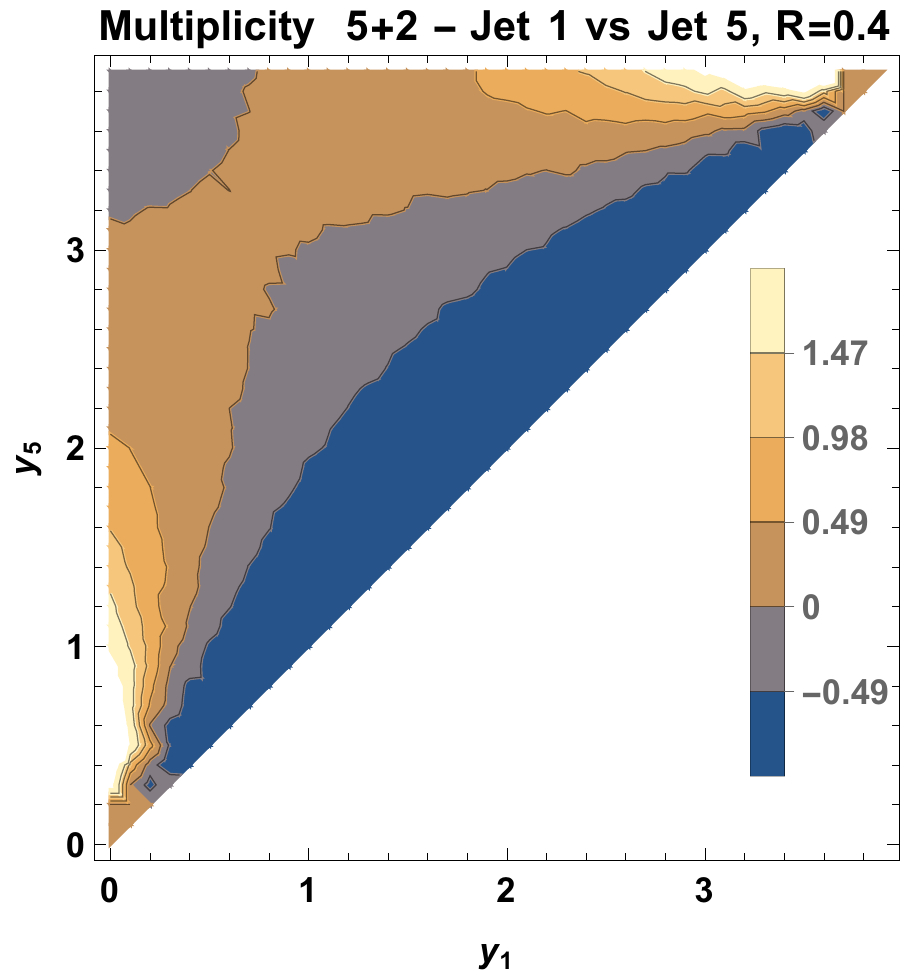}
  \caption{}
  \label{fig:sfig1}
\end{subfigure}%
\begin{subfigure}{.5\textwidth}
  \centering
  \includegraphics[width=.7\linewidth]{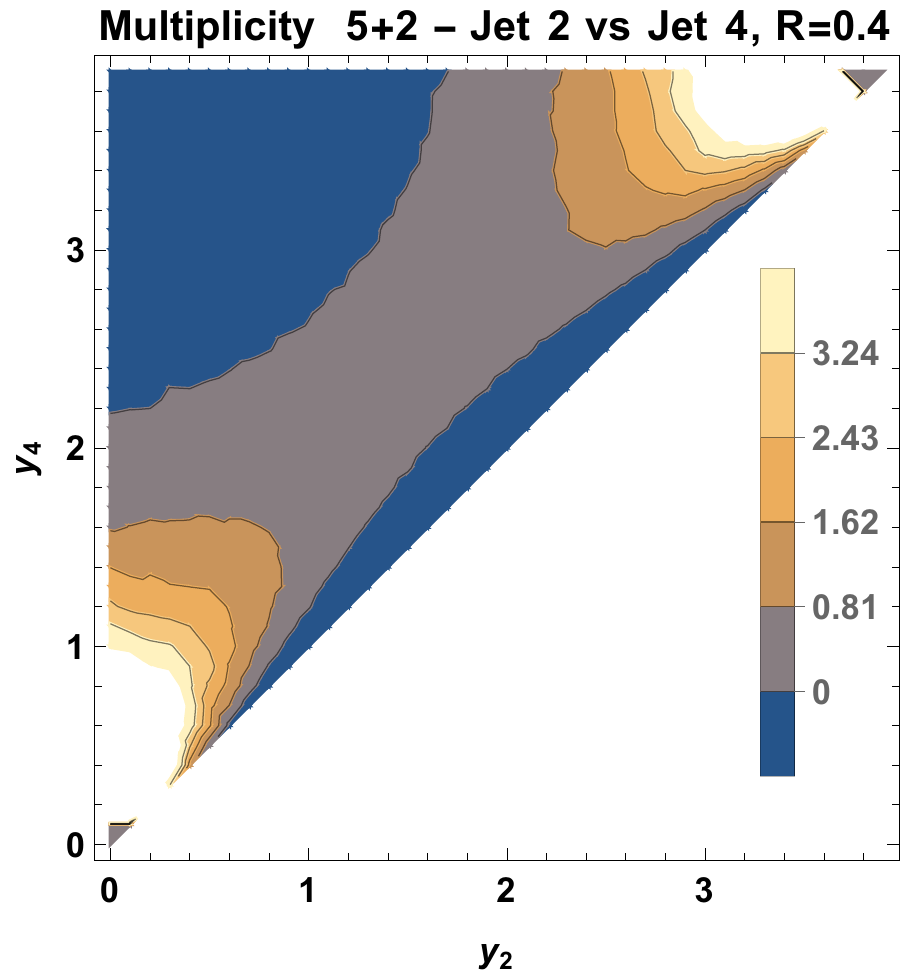}
  \caption{}
  \label{fig:sfig2}
\end{subfigure}
\\
\begin{subfigure}{.5\textwidth}
  \centering
  \includegraphics[width=.7\linewidth]{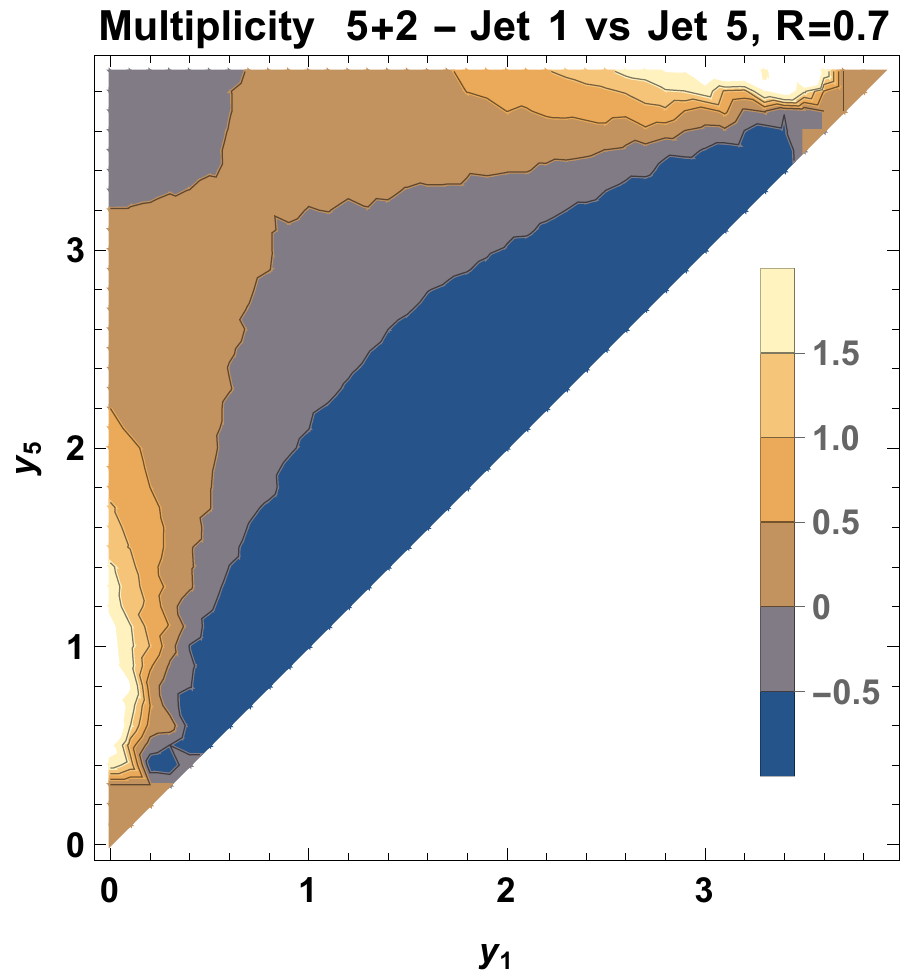}
  \caption{}
  \label{fig:sfig1}
\end{subfigure}%
\begin{subfigure}{.5\textwidth}
  \centering
  \includegraphics[width=.7\linewidth]{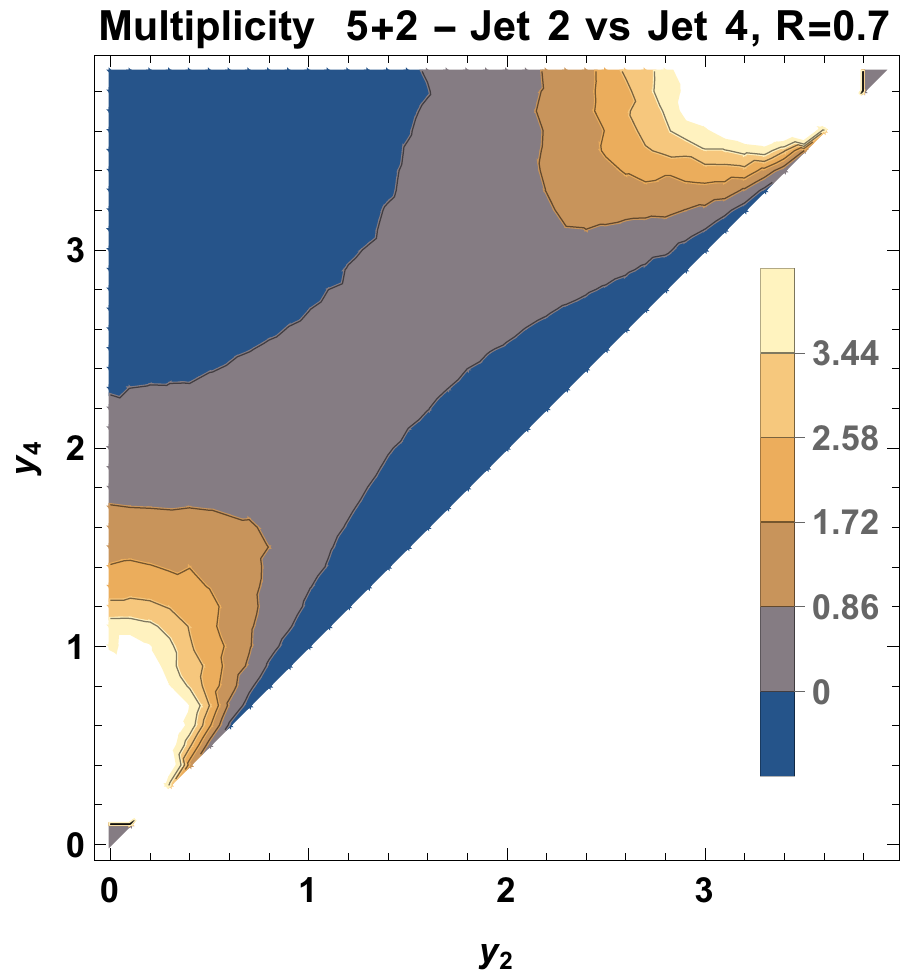}
  \caption{}
  \label{fig:sfig2}
\end{subfigure}
\caption{Top: The correlation functions of Fig.~\ref{R7-1524} with the collinear BFKL model and for jet radius $R = 0.4.$
Bottom: The same but for jet radius $R = 0.7$.}
\label{fig:coll2}
\end{figure}

This concludes our exploratory study of basic distributions which could be 
extracted from LHC data in multi-jet events. The analysis is simple since it only takes into account the longitudinal sector of the phase space and does not introduce relevant dynamics in transverse momentum space. This nevertheless should be enough to capture the gross features of the proposed observables which focus on single and double rapidity correlations. These features should be global and independent of the selected $p_T$ in the twin-jets. In future works we plan to introduce the more refined transverse space information stemming from effective theories based on the concept of clusters in multi-particle production. In this context we will investigate whether the ideas here presented are connected to predictions derived from first principles in Yang-Mills theories such as the BFKL formalism either analytically (see the next section for a treatment in the collinear region) or via the Monte Carlo code 
{\tt BFKLex}~\cite{Chachamis:2011rw,Chachamis:2011nz,Chachamis:2012fk,Chachamis:2012qw,Chachamis:2015zzp,Chachamis:2015ico}.

\section{The Chew-Pignotti model and BFKL dynamics}

In this section we would like to further motivate the use of the Chew-Pignotti model for the approximate description of the observables here discussed in a more modern language. It is not our target to evaluate them with general purpose Monte Carlo event generators since these are not optimized for the particular kinematics under study. It is more natural to evaluate these quantities within the BFKL formalism which we will show that in its simplest form is equivalent to the Chew-Pignotti model for rapidity correlations. After doing so, we will provide some kinematical cuts needed for a first  experimental investigation of our results. We leave for a future work a more complicated and detailed study of the full BFKL dynamics in this context. 

At a hadronic collider running at center-of-mass energy $\sqrt{s}$, in a leading logarithmic approximation where terms of the form ${\bar \alpha}_s^n \ln^n{s}$ are resummed (with the QCD coupling ${\bar \alpha}_s= {\bar \alpha}_s N_c  / \pi$), the differential partonic cross section for the production of two tagged jets in forward and backward directions with transverse momentum $\vec{p}_{i=1,2}$ and generated by gluons with longitudinal momentum fractions $x_{i=1,2}$ reads (with rapidity separation between the tagged jets $Y \sim \ln{x_1 x_2 s / \sqrt{\vec{p}_1^{~2} \vec{p}_2^{~2}}}$)
\begin{eqnarray}
\frac{d \hat{\sigma}}{d^2 \vec{p}_1 d^2 \vec{p}_2} &=& 
\frac{\pi^2 {\bar \alpha}_s^2}{2} \frac{f(\vec{p}_1^{~2}, \vec{p}_2^{~2},Y)}{\vec{p}_1^{~2} \vec{p}_2^{~2}} \,.
\end{eqnarray}
In the BFKL formalism, an effective field theory valid when $Y$ is large,  the gluon Green's function $f$ can be shown to follow, in a collinear approximation, the integral equation
\begin{eqnarray}
\frac{\partial f (K^2,Q^2,Y) }{\partial Y} &=& \delta (K^2-Q^2) \nonumber \\
&&\hspace{-1cm}+ \, 
 {\bar \alpha}_s \int_0^\infty d q^2 \left(\frac{\theta (K-q)}{K^2}+\frac{\theta (q-K)}{q^2} + 4 (\ln{2}-1) \delta(q^2-K^2)\right)
f (q^2,Q^2,Y) \, .
\end{eqnarray}
Its solution can be written in the iterative form
\begin{eqnarray}
f (K^2,Q^2,Y) &=& e^{4(\ln{2}-1){\bar \alpha}_s Y} \Bigg\{\delta (K^2-Q^2)    \nonumber\\
&&\hspace{-1cm}+ \,    \sum_{N=1}^\infty 
 \frac{({\bar \alpha}_s Y)^N}{N!} \left[ \prod_{L=1}^N
\int_0^\infty d x_L \left(\frac{\theta(x_{L-1}-x_L)}{x_{L-1}} + \frac{\theta(x_L-x_{L-1})}{x_{L}}\right) \right]
\delta (x_N-Q^2) \Bigg\} 
\label{FCSumMC}
\end{eqnarray}
with $x_0 = K^2$. To connect with more standard representations we use 
$\delta(K^2-Q^2)=\int \frac{d \gamma}{2 \pi i Q^2} \left(\frac{K^2}{Q^2}\right)^{\gamma-1}$ 
and 
\begin{eqnarray}
&&\int_0^\infty d x_N 
\left(
\frac{\theta(x_{N-1}-x_N)}{x_{N-1}} 
+ \frac{\theta(x_N-x_{N-1})}{x_{N}}\right) 
\left(\frac{x_N}{K^2}\right)^{\gamma-1} = \left(\frac{1}{\gamma}+\frac{1}{1-\gamma}\right) \left(\frac{x_{N-1}}{K^2}\right)^{\gamma-1},
\end{eqnarray}
valid for $0<\gamma<1$, to write
\begin{eqnarray}
\left[ \prod_{L=1}^N \int_0^\infty d x_L 
\left(\frac{\theta(x_{L-1}-x_L)}{x_{L-1}} 
+ \frac{\theta(x_L - x_{L-1})}{x_{L}}\right) \right]
\left(\frac{x_N}{K^2}\right)^{\gamma-1}  
= \left(\frac{1}{\gamma}+\frac{1}{1-\gamma}\right)^N.
\end{eqnarray}
We then have
\begin{eqnarray}
f (K^2,Q^2,Y) &=&  \frac{e^{4(\ln{2}-1) {\bar \alpha}_s Y}}{Q^2} \int \frac{d \gamma}{2 \pi i} \left(\frac{K^2}{Q^2}\right)^{\gamma-1}
 \sum_{N=0}^\infty \frac{\left({\bar \alpha}_s Y\right)^N}{N!}
\left(\frac{1}{\gamma}+\frac{1}{1-\gamma}\right)^N  \nonumber\\
&=&  \int \frac{d \gamma}{2 \pi i Q^2} \left(\frac{K^2}{Q^2}\right)^{\gamma-1} e^{{\bar \alpha}_s Y \chi\left( \gamma\right)} \, ,
\label{FCMM}
\end{eqnarray}
where $\chi(\gamma) = 4 (\ln{2}-1)+ \gamma^{-1}+ (1-\gamma)^{-1}$. 

The partonic cross section is calculated after integrating over the phase space of the two tagged jets weighted by jet vertices $\Phi_{{\rm jet }_{i}}=\Phi_{{\rm jet }_{\mathrm{i}}}^{(0)}+\bar{\alpha}_{s} \Phi_{ {\rm jet }_{\mathrm{i}}}^{(1)}+\ldots$, i.e.
\begin{eqnarray}
\hat{\sigma}\left(\alpha_{s}, \mathrm{Y}, p_{1,2}^{2}, l_{1,2}^{2}\right)=\int d^{2} \vec{q}_{1} \int d^{2} \vec{q}_{2} \Phi_{\mathrm{jet}_{1}}\left(\vec{q}_{1}, p_{1}^{2}, l_{1}^{2}\right) \Phi_{\mathrm{jet}_{2}}\left(\vec{q}_{2}, p_{2}^{2}, l_{1}^{2}\right) \frac{d \hat{\sigma}}{d^{2} \vec{q}_{1} d^{2} \vec{q}_{2}}
\, .
\end{eqnarray}
In the context of the observables here studied, at leading order, $\Phi_{\mathrm{jet}_{\mathrm{i}}}^{(0)}\left(\vec{q}, p_{i}^{2},l_{i}^{2}\right)=\theta\left(l_{i}^{2}-q^{2}\right) \theta\left(q^{2}-p_{i}^{2}\right) $ is a product of two Heaviside step functions. Since the tagged jets are produced collinearly, this parton level expression can be convoluted through the variables $x_i$ with standard parton distribution functions to generate hadron-level results. In order to reduce the influence from these pdfs and possible theoretical uncertainties, it is convenient to work with normalised quantities in a preliminary analysis.

At this point we can highlight that the BFKL formalism operates in the so-called multi-Regge kinematics for which $Y \gg 1$ and the momentum transferred at each sub-channel associated to a pair of mini-jets does not grow with energy. In terms of emitted transverse momenta it is dominated by low $p_T$ emissions. If we keep $|p_i| \sim |l_i|$ this constrains the multiparticle phase space within the class of multiperipheral final states here investigated. Therefore, for the class of two particle rapidity correlations studied in the previous sections, the predictions from BFKL should be very similar to those in the Chew-Pignotti simple model. 

{To highlight this, we have implemented the BFKL collinear model in a Monte Carlo code, fixing the transverse momenta of the twin jets to take the values 30 and 35 GeV and their rapidity separation $Y=4$. We also set the multiplicity of the minijets to be 5 and we use the anti-kt jet clustering algorithm as implemented in {\tt fastjet}~\cite{Cacciari:2011ma}. In Figs.~\ref{fig:coll1} and~\ref{fig:coll2} we present the collinear model results with jet radii $R = 0.4$ and $R = 0.7$ for the corresponding correlation functions of Figs.~\ref{R7-1425} and~\ref{R7-1524}. It is remarkable how similar the collinear model results are to the Chew-Pignotti ones as well as the fact that the actual jet radius R does not appear to have a significant impact. We decided in Figs.~\ref{fig:coll1} and~\ref{fig:coll2} to keep the proper rapidity ranges unaltered to make easier the association to a realistic experimental data setup.}

In Eq.~(\ref{FCSumMC}) each $N$ accounts for the emission of a new minijet and the three components of the kernel drive the virtuality of $t$-channel particles into infrared or ultraviolet regions with the delta function term being a non-Sudakov no-emission contribution. We can make use of Eq.~(\ref{sigma_N2}) to write
\begin{eqnarray}
f (K^2,Q^2,Y) &=&  \sum_{N=0}^\infty  {\bar \alpha}_s^{N} 
\int_{0}^{Y} \prod_{i=1}^{N+1} dz_i \delta \left(Y-
\sum_{s=1}^{N+1} z_s \right) \xi^{(N)} (K^2,Q^2) 
\end{eqnarray}
where
\begin{eqnarray}
\xi^{(N)} (K^2,Q^2) &=& \int \frac{d \gamma}{2 \pi i Q^2} \left(\frac{K^2}{Q^2}\right)^{\gamma-1}  \chi^N\left( \gamma\right) \, .
\end{eqnarray}
To single out particular emissions we apply the logic of Eqs.~(\ref{dsdy}) and 
(\ref{d2sdydy}), {\it i.e.}
\begin{eqnarray}
\frac{d f_{N}^{(l)} (K^2,Q^2,Y, y_l) }{d y_l} &=&  {\bar \alpha}_s^{N} \frac{\left(\frac{Y}{2}-y_l \right)^{N-l}}{(N-l)!} \frac{\left(y_l+\frac{Y}{2}\right)^{l-1}}{(l-1)!} \xi^{(N)} (K^2,Q^2)  \, ,  \\
 \frac{d f_{N}^{(l,m)} (K^2,Q^2,Y, y_l,y_m) }{d y_l d y_m} &=& 
 {\bar \alpha}_s^{N} \frac{\left(\frac{Y}{2}-y_l \right)^{N-l}}{(N-l)!}
\frac{(y_l-y_m)^{l-m-1}}{(l-m-1)!} 
\frac{\left(y_m+\frac{Y}{2}\right)^{m-1}}{(m-1)!} \xi^{(N)} (K^2,Q^2)   \, .
\end{eqnarray}
We can see that  for normalized quantities the 
$\xi^{(N)}$ factor cancels out and we obtain the same predictions as for 
the Chew-Pignotti model. It is needless to stress that the full BFKL formalism carries more subtle dependences in rapidity, transverse momentum and azimuthal angles which we will study in detail in future works. Nevertheless we find it interesting that in rapidity space it is not that far from the old multiperipheral cluster approaches. 

Let us conclude with a more concrete discussion of the range of parameters needed for a first experimental study of these effects in the 13 TeV data recorded by the LHC as well as the data from the previous run at 7 and 8 TeV. One should look at the relevant dijet experimental analyses for 7 TeV data from both ATLAS and CMS~\cite{Aad:2011jz,Chatrchyan:2012pb,Aad:2014pua,Khachatryan:2016udy}. There, more than one million of events per experiment covering a rapidity span of more than 9 units was selected by tagging a dijet configuration in multi-jet final states after imposing a jet veto scale $Q_0 = 20$ GeV for the crowd jets\footnote{Here we follow the terminology we introduced in the present paper.}. Any other jet apart from the dijet system contributed to the jet multiplicity if it had a $p_T$ larger than $Q_0$. The outermost jets had a bit different kinematical cuts in the two experiments but one can safely assume that in the analyses for 13 TeV they can have a $p_T$ as low as 35 GeV and probably lower, down to 20 GeV. This makes us confident that one will have good statistics when isolating events with $N=$ 3, 4, 5 and 6 jets in the final state (including the two most forward and most backward ones) although the drop in selected events for $N = 6$ will be considerable. 

\section{Conclusions}

With the advent of the 13 TeV data from the LHC at low luminosity 
numerous studies of multi-jet physics can be performed. In the present work we suggest to investigate a particular set of events 
where both a forward and a backward jet are clearly identified with similar  transverse momenta, ``twin-jets". As the available phase space increases a ``crowd" of low transverse energy mini-jet populates the gap between them. These are specific events within the class of Mueller-Navelet configurations but quite restricted in windows of $p_T$ in order to ensure that they belong to multiperipheral regions of phase space with low momentum transfer (these have been detailed at the end of section 5). We propose their experimental study since in this set up it will be possible to distinctly identify features of different multi-particle production models such as those predicted by the BFKL formalism (a discussion of its collinear limit has been introduced). 
We have performed a simple analysis based on a total decoupling of the longitudinal and transverse degrees of freedom and keeping the latter as a constant non-dynamical contribution. Within this context we have provided analytic predictions for single and double differential distributions in rapidity which can be compared with experimental results analyzed with equivalent fits to finite expansions on a basis of Chebyshev polynomials. This will provide important information on the degree of correlations of this class of 
multi-particle final states in  proton-proton interactions at very high center-of-mass energies.

Future works will necessarily include the study of observables 
proving the QCD dynamics in transverse momentum and azimuthal angle space and their role in introducing short and, possibly, long range jet-jet correlations.

\section*{Acknowledgements}

This work has been supported by the Spanish Research Agency (Agencia Estatal de Investigaci{\'o}n) through the grant IFT Centro de Excelencia Severo Ochoa SEV-2016-0597 and the Spanish Government grant FPA2016-78022-P.  It has also received funding from the European Union’s Horizon 2020 research
and innovation programme under grant agreement No. 824093. The work of GC was supported by the Funda\c{c}{\~ a}o para a Ci{\^ e}ncia e a Tecnologia (Portugal) under project CERN/FIS-PAR/0024/2019 and contract ‘Investigador FCT - Individual Call/03216/2017’.

\end{document}